\documentclass[journal]{IEEEtran}
\usepackage[keeplastbox]{flushend}
\usepackage{amsfonts}
\usepackage{amsopn}
\usepackage{amsmath}
\usepackage{mathrsfs}
\usepackage{multirow}
\usepackage{setspace}
\usepackage{algorithm}
\usepackage{algorithmic}
\usepackage{booktabs} 
\usepackage[caption=false,font=footnotesize]{subfig}
\usepackage{graphicx}
\usepackage{epstopdf}
\usepackage{booktabs}
\usepackage{boldline,multirow}
\usepackage{cite}
\usepackage{threeparttable}
\usepackage{multirow}
\usepackage{boldline}

\begin{document}
\title{Pixel-Wise PolSAR Image Classification via a Novel Complex-Valued Deep Fully Convolutional Network}

\author{\IEEEauthorblockN{Yice~Cao,
Yan~Wu,~\IEEEmembership{Member,~IEEE,}
Peng~Zhang,~\IEEEmembership{Member,~IEEE,}
Wenkai~Liang, 
and~Ming~Li,~\IEEEmembership{Member,~IEEE}}
\thanks{This work was supported in part by the Natural Science Foundation of China under Grant 61772390, Grant 61871312, and in part by the Natural Science Basic Research Plan in Shaanxi Province of China under Grant 2019JZ14. \emph{(Corresponding author: Yan Wu.)}

Y. Cao, Y. Wu and W. Liang are with the Remote Sensing Image Processing and Fusion Group, School of Electronic Engineering, Xidian University, Xi'an 710071, China (e-mail: ywu@mail.xidian.edu.cn)

P. Zhang and M. Li are with the National Key Laboratory of Radar Signal Processing, Xidian University, Xi'an 710071, China, and also with the Collaborative Innovation Center of Information Sensing and Understanding, Xidian University, Xi’an 710071, China.}}
\markboth{}%
{Shell \MakeLowercase{\textit{et al.}}: Bare Demo of IEEEtran.cls for IEEE Transactions on Magnetics Journals}
\IEEEtitleabstractindextext{%
\begin{abstract}
Although complex-valued (CV) neural networks have shown better classification results compared to their real-valued (RV) counterparts for polarimetric synthetic aperture radar (PolSAR) classification, the extension of pixel-level RV networks to the complex domain has not yet thoroughly examined. This paper presents a novel complex-valued deep fully convolutional neural network (CV-FCN) designed for PolSAR image classification. Specifically, CV-FCN uses PolSAR CV data that includes the phase information and utilizes the deep FCN architecture that performs pixel-level labeling. It integrates the feature extraction module and the classification module in a united framework. Technically, for the particularity of PolSAR data, a dedicated complex-valued weight initialization scheme is defined to initialize CV-FCN. It considers the distribution of polarization data to conduct CV-FCN training from scratch in an efficient and fast manner. CV-FCN employs a complex downsampling-then-upsampling scheme to extract dense features. To enrich discriminative information, multi-level CV features that retain more polarization information are extracted via the complex downsampling scheme. Then, a complex upsampling scheme is proposed to predict dense CV labeling. It employs complex max-unpooling layers to greatly capture more spatial information for better robustness to speckle noise. In addition, to achieve faster convergence and obtain more precise classification results, a novel average cross-entropy loss function is derived for CV-FCN optimization. Experiments on real PolSAR datasets demonstrate that CV-FCN achieves better classification performance than other state-of-art methods.    
\end{abstract}

}
\maketitle
\IEEEdisplaynontitleabstractindextext
%
\IEEEpeerreviewmaketitle

\section{Introduction}
\IEEEPARstart{P}{olarimetric} synthetic aperture radar (PolSAR) images have received a lot of attention as they can provide more comprehensive and abundant information compared with SAR images \cite{PolSAR book}. In the process of PolSAR image analysis and interpretation, PolSAR image classification is arguably rather typical and important. Until now, numerous traditional schemes have been developed for PolSAR image classification, such as Wishart classifiers \cite{Wishart94,Wishart04,Wishart-Chernoff distance}, target decompositions (TDs) \cite{review(96)-TD,Cloude(97)-TD,AWT(10)-TD,Arii(11)-TD} and random fields (RFs) \cite{MRF08,TMF,SWY-MRF}. However, these traditional methods focus on extracting features which are not only mostly low-level and hand-crafted but also involve a considerable amount of manual trial and error \cite{RF-18}. Besides, hand-engineered features such as TDs features heavily rely on the complex analysis of PolSAR data. Meanwhile, the selection of descriptive feature sets is a burden regarding computation time.

With the rapid development of learning algorithms, several machine learning tools do perform feature learning (or at least feature optimization), such as support vector machines (SVMs) \cite{SVM09,texture-HC} and random forest (RF) \cite{RF-18}. However, they are still shallow models that focus on a large number of input features and may not be robust to nonlinear data \cite{DL in RS}. Recently, deep learning (DL) has achieved remarkable results in the remote sensing community \cite{PolSAR-Fuzzy network,SAR-CD-HTNN,SAR-CNN}. Compared with the aforementioned conventional methods, DL techniques can automatically learn discriminative features and perform advanced tasks by multiple neural layers in an end-to-end manner, thereby reducing manual error and achieving promising results \cite{ohter-NN}. In recent years, some better DL-based algorithms have significantly improved the performance of PolSAR image classification, such as sparse autoencoder (SAE) \cite{RV-MLP(18)}, deep belief network (DBN) \cite{JLC-WDBN}, convolutional neural network (CNN) \cite{XF-RV-SCNN,BHX-GNN}, deep fully convolutional neural network (FCN) \cite{HC-FCN,JLC-PCN,FM-FCN}, and so on.

Notably, most studies on DL methods for PolSAR classification tasks predominantly focus on the case of real-valued neural networks (RV-NNs). In RV-NNs, input, weights, and output are all modeled as real-valued (RV) numbers. This means that projections are required to convert the PolSAR complex-valued (CV) data to RV features as RV-NNs input. Although RV-NNs have demonstrated excellent performance in PolSAR image classification tasks, there are a couple of problems stated by RV features. Firstly, it is unclear which projection yields the best performance towards a particular PolSAR image. Although the descriptive feature set generated by multi-projection has achieved remarkable results, a larger feature set will increase computing time and memory consumption, and may even cause data redundancy problems \cite{RF-18}. Secondly, projection sometimes means a loss of valuable information, especially the phase information, which may lead to unsatisfactory results. In fact, the phase of multichannel SAR data can provide useful information in the interpretation of SAR images. Especially for PolSAR systems, phase differences between polarizations have received significant attention for a couple of decades \cite{Lee(94)-phase,Lee(01)-phase,Lee(09)-phase,Turkar(12)-phase}.

In view of the aforementioned problems, some researchers have begun to investigate networks which are tailored to CV data of PolSAR images rather than requiring any projection to classify PolSAR images. Hansch et al. \cite{CV-MLP} first proposed the complex-valued MLPs (CV-MLPs) for land use classification in PolSAR images. Shang et al. \cite{Hirose-QNN(14)} suggested a complex-valued feedforward neural networks in the Poincare sphere parameter space. Moreover, an improved quaternion neural network \cite{Hirose-QNN(18)} and a quaternion autoencoder \cite{Hirose-QA} have been proposed for PolSAR land classification. Recently, a complex-valued CNN (CV-CNN) specifically designed for PolSAR image classification has been proposed by Zhang et al. \cite{XF-CV-SCNN}, where the authors derived a complex backpropagation algorithm based on stochastic gradient descent for CV-CNN training.

Although CV-NNs have achieved remarkable breakthroughs for PolSAR image classification, they still suffer some challenges. Firstly, we find that relatively deep networks architectures have not received considerable attention in the complex domain. Structures of the above CV-NNs are relatively simple with limited feature extraction layers. This results in limited learning characteristics, which may yield the risk of sub-optimal classification results. Secondly, these networks fail to sufficiently take spatial information into account to effectively reduce the impact of speckles on classification results. Due to the inherent existence of speckle in PolSAR images, the pixel-based classification accuracy is easily affected and even leads to incorrect results. In this case, those CV-NNs would be ineffective to explicitly distinguish complex classes, since only local contexts caused by small image patches are considered. Thirdly, it is necessary to construct a CV-NN for direct pixel-wise labeling to predict fast and effectively. Actually, the image classification is a dense (pixel-level) problem that aims at assigning a label to each pixel in the input image. However, existing CV-NNs usually assign an entire input image patch to a category. This results in a large amount of redundant processing and leads to seriously repetitive computation.

In response to the above challenges, this paper explores a complex-valued deep FCN architecture, which is an extension of FCN to the complex domain. The FCN is first proposed in \cite{Long(15)-FCN} and is an excellent pixel-level classifier for semantic labeling. Typically, FCN outputs a 2-dimensional (2D) spatial image and can preserve certain spatial context information for accurate labeling results. Recently, FCNs have demonstrated remarkable classification ability in the remote sensing community \cite{road(17)-FCN,RSI(17)-FCN}. However, to utilize FCN in the complex domain (i.e., CV-FCN) for PolSAR image classification, some tricky problems need to be tackled. Firstly, the CV-FCN tailored to PolSAR data requires a proper scheme for complex-valued weight initialization. Generally, FCNs are often pre-trained on VGG-16 \cite{MDZ-Visualize}, whose parameters are first trained using optical images and are all real-valued numbers. However, those parameters are not appropriate to initialize CV weights for CV-FCN and are ineffective for PolSAR images since they cannot preserve polarimetric phases information. Therefore, a proper complex-valued weight initialization scheme not only can effectively initialize CV weights but also has the potential to reduce the risks of vanishing or exploding gradients, thereby training rapidly and improving the performance of networks. Secondly, layers in the upsampling scheme of CV-FCN should be constructed in the complex domain. Although some works have extended some layers to the complex domain \cite{CV-MLP,XF-CV-SCNN,DCN(18)}, upsampling layers have not yet thoroughly examined in such domain. Finally, in the training processing of CV-FCN, it is necessary to select a loss function for CV predicted labeling. The aim is to achieve faster convergence during CV-FCN optimization and obtain higher classification accuracy. Thus, how to design a reasonable loss function in the complex domain that is suitable for PolSAR images classification needs to be solved.

In view of the above-involved limitations, we present a novel complex-valued deep fully convolutional network (CV-FCN) for classification of PolSAR imagery. The proposed deep CV-FCN adopts the complex downsampling-then-upsampling scheme to achieve pixel-wise classification results. To this end, this paper focuses on four works: 1) complex-valued weights initialization for faster PolSAR feature learning; 2) multi-level CV features extraction for enriching discriminative information; 3) more spatial information recovery for stronger speckle noise immunity; 4) average cross-entropy loss function for more precise labeling results. Specifically, CV weights for CV-FCN are first initialized by a new complex-valued weight initialization scheme. This scheme explicitly focuses on the statistical characteristic of PolSAR data for training. Thus, it  is very effective for faster training. Then, different-level CV features that retain more polarization information are extracted via the complex downsampling section. Those CV features have a powerful discriminative capacity for various classes. Subsequently, the complex upsampling section upsamples low-resolution CV feature maps and generates dense labeling. Notably, for more spatial information retaining, the complex max-unpooling layers are used in the upsampling section. Those layers recover more spatial information by the max locations maps to reduce the effect of speckles on coherent labeling results as well as improve boundary delineation. In addition, to promote CV-FCN training more effectively, an average cross-entropy loss function is employed to update CV-FCN parameters. The loss function performs cross-entropy operations on the real and imaginary parts of CV predicted labeling, respectively. In this way, the phase information is also taken into account during parameters updating, resulting in more precise classification for PolSAR images. Extensive experimental results evidently reflect the effectiveness of CV-FCN for classification of PolSAR imagery. In summary, the major contributions of this paper can be highlighted as follows: 

\begin{figure*}[ht]
\centering
\subfloat{\includegraphics[width=17.8cm,height=8.5cm]{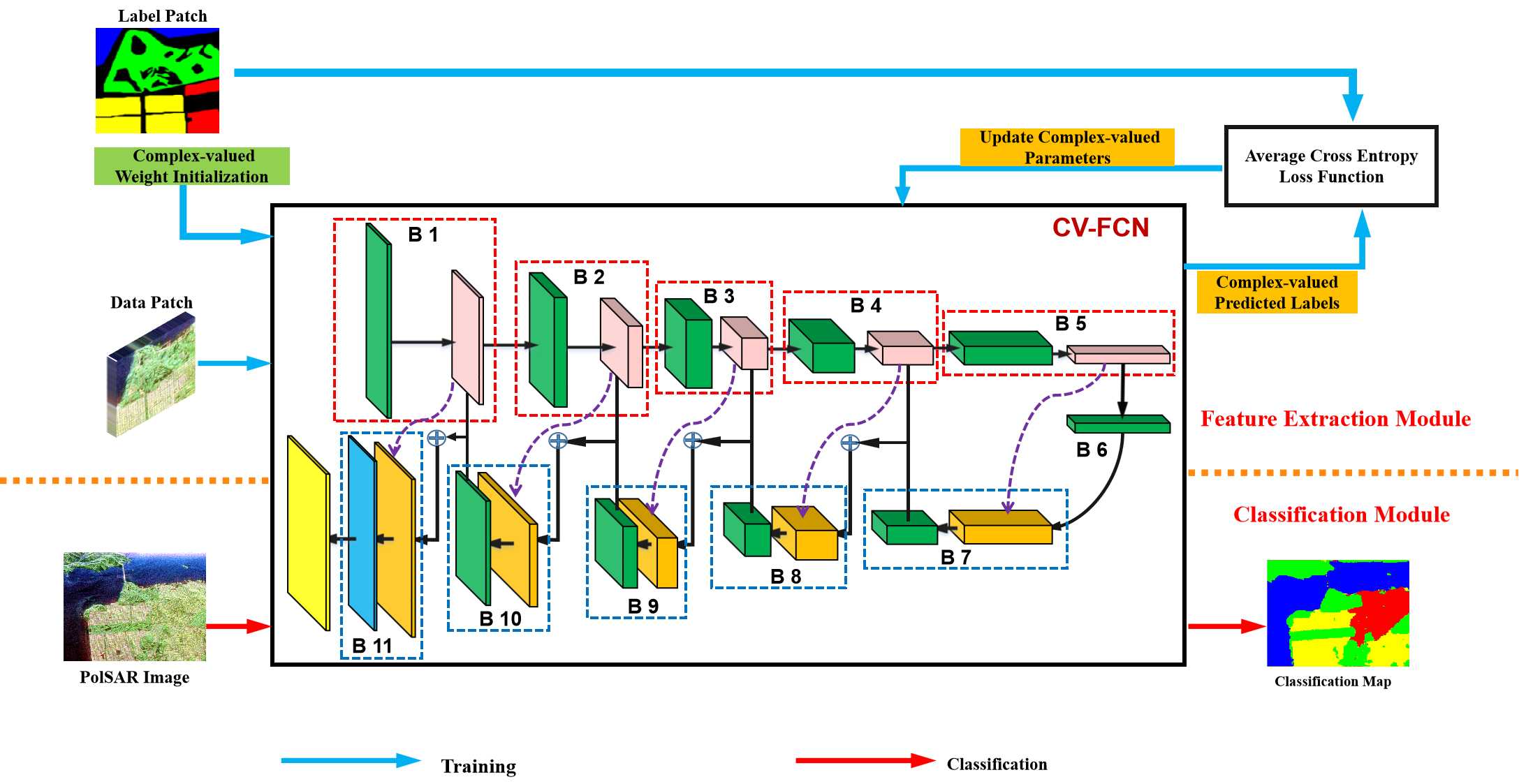}}

\subfloat{\includegraphics[width=18cm,height=1cm]{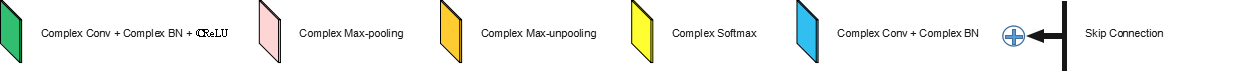}}
\caption{The deep CV-FCN framework for PolSAR image classification, where includes two modules: the feature extraction module (in the upper part) and the classification module (in the lower part). 'Complex Conv' denotes the complex convolution layer, 'Complex BN' denotes the complex batch normalization layer, '$\mathbb{C}$ReLU' denotes the complex-valued ReLU activation. The purple dotted arrows denotes the transfer of the max locations maps.}
\label{fig:CV-FCN framework}
\end{figure*}
In summary, the major contributions of this paper can be highlighted as follows:

\begin{enumerate}
\item The CV-FCN structure is proposed for PolSAR image classification, which weights, biases, input, and output are all modeled as complex values. The CV-FCN directly utilizes PolSAR CV data as input without any data projection, in which case it can extract multilevel and more robust CV features, which can retain more polarization information and have the powerful discriminative capacity for various categories.

\item A new complex-valued weight initialization scheme is employed to initialize CV-FCN parameters and conduct CV-FCN training from scratch. It allows CV-FCN to mine polarimetric features after only relatively few tuning. Thus, it can make CV-FCN training faster and save computation time.

\item A complex upsampling scheme for CV-FCN is proposed to capture more spatial information by max-unpooling layers. This scheme can not only eliminate the upsampling learning simplifying optimization but also recover more spatial information by max locations maps to reduce the impact of speckles. Thus, smoother and more coherent classification results can be achieved. 

\item A new average cross-entropy loss function in the complex domain is employed for CV-FCN optimization. It takes the phase information into account during parameters updating by average cross-entropy operation of CV predicted labels. Therefore, the new loss function enables CV-FCN optimization more precise while boosting the labeling accuracy.  
\end{enumerate}
The remainder of this paper is organized as follows. Section II formulates a detailed theory for the classification method of CV-FCN. In section III, we conduct experiments on real benchmark PolSAR images and give detailed comparisons and analyses. Finally, the conclusion and future works are discussed in Section IV.

\section{Proposed CV-FCN for Classification of PolSAR Imagery}
In this work, a deep CV-FCN is proposed to conduct PolSAR image classification. The CV-FCN method integrates the feature extraction module and the classification module in a unified framework. Thus, features extracted through CV-FCN that is trained by PolSAR data are more able to distinguish various categories for PolSAR classification tasks. In the following, we first give the framework of the deep CV-FCN classification method in Section II-A. Then, to learn more discriminative features for classification faster and more accurately, it is critical to train CV-FCN suitable for PolSAR images. Thus, we highlight and introduce four critical works for CV-FCN training in Section II-B, C, D. They include CV weight initialization, deep and multi-level CV feature extraction, more spatial information recovery, and loss function for more precise optimization. Finally, the CV-FCN classification algorithm is summarized in Section II-E.

\subsection{Framework of the Deep CV-FCN Classification Method}
The framework of CV-FCN classification method is shown in Fig. \ref{fig:CV-FCN framework}, which is composed of two separate modules: feature extraction module and classification module. In the feature extraction module, CV-FCN is trained to exploit the discriminative information. Then, the trained CV-FCN is used to classify PolSAR images in the classification module. 

The data patches set and the corresponding label patches set are first prepared as input to CV-FCN before training. The two sets are generated from the PolSAR data set and the corresponding ground truth mask, respectively. Let the CV PolSAR dataset be $\mathcal{H} \in \mathbb{C}^{h \times w \times B}$, where $h$ and $w$ are the height and width of the spatial dimensions respectively, and $B$ is the number of complex bands, $\mathbb{C}$ is the complex domain. The corresponding ground truth mask is denoted as $\mathcal{G} \in \mathbb{R}^{h \times w}$. The set of all data patches cropped from the given data set is denoted as $\mathcal{I}=\left\{I_{1}, I_{2}, \ldots, I_{n}\right\}$, and the corresponding label patches set is $\mathcal{T}=\left\{T_{1}, T_{2}, \ldots, T_{n}\right\}$, where $I_{i} \in \mathbb{C}^{H \times W \times B}$ and $T_{i} \in \mathbb{R}^{H \times W}$ ($i\in [1,n]$) represent one data patch and corresponding label patch, respectively. Here $n$ is the total number of patches, $H$ and $W$ are the patch size in the spatial dimension.

In the feature extraction module, the CV-FCN is mainly trained. A novel complex-valued weight initialization scheme is first adopted to initialize CV-FCN. Then, a certain percent of patches from the set $\mathcal{I}$ are randomly chosen as the training data patches $\mathcal{I}_{train}$ to the network. These data patches are forward-propagated through the complex downsampling section of CV-FCN [marked by red dotted boxes in Fig. \ref{fig:CV-FCN framework}] to extract multi-level CV feature maps. Then those low-resolution feature maps are upsampled by the complex upsampling section [marked by blue dotted boxes in Fig. \ref{fig:CV-FCN framework}] to generate predicated label patches. Subsequently, calculate the error between predicated label patches and the corresponding label patches $\mathcal{T}_{train}$ according to a novel loss function, and then iteratively updating CV parameters in CV-FCN. According to some certain conditions, the updating iteration will terminate when the error value does not substantially change.

In the classification module, we feed the entire PolSAR dataset $\mathcal{H}$ to the trained network. The label of every pixel in this PolSAR image is predicted based on the output of the last complex softmax layer. Notably, compared with a CNN model which predicts a single label for the center of each image patch, the CV-FCN model can predict all pixels in the entire image at one time. Thus, this enables pixel-level labeling and can decrease the computation time during the prediction.
\begin{table}[!t] 
\scriptsize
\renewcommand\arraystretch{1.2}  
\fontsize{5}{7}\selectfont
\centering
\caption{Detailed Configuration of the CV-FCN. $K$ Denotes the Total \protect\\ Number of Classes. The Complex BN Layers and $\mathbb{C}$ReLU\protect\\ Layers are Omitted for Brevity}
\label{tab:CV-FCN configuration}
\begin{tabular}{|c|c|c|c|c|c|}
\hline
\bf Section & \bf Block & \bf Module \bf type & \bf Dimension & \bf Stride & \bf Pad \\
\hline
\multirow{10}{*}{\shortstack{Downsampling\\Section}} & \multirow{2}{*}{B 1} & Complex Convolution & 3$\times$3$\times$12$\times$12 & 1 & 1 \\ \clineB{3-6}{1}
&  & Complex Max-Pooling & 2$\times$2 & 2 & 0 \\ \clineB{2-6}{1}

& \multirow{2}{*}{B 2} & Complex Convolution & 3$\times$3$\times$12$\times$24 & 1 & 1 \\ \clineB{3-6}{1}
&  & Complex Max-Pooling & 2$\times$2 & 2 & 0 \\ \clineB{2-6}{1}

& \multirow{2}{*}{B 3} & Complex Convolution & 3$\times$3$\times$24$\times$48 & 1 & 1 \\ \clineB{3-6}{1}
&  & Complex Max-Pooling & 2$\times$2 & 2 & 0 \\ \clineB{2-6}{1}

& \multirow{2}{*}{B 4} & Complex Convolution & 3$\times$3$\times$48$\times$96 & 1 & 1 \\ \clineB{3-6}{1}
&  & Complex Max-Pooling & 2$\times$2 & 2 & 0 \\ \clineB{2-6}{1}

& \multirow{2}{*}{B 5} & Complex Convolution & 3$\times$3$\times$96$\times$192 & 1 & 1 \\ \clineB{3-6}{1}
&  & Complex Max-Pooling & 2$\times$2 & 2 & 0 \\ \clineB{1-6}{1}

& \multirow{1}{*}{B 6} & Complex Convolution & 1$\times$1$\times$192$\times$192 & 1 & 1 \\
\hline

\multirow{10}{*}{\shortstack{Upsampling\\Section}} & \multirow{2}{*}{B 7} & Complex Up-Pooling & 2$\times$2 & 1 & 0 \\ \clineB{3-6}{1}
&  & Complex Convolution & 3$\times$3$\times$192$\times$96 & 1 & 1 \\ \clineB{2-6}{1}

& \multirow{2}{*}{B 8} & Complex Up-Pooling & 2$\times$2 & 1 & 0 \\ \clineB{3-6}{1}
&  & Complex Convolution & 3$\times$3$\times$96$\times$48 & 1 & 1 \\ \clineB{2-6}{1}

& \multirow{2}{*}{B 9} & Complex Up-Pooling & 2$\times$2 & 1 & 0 \\ \clineB{3-6}{1}
&  & Complex Convolution & 3$\times$3$\times$48$\times$24 & 1 & 1 \\ \clineB{2-6}{1}

& \multirow{2}{*}{B 10} & Complex Up-Pooling & 2$\times$2 & 1 & 0 \\ \clineB{3-6}{1}
&  & Complex Convolution & 3$\times$3$\times$24$\times$12 & 1 & 1 \\ \clineB{2-6}{1}

& \multirow{2}{*}{B 11} & Complex Up-Pooling & 2$\times$2 & 1 & 0 \\ \clineB{3-6}{1}
&  & Complex Convolution & 3$\times$3$\times$12$\times$2$K$ & 1 & 1 \\ \clineB{1-6}{1}

& \multirow{1}{*}{} & Complex Softmax & & & \\

\hline
\end{tabular}
\end{table}
\begin{figure*}[ht]
\centering
\subfloat{\includegraphics[width=14cm,height=5cm]{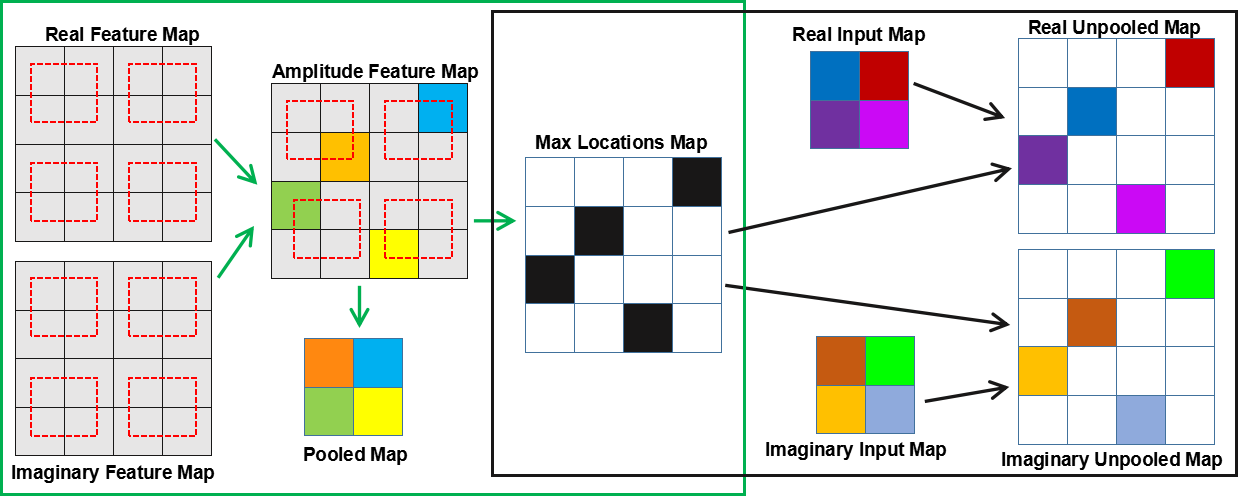}}
\caption{An simple illustration of complex max-pooling operator and complex max-unpooling operator. Where the green box and the black box are the structures of the complex max-pooling operator and the complex max-unpooling operator, respectively.}
\label{fig:max pooling layer}
\end{figure*} 
\subsection{New Complex-valued Weight Initialization Scheme Using Polarization Data Distribution for Faster Feature Learning}
The CV-FCN architecture for PolSAR image classification task has been systematically built, the weight initialization problem will arise when training the network. Generally, deep ConvNets can update from pre-trained weights generated by the transfer learning technique. However, those weights are all real-valued numbers and only reflect the backscattering intensities, while the loss of the polarimetric phase \cite{WW-convnet}. Here, based on the distributions of polarization data, a new complex-valued (CV) weight initialization scheme is employed for faster network learning.

For RV networks that process PolSAR images, learned weights, commonly known as kernels, can well characterize scattering patterns, particularly in high-level layers \cite{XF-RV-SCNN}. In \cite{XF-CV-SCNN}, the initialization scheme is just to initialize the real and imaginary parts of a CV weight separately with a uniform distribution. Fortunately, for a reciprocal medium, a complex scattering vector $\mathrm{u}=\left[S_{1}, S_{2}, S_{3}\right]^{T}$ can be modeled by a multivariate complex Gaussian distribution, where individual complex scattering coefficient $S_{i}(i=1,2,3)$ of $\mathrm{u}$ is assumed to have complex Gaussian distribution \cite{PolSAR book}. Thus, we utilize this distribution to initialize complex weights in CV-FCN.

Suppose that a CV weight is denoted as $\mathrm{W}=\mathfrak{R}(\mathrm{W})+j . \mathfrak{J}(\mathrm{W})$, where the real component $\mathfrak{R}(\mathrm{W})$ and the imaginary component $\mathfrak{J}(\mathrm{W})$ are all identically Gaussian distributed with 0 mean and variance $\sigma^{2} / 2$. Here, the initialization criterion proposed by He et al. \cite{He(15)-criterion} is used to calculate the variance of $\mathrm{W}$, i.e., $\operatorname{Var}(\mathrm{W})=2 / n_{i n}$, where $n_{i n}$ is the number of input units, since this criterion provides the current best practice when the activation function is ReLU.

Notably, the CV weight can also be denoted as $\mathrm{W}=|\mathrm{W}| e^{j \theta}$, where the magnitude $|\mathrm{W}|$ follows the Rayleigh distribution. The expectation and the variance are given by
\begin{equation}
\mathbb{E}[|\mathrm{W}|]=\frac{\sigma \sqrt{\pi}}{2},
\label{W-expectation}
\end{equation}
\begin{equation}
\operatorname{Var}(|\mathrm{W}|)=\frac{4-\pi}{2} \sigma^{2},
\label{W-variance}
\end{equation}
where $\sigma$ is the single parameter in the Rayleigh distribution. In addition, the variance $\operatorname{Var}(|\mathrm{W}|)$ and the variance $\operatorname{Var}(\mathrm{W})$ can be defined as
\begin{equation}
\begin{aligned} \operatorname{Var}(|\mathrm{W}|) &=\mathbb{E}\left[|\mathrm{W}||\mathrm{W}|^{*}\right]-(\mathbb{E}[|\mathrm{W}|])^{2} \\ &=\mathbb{E}\left[|\mathrm{W}|^{2}\right]-(E[|\mathrm{W}|])^{2} \end{aligned},
\end{equation}
\begin{equation}
\begin{aligned} \operatorname{Var}(\mathrm{W}) &=E\left[\mathrm{WW}^{*}\right]-(E[\mathrm{W}])^{2} \\ &=E\left[|\mathrm{W}|^{2}\right]-(E[\mathrm{W}])^{2} \end{aligned}.
\end{equation}

According to the initialization rules of \cite{DCN(18)}, in the case of $\mathrm{W}$ symmetrically distributed around 0, $\operatorname{Var}(\mathrm{W})=\mathbb{E}\left[|\mathrm{W}|^{2}\right]$. Thus, $\operatorname{Var}(\mathrm{W})$ can be formulated as
\begin{equation}
\operatorname{Var}(\mathrm{W})=\operatorname{Var}(|\mathrm{W}|)+(\mathbb{E}[|\mathrm{W}|])^{2}.
\end{equation}

Taking Equation (\ref{W-expectation}) and Equation (\ref{W-variance}) into account, $\operatorname{Var}(\mathrm{W})$ is calculated as
\begin{equation}
\operatorname{Var}(\mathrm{W})=\left(\frac{\sigma \sqrt{\pi}}{2}\right)^{2}+\frac{4-\pi}{2} \sigma^{2}=2 \sigma^{2}.
\label{val-W}
\end{equation}

According to He’s initialization criterion and Equation (\ref{val-W}), the single parameter in the Rayleigh distribution can be computed as $\sigma=\sqrt{2/n_{i n}}$. At this point, the Rayleigh distribution can be used to initialize the amplitude $|\mathrm{W}|$. In addition, the phase $\theta$ is initialized by using the uniform distribution between $-\pi$ and $\pi$. Thus, the initialization of the complex weight is finished.

It is worth noting that our initialization scheme is quite different from the random initialization on both the real and imaginary parts of a CV weight \cite{XF-CV-SCNN}. The most notable superiority of the new initialization scheme will be explicitly focusing on the statistical characteristic of training data, which makes it possible to learn a CV network suitable for PolSAR images after a small amount of fine-tuning. We can understand that the network exhibits some of the same properties as the data to be learned at the beginning, which seems to give a priori rather than the initial random information. Thus, it is possible to increase the potential chance of learning some special property of PolSAR datasets and is much effective for faster training.

\subsection{Deep CV-FCN for Dense Feature Extraction} 
\label{sec:CV-FCN introduce}
In the forward propagation of CV-FCN training procedure, dense features are extracted through the complex downsampling-then-upsampling scheme. The detailed configuration of CV-FCN is shown in Table \ref{tab:CV-FCN configuration}. The complex downsampling section first extracts effective multi-level CV features through downsampling blocks (i.e., B1-B5 in Fig. \ref{fig:CV-FCN framework}). Then, the complex upsampling section recovers more spatial information in a simple manner and produces dense labeling through a series of upsampling blocks (i.e., B7-B11 in Fig. \ref{fig:CV-FCN framework}). In particular, fully skip connections between the complex downsampling section and the complex upsampling section fuse shallow, fine features and deep, coarse features to preserve sufficient detailed information for complex classes distinction.

\subsubsection{Multi-level Complex-valued Feature Extraction via the Complex Downsampling Section}
\label{subsec:forward}
\ The complex downsampling section consisting of downsampling blocks extracts 2-D CV features of different levels. In CV-FCN, five downsampling blocks are employed to extract more abstract and extensive features. Each of them contains four layers, including a complex convolution layer, a complex batch normalization layer, a complex activation layer, and a complex max-pooling layer. Among these layers, the main feature extraction work is performed in the complex convolution layer. Compared with the real convolution layer, it extracts CV features retaining more polarization information and discriminative information through the complex convolution operation.

In the $l$th complex convolution layer, given complex filters $\mathrm{W}^{l} : \mathrm{W}^{l} \in \mathbb{C}^{H \times H \times M_{l-1} \times N_{l}}$ and complex bias $b^{l} \in \mathbb{C}^{N_{l}}$, where $M_{l-1}$ is the number of input channel and $N_{l}$ is the number of output channel. The output complex feature maps $\mathrm{Y}^{l} \in \mathbb{C}^{H_{l} \times M_{l} \times N_{l}}$ outputted by the complex convolution layer is computed by
\begin{equation}
Y^{l}=W^{l} \otimes X^{l-1}+b^{l},
\end{equation}
where $X^{l-1} \in \mathbb{C}^{H_{l-1} \times W_{l-1} \times M_{l-1}}$ is the given input complex feature maps, and ${H_{l-1} \times W_{l-1}}$ is the input feature map size. $\otimes$ is the convolution operation in the complex domain.  The matrix notation of the $n$th output complex feature map $Y_{n}^{l}\left(n \in\left[1, N_{l}\right]\right)$ is given by 
\begin{footnotesize}
\begin{equation}
\begin{aligned}
\left[ \begin{array}{l}{\mathfrak{R}\left(Y_{n}^{l}\right)} \\ {\mathfrak{I}\left(Y_{n}^{l}\right)}\end{array}\right]&\\ 
&=\sum_{m=1}^{M_{l-1}} \left\{\left[ \begin{array}{cc}{\mathfrak{R}\left(\mathrm{W}_{n m}^{l}\right)} & {-\mathfrak{J}\left(\mathrm{W}_{n m}^{l}\right)} \\ {\mathfrak{J}\left(\mathrm{W}_{n m}^{l}\right)} & {\mathfrak{R}\left(\mathrm{W}_{n m}^{l}\right)}\end{array}\right] \odot \left[ \begin{array}{l}{\mathfrak{R}\left(X_{m}^{l-1}\right)} \\ {\mathfrak{I}\left(X_{m}^{l-1}\right)}\end{array}\right] \right\}&\\
&+\left[ \begin{array}{l}{\mathfrak{R}\left(b_{n}^{l}\right)} \\ {\mathfrak{I}\left(b_{n}^{l}\right)}\end{array}\right],
\end{aligned}
\end{equation}
\end{footnotesize}
where $\mathrm{W}_{n m}^{l}=\mathfrak{R}\left(\mathrm{W}_{n m}^{l}\right)+j.\mathfrak{I}\left(\mathrm{W}_{n m}^{l}\right)$, and $\mathfrak{R}\left(\mathrm{W}_{n m}^{l}\right)$ and $\mathfrak{I}\left(\mathrm{W}_{m n}^{l}\right)$ are respectively the real part and the imaginary part of $\mathrm{W}_{n m}^{l}$. $\odot$ is the convolution operation in the real domain. Thus the $n$th output complex feature map can be represented as
\begin{equation}
Y_{n}^{l}=\mathfrak{R}\left(Y_{n}^{l}\right)+j.\mathfrak{I}\left(Y_{n}^{l}\right).
\end{equation}

The complex batch normalization (BN) layer \cite{DCN(18)} is performed for normalization after complex convolution, which holds great potential to relieve networks from overfitting. For the non-linear transformation of CV features, we find that the complex-valued ReLU ($\mathbb{C}$ReLU) as the complex activation can provide us good results. The $\mathbb{C}$ReLU is defined as
\begin{equation}
\mathbb{C} \operatorname{Re} \operatorname{LU}(x)=\operatorname{Re} \operatorname{LU}(\mathfrak{R}(x))+j.\operatorname{Re} \operatorname{LU}(\mathfrak{I}(x)),
\end{equation}
where $(x=\mathfrak{R}(x)+j.\mathfrak{I}(x)) \in \mathbb{C}$. Then the output $Z^{l+1} \in \mathbb{C}^{H_{l} \times W_{l} \times N_{l}}$ in the $(l+1)$th complex nonlinear layer can be given
\begin{equation}
\begin{aligned} Z^{l+1} &=\mathbb{C} \operatorname{Re} \operatorname{LU}\left(Y^{l}\right) \\ &=\mathbb{C} \operatorname{Re} \operatorname{LU}\left[\mathfrak{R}\left(Y^{l}\right)+j.\mathfrak{I}\left(Y^{l}\right)\right] \\ &=\operatorname{Re} \operatorname{LU}\left(\mathfrak{R}\left(Y^{l}\right)\right)+j.\operatorname{Re} \operatorname{LU}\left(\mathfrak{I}\left(Y^{l}\right)\right) \end{aligned}.
\end{equation}

Furthermore, the complex max-pooling layer \cite{XF-CV-SCNN} is adopted to generalize features into a higher level. In this way, features are more robust and CV-FCN can converge well. After five downsampling blocks, the block 6 (B6 in Fig. \ref{fig:CV-FCN framework}) including a complex convolution layer with 1$\times$1 kernels and a complex batch normalization layer densifies its sparse input and extracts complex convolution features.

\subsubsection{Using Complex Upsampling Section for More Spatial Information Recovery to stronger Speckle Noise Immunity}
\label{subsec:upsample}
After the complex downsampling section for multi-level CV features extraction, a complex upsampling section is implemented to upsample those CV feature maps. Specifically, the new complex max-unpooling layers are employed in the complex upsampling section. The reason is two-fold. On the one hand, compared with the complex deconvolution layer which is another upsampling operation, the complex max-unpooling layer reduces the number of trainable parameters and mitigates information loss due to complex pooling operations. On the other hand, owing to the inherent existence of speckle in PolSAR images, obtaining smooth labeling results is not easy. This issue can be addressed by the complex max-unpooling layer that recovers more spatial information by the max locations maps [represented by purple dotted arrows in Fig. \ref{fig:CV-FCN framework}]. The spatial information is a critical indicator for confusing categories classification, which captures more wider visual cues to stronger speckle noise immunity.

To be more intuitive, Fig. \ref{fig:max pooling layer} illustrates an example of the complex max-unpooling operation. The green and black boxes are simple structures of the complex max-pooling and complex max-unpooling, respectively. As shown in the green box, the amplitude feature map is formed by the real and imaginary feature maps where the red dotted box represents 2$\times$2 pooling window with a stride of 2. In the amplitude feature map, four maximum amplitude values are chosen by corresponding pooling windows which are marked by orange, blue, green, and yellow, respectively. They construct the pooled map. At the same time, locations of those maxima are recorded in a set of switch variables which is visualized by the so-called max locations map. On the other hand, within the black box, the real and imaginary input maps are upsampled by the usage of this max locations map, respectively. Then the real and imaginary unpooled maps are produced. Here, those unpooled maps are sparse wherein white regions have the values of 0. This will ensure that the resolution of the output is higher than the resolution of its input.

In particular, we perform fully skip connections which can fuse multi-level features to preserve sufficient discriminative information for the classification of complex classes. Finally, the complex output layer with the complex softmax function is used to calculate the prediction probability map. Thus, the output of CV-FCN can be formulated as 
\begin{equation}
\begin{aligned} 
O^{l}&=g\left(Y^{l-1}\right)\\
&=g\left(\mathfrak{R}\left(Y^{l-1}\right)\right)+j.g\left(\mathfrak{I}\left(Y^{l-1}\right)\right)\\
&=\frac{1}{1+e^{-\mathfrak{R}\left(Y^{l-1}\right)}}+j.\frac{1}{1+e^{-\mathfrak{I}\left(Y^{l-1}\right)}},
\end{aligned}
\end{equation}
where $Y^{l-1} \in \mathbb{C}^{H \times W \times K}$ is the inputs of the complex output layer, $g(.)$ is the softmax function in the real domain. In this layer, output feature maps are the same size as the data cubic fed into CV-FCN. This enables pixel-to-pixel training. After the complex downsampling section and the complex upsampling section, the complex forward propagation process of the training phase is completed.
\subsection{Average Cross-entropy Loss Function for Precise CV-FCN Optimization}
\label{sec:loss}
To promote CV-FCN training more effectively and achieve more precise results, a novel loss function is used as the learning objective to iteratively update CV-FCN parameters $\Theta$ during the backpropagation. $\Theta$ includes $\mathrm{W}^{l}$ and ${b}^{l}$. Usually, for multi-class classification tasks, the cross-entropy loss function performs well to update parameters. Compared with the quadratic cost function, it can increase the training speed and promote the training of NNs more effectively. Thus, a novel average cross-entropy loss function is employed for CV predicted labels in PolSAR classification tasks, which is based on the definition of the popular cross-entropy loss function. Formally, the average cross-entropy (ACE) loss function is defined as
\begin{footnotesize}
\begin{equation}
\begin{aligned} 
J^{ACE}=-\frac{1}{2} \cdot \frac{1}{N} \cdot \sum_{n=1}^{N} &\left\{\left[\mathfrak{R}\left(R[n]\right) \cdot \ln \left(\mathfrak{R}\left(O[n]\right)\right)\right.\right.&\\
&+\left.\left(1-\mathfrak{R}\left(R[n]\right)\right) \cdot \ln \left(1-\mathfrak{R}\left(O[n]\right)\right)\right] \\ &+\left[\mathfrak{I}\left(R[n]\right) \cdot \ln \left(\mathfrak{I}\left(O[n]\right)\right)\right.&\\
&\left.\left.+\left(1-\mathfrak{I}\left(R[n]\right)\right) \cdot \ln \left(1-\mathfrak{I}\left(O[n]\right)\right)\right]\right\}, 
\end{aligned}
\end{equation}
\end{footnotesize}
where $O \in \mathbb{C}^{H \times W \times K}$ indicates the output data cubic in the last complex softmax layer and $K$ is the total number of classes. $R \in \mathbb{C}^{H \times W \times K}$ is the sparse representation of the true label patch $T \in \mathbb{R}^{H \times W}$, which is converted by one-hot encoding. Notably, non-zeros positions within $R \in \mathbb{C}^{H \times W \times K}$ are $1+1 \cdot j$ instead of $1+0 \cdot j$. This means that we also take the phase information into account during parameters updating. As a result, the updated CV-FCN can work effectively, leading to more precise classification results for PolSAR images.

$\Theta$ can be updated iteratively by $J$ and learning rate $\eta$ according to
\begin{equation}
\mathrm{W}_{n m}^{l}[t+1]=\mathrm{W}_{n m}^{l}[t]-\eta \frac{\partial J[t]}{\partial \mathrm{W}_{n m}^{l}[t]},
\label{updata W}
\end{equation}
\begin{equation}
{b}_{n}^{l}[t+1]={b}_{n}^{l}[t]-\eta \frac{\partial J[t]}{\partial {b}_{n}^{l}[t]}.
\label{updata b}
\end{equation}

To calculate (\ref{updata W}) and (\ref{updata b}), the key point is computing the partial derivatives. Note $J$ is a real-valued loss function, it can be back-propagated through CV-FCN according to the generalized complex chain rule in \cite{CV-MLP}. Thus, the partial derivatives can be calculated as follows:
\begin{footnotesize}
\begin{equation}
\begin{aligned} 
\frac{\partial J}{\partial \mathrm{W}_{n m}^{l}}&=\frac{\partial J}{\partial \mathfrak{R}\left(\mathrm{W}_{n m}^{l}\right)}+\frac{\partial J}{\partial \mathfrak{I}\left(\mathrm{W}_{n m}^{l}\right)}\\
&=\left(\frac{\partial J}{\partial \mathfrak{R}\left(\mathrm{Y}_{n}^{l}\right)}\frac{\partial \mathfrak{R}\left(\mathrm{Y}_{n}^{l}\right)}{\partial \mathfrak{R}\left(\mathrm{W}_{n m}^{l}\right)}+\frac{\partial J}{\partial \mathfrak{I}\left(\mathrm{Y}_{n}^{l}\right)}\frac{\partial \mathfrak{I}\left(\mathrm{Y}_{n}^{l}\right)}{\partial \mathfrak{R}\left(\mathrm{W}_{n m}^{l}\right)}\right)&\\
&+j.\left(\frac{\partial J}{\partial \mathfrak{R}\left(\mathrm{Y}_{n}^{l}\right)}\frac{\partial \mathfrak{R}\left(\mathrm{Y}_{n}^{l}\right)}{\partial \mathfrak{I}\left(\mathrm{W}_{n m}^{l}\right)}+\frac{\partial J}{\partial \mathfrak{I}\left(\mathrm{Y}_{n}^{l}\right)}\frac{\partial \mathfrak{I}\left(\mathrm{Y}_{n}^{l}\right)}{\partial \mathfrak{I}\left(\mathrm{W}_{n m}^{l}\right)}\right),
\end{aligned} 
\end{equation}
\end{footnotesize} 
\begin{footnotesize}
\begin{equation}
\begin{aligned} 
\frac{\partial J}{\partial {b}_{n}^{l}}
&=\frac{\partial J}{\partial \mathfrak{R}\left({b}_{n}^{l}\right)}+\frac{\partial J}{\partial \mathfrak{I}\left({b}_{n}^{l}\right)}\\
&=\left(\frac{\partial J}{\partial \mathfrak{R}\left(\mathrm{Y}_{n}^{l}\right)}\frac{\partial \mathfrak{R}\left(\mathrm{Y}_{n}^{l}\right)}{\partial \mathfrak{R}\left({b}_{n}^{l}\right)}+\frac{\partial J}{\partial \mathfrak{I}\left(\mathrm{Y}_{n}^{l}\right)}\frac{\partial \mathfrak{I}\left(\mathrm{Y}_{n}^{l}\right)}{\partial \mathfrak{R}\left({b}_{n}^{l}\right)}\right)&\\
&+j.\left(\frac{\partial J}{\partial \mathfrak{R}\left(\mathrm{Y}_{n}^{l}\right)}\frac{\partial \mathfrak{R}\left(\mathrm{Y}_{n}^{l}\right)}{\partial \mathfrak{I}\left({b}_{n}^{l}\right)}+\frac{\partial J}{\partial \mathfrak{I}\left(\mathrm{Y}_{n}^{l}\right)}\frac{\partial \mathfrak{I}\left(\mathrm{Y}_{n}^{l}\right)}{\partial \mathfrak{I}\left({b}_{n}^{l}\right)}\right).
\end{aligned} 
\end{equation}
\end{footnotesize}       

When the value of loss function no longer decreases, the parameters update is suspended and the training phase is completed. Then the trained network will be used to predict the entire PolSAR image in the classification phase. 
\begin{algorithm} 
\caption{CV-FCN Classification Algorithm for PolSAR Imagery}  
\label{alg:CV-FCN algorithm} 
{\bf Input:}
PolSAR dataset $\mathcal{H} \in \mathbb{C}^{h \times w \times B}$, learning rate $\eta$, batch size, momentum parameter.

{\bf Output:} Dense label $\mathbf{k}$.
\begin{algorithmic}[1] 
\STATE Construct the data patches set $\mathcal{I}$ and the label patches set $\mathcal{T}$ using $\mathcal{H}$; 
\STATE Initialize CV-FCN parameters $\Theta$ by Section II-B; 
\STATE Choose the entire training set $D_{train}=\{\mathcal{I}_{train},\mathcal{T}_{train}\}$ from $\mathcal{I}$ and $\mathcal{T}$; 
\STATE {\bf Repeat:}
\STATE Forward pass the complex downsampling section to obtain multi-level feature maps by Section II-C1;
\STATE Call the complex upsampling section to recover more spatial information by Section II-C2;
\STATE Calculate loss function $E$ by Section II-D; 
\STATE Update $\Theta$ by $J$ and $\eta$.
\STATE {\bf Until:}meet the termination criterion.
\STATE Classify the entire PolSAR image by forward passing the trained network to obtain $\mathbf{k}$. 
\STATE {\bf End}
\end{algorithmic}
\end{algorithm} 
\subsection{CV-FCN PolSAR Classification Algorithm}
For more intuitive, the proposed CV-FCN PolSAR Classification algorithm is illustrated by Algorithm \ref{alg:CV-FCN algorithm}. Specially, we first construct the entire training set for CV-FCN and employ the new complex-valued weight initialization scheme to initialize the network. And then, we train CV-FCN by alternately updating CV-FCN parameters using the average cross-entropy loss function. Finally, the entire PolSAR image is classified using the trained network.

\section{Experimental Analysis and Evaluation}
\label{sec:experiment}
In this section, experimental datasets description and evaluation metrics are first presented. Then, input data vector and experimental settings are listed for CV-FCN training. Moreover, the effectiveness of some strategies for CV-FCN is analyzed in detail by a series of special experiments. Finally, comparisons with other classification methods on three PolSAR datasets are presented to demonstrate the superiority of the proposed CV-FCN. 
\subsection{Experimental Datasets Description}
We use three benchmark PolSAR datasets for experiments. Details about these datasets are listed as follows.
\begin{figure}[ht]
\centering
\subfloat[]{\includegraphics[width=0.4\linewidth,height=3cm]{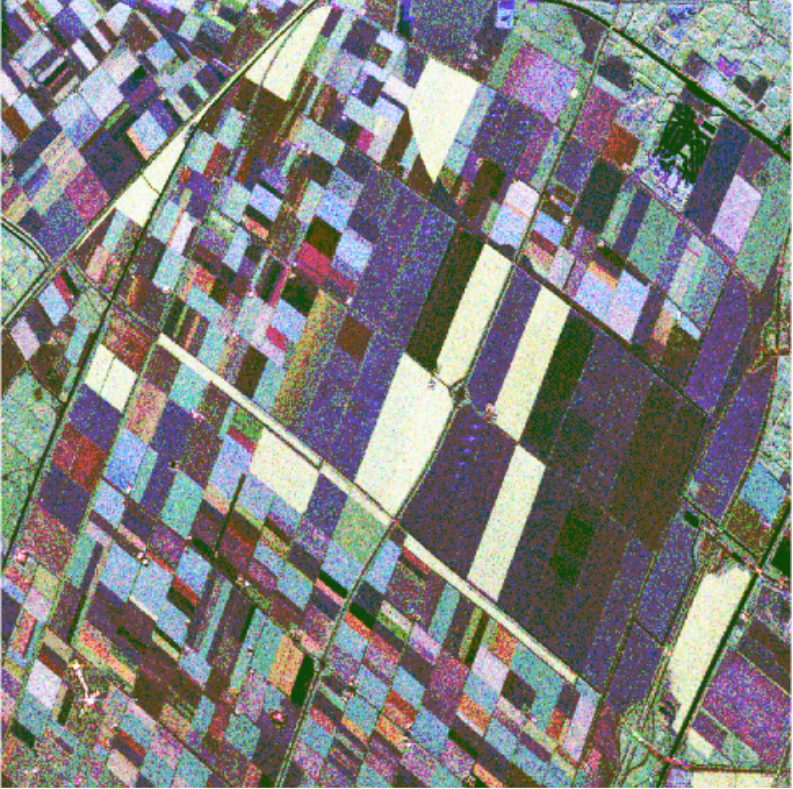}}
\quad
\subfloat[]{\includegraphics[width=0.4\linewidth,height=3cm]{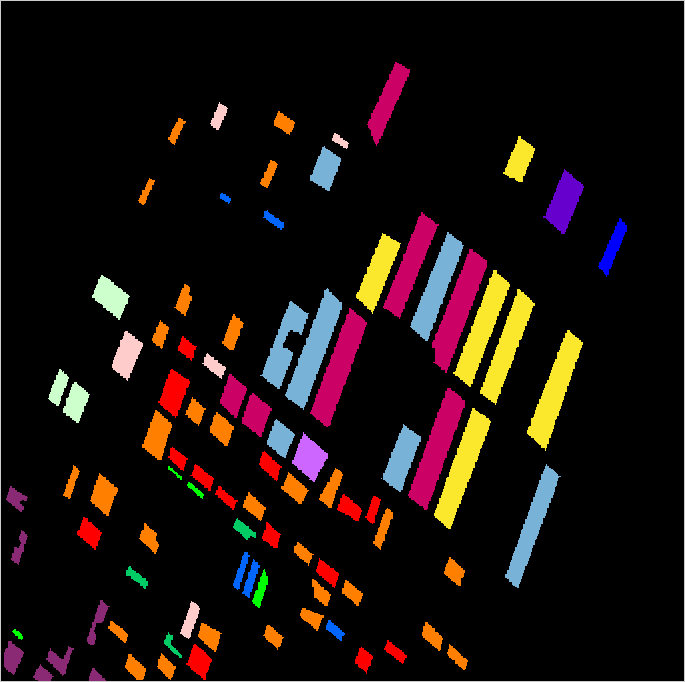}}

\subfloat[]{\includegraphics[width=0.8\linewidth]{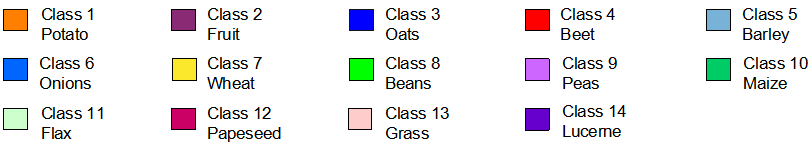}}
\caption{Flevoland Benchmark PolSAR image and related ground truth categorization information. (a) The PauliRGB image. (b) The ground truth categorization map. (c) Color code of different classes.}
\label{fig:Flevoland-dataset}
\end{figure} 
\begin{figure}[ht]
\centering
\subfloat[]{\includegraphics[width=0.4\linewidth,height=3cm]{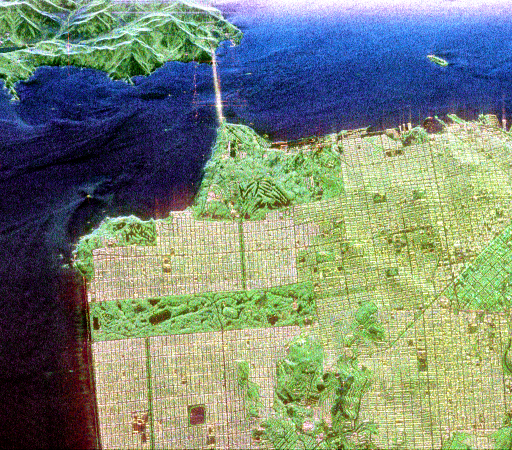}}
\quad
\subfloat[]{\includegraphics[width=0.4\linewidth,height=3cm]{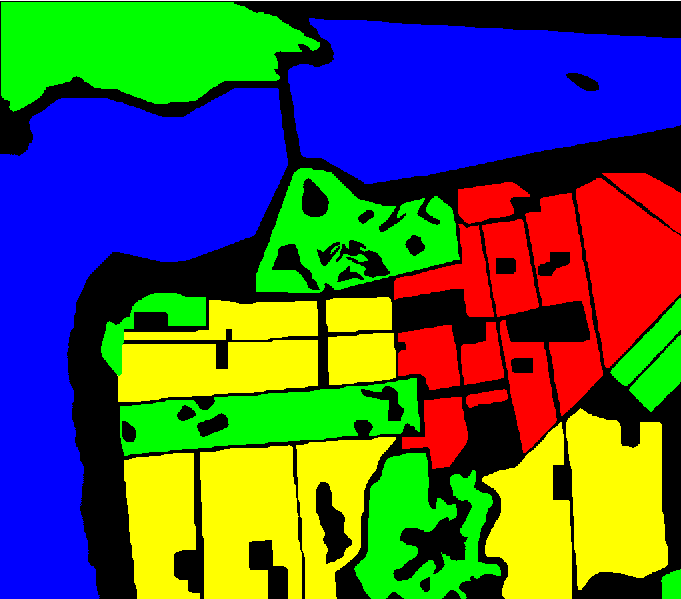}}

\subfloat[]{\includegraphics[width=0.8\linewidth]{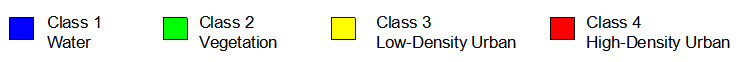}}
\caption{San Francisco PolSAR image and related ground truth categorization information. (a) The PauliRGB image. (b) The ground truth categorization map. (c) Color code of different classes.}
\label{fig:SanFrancisco-dataset}
\end{figure} 
\begin{figure}[ht]
\centering
\subfloat[]{\includegraphics[width=0.35\linewidth,height=3cm]{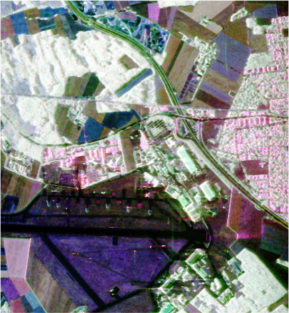}}
\quad
\subfloat[]{\includegraphics[width=0.35\linewidth,height=3cm]{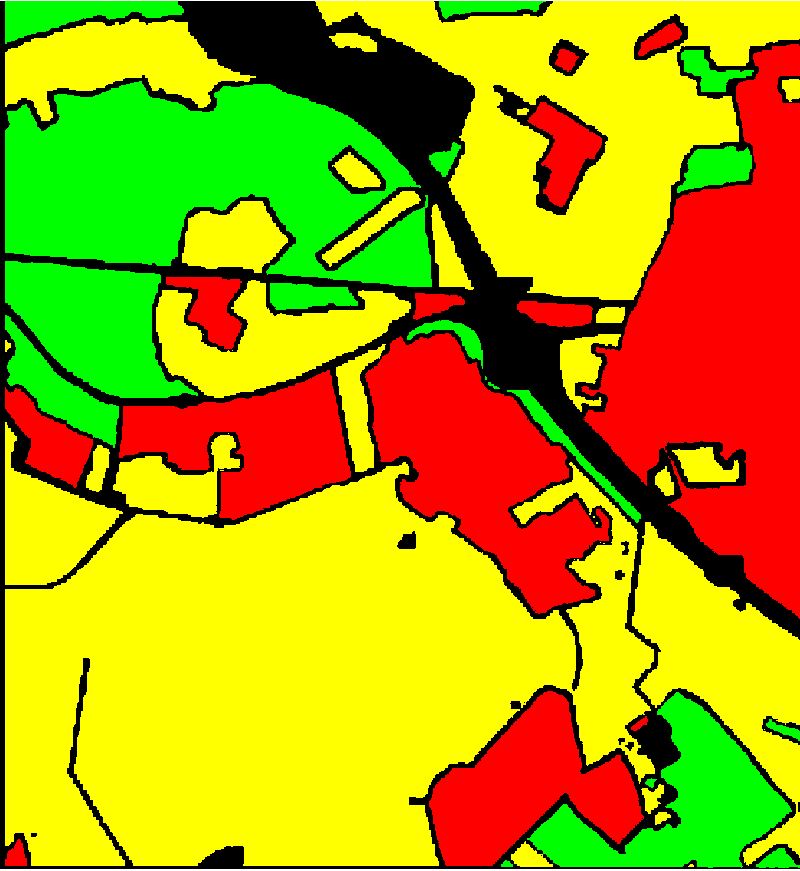}}
\quad
\subfloat[]{\includegraphics[width=0.2\linewidth,height=3cm]{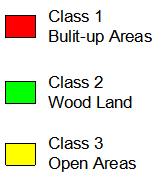}}
\caption{Oberpfaffenhofen PolSAR image and related ground truth categorization information. (a) The PauliRGB image. (b) The ground truth categorization map. (c) Color code of different classes.}
\label{fig:Germany-dataset}
\end{figure} 
\subsubsection{Flevoland Benchmark dataset}
Fig. \ref{fig:Flevoland-dataset}(a) shows the PauliRGB image of Flevoland Benchmark data, which was acquired by NASA/JPL AIRSAR in 1991. The size of the image is 1020$\times$1024. The ground-truth class labels and the corresponding color codes are shown in Fig. \ref{fig:Flevoland-dataset}(b) and Fig. \ref{fig:Flevoland-dataset}(c), respectively. There are 14 classes in the image including potato, fruit, oats, beet, barley, onions, wheat, beans, peas, maize, flax, rapeseed, grass, and lucerne.
\subsubsection{San Francisco dataset}
This AIRSAR full PolSAR image provides good coverage of four targets including water, vegetation, low-density urban and high-density urban. The original data has a dimension of  900$\times$1024 pixels with a spatial resolution of 10 m, as shown in Figure \ref{fig:SanFrancisco-dataset}(a). The ground-truth class labels and color codes are shown in Fig. \ref{fig:SanFrancisco-dataset}(b) and Fig. \ref{fig:SanFrancisco-dataset}(c).
\subsubsection{Oberpfaffenhofen dataset}
This data is an ESAR data of Oberpfaffenhofen area in Germany,  provided by the German Aerospace Center, has a size of 1300$\times$1200 pixels. The Pauli- RGB image, the ground-truth class labels, and color codes are respectively shown in Fig. \ref{fig:Germany-dataset}(a)-(c). There are three classes in the image: built-up areas, wood land, and open areas. 
\subsection{Evaluation Metrics}
With the hand-marked ground-truth images, the overall accuracy (OA), average accuracy (AA) and Kappa coefficient ($\kappa$) are used as the evaluation measures for classification performance evaluation. Where OA represents the ratio of the number of correctly labeled pixels divided by the total number of test pixels; AA is defined as the average of individual class accuracy; Kappa which does not consider the successful classification that obtained by chance gives a good representation of the overall performance of the classifiers. The larger values of three criteria, the better classification performance.
\subsection{Preparing for Classifier Model Training}
\subsubsection{Complex-valued Input Vector for CV-FCN}
Before training CV-FCN, the CV input vector needs to be determined. CV-FCN works directly on the PolSAR CV data without any data projection from the complex to the real domain. Since the coherency matrix or covariance matrix completely describes the distributed target \cite{PolSAR book}, the PolSAR data is usually presented in these formats. The polarimetric coherency matrix $\mathrm{T}$ is calculated as
\begin{equation}
\mathrm{T}=\frac{1}{L} \sum_{\mathrm{i=1}}^{L} \mathrm{u}_{\mathrm{i}} \mathrm{u}_{\mathrm{i}}^{\mathrm{H}}=\left[ \begin{array}{ccc}{T_{11}} & {T_{12}} & {T_{13}} \\ {T_{21}} & {T_{22}} & {T_{23}} \\ {T_{31}} & {T_{32}} & {T_{33}}\end{array}\right],
\end{equation}
where the superscript $\mathrm{H}$ denotes the complex conjugate transpose, ${L}$ is the number of looks, and $\mathrm{u}_{\mathrm{i}}$ denotes the ${\mathrm{i}}$th scattering vector in the multi-look processing window.

The coherency matrix $\mathrm{T}$ is a Hermitian positive semidefinite matrix which implies that main diagonal elements are RV and other CV elements are conjugate symmetric about the main diagonal. Therefore, the six elements of the upper triangular matrix of $\mathrm{T}$ can be used to fully represent the PolSAR data \cite{PolSAR book}. So we utilize these six elements to construct the CV input vector for CV-FCN, which is represented by
\begin{equation}
\left\{T_{11}+0 . j, T_{22}+0 . j, T_{33}+0 . j, T_{12}, T_{13}, T_{23}\right\}.
\end{equation}

Here, imaginary parts of $T_{11}, T_{22}, T_{33}$ in CV input feature vector are all expanded with a value of 0. On the other hand, compared with CV input feature vector with phase information, the RV input feature vector without phase information can be represented by
\begin{footnotesize}
\begin{equation}
\{ T_{11}, T_{22}, T_{33},\mathfrak{R}\left(T_{12}\right), \mathfrak{R}\left(T_{13}\right), \mathfrak{R}\left(T_{23}\right),\mathfrak{I}\left(T_{12}\right), \mathfrak{I}\left(T_{13}\right), \mathfrak{I}\left(T_{23}\right) \}.
\label{RV-data}
\end{equation}
\end{footnotesize} 
\subsubsection{Parameter Settings}
Relevant parameter settings are required before CV-FCN training. For PolSAR image classification, some works of literature have discussed the sampling rate and parameter settings of NNs structures \cite{BHX-GNN,HC-FCN,JLC-PCN} in detail. Hence, we no longer spend time discussing again and will choose them through experiments.

In this paper, the sliding window operation in \cite{LY(18)-FCN,WS(18)-FCN} is used to generate the data patches set from experimental images and corresponding label patches set from ground truth images. Here, we choose 128 as the default setting of sliding windows size and 40 as the default setting of stride for all experimental datasets, which is a trade between classification performance and computational burden. Additionally, to mitigate overfitting on datasets, the data augmentation strategy \cite{FM-FCN} was carried out by vertically and horizontally flipping all patches. Then all these patches are the input data of the proposed CV-FCN, where 90$\%$ for training and 10$\%$ for validation, respectively. It is worth noting that only labeled pixels in individual label patch are considered in modifying parameters of the network during the training \cite{JLC-PCN}.

Moreover, Adam with momentum 0.9 is used to update CV-FCN parameters. The learning rate $\eta$ is 0.0001. The size of mini-batch is empirically set to 30. The training epoch number is 200 until the objective function converges. Additionally, dropout regularization is adopted to reduce overfitting. In this paper, all non-deep methods are run on Matlab R2014b, and DL-based methods are implemented in the Keras framework with TensorFlow as the back end. The machine used for experiments is a Lenovo Y720 cube gaming PC with an Intel Core i7-7700 CPU, an Nvidia GeForce GTX 1080 GPU, and 16GB RAM under Ubuntu 18.04 LTS operating system. To make comparisons as fair as possible, we take the average of 10 experiments as the final result.

\subsection{ CV-FCN Model Analysis and Discussions}
To evaluate the performance of some aspects in CV-FCN model, two ablation experiments and two comparison experiments are conducted as follows. Notably, the most perspective of the proposed CV-FCN is the complex-valued upsampling scheme, in which the fully skip connections and the max location maps are the two important strategies. Therefore, two ablation experiments are designed for comparison and evaluation. Specifically, the impact of fully skip connections in CV-FCN structure is investigated firstly. Then, the effect of max locations maps on classification performance is evaluated on all datasets. Additionally, a comparison experiment about the new weight initialization scheme is conducted. Finally, the effectiveness of different loss functions on precise classification is compared. The two ablation experiments are all conducted on all three datasets. Moreover, the third experiment and the fourth experiment is conducted on the Flevoland Benchmark dataset and the Oberpfaffenhofen dataset, separately. For the Flevoland Benchmark dataset, 5$\%$ labeled pixels per class are randomly chosen for training, and the rest for the test. For the San Francisco image and the Oberpfaffenhofen image, about 1$\%$ of the whole labeled reference pixels are randomly selected for training, and the rest is used to evaluate experimental performance.
\begin{table}[tp] 
\centering
\fontsize{5.6}{7}\selectfont 
\caption{Overall, Average Accuraties (\%) and Kappa Coefficient of \protect\\the NS$\_$CV-FCN Method and the CV-FCN Method}
\label{tab:skip connection} 
\begin{tabular}{c|c|ccc}  
\hline
\toprule[0.3pt]
\hline 
\toprule[0.3pt]
Dataset & Methods & OA & AA & $\kappa$ \\
\hline
\multirow{2}{*}{Flevoland}& NS$\_$CV-FCN & 98.73\% & 94.97\% & 0.9851 \\ \clineB{2-5}{1}
& CV-FCN & {\bf99.72\%} & {\bf98.14\%} & {\bf0.9967} \cr
\hline
\multirow{2}{*}{SanFrancisco}& NS$\_$CV-FCN & 98.37\% & 98.12\% & 0.9774 \\ \clineB{2-5}{1}
& CV-FCN & {\bf99.69\%} & {\bf99.65\%} & {\bf0.9957} \cr
\hline
\multirow{2}{*}{Oberpfaffenhofen}& NS$\_$CV-FCN & 93.36\% & 90.51\% & 0.8831 \\ \clineB{2-5}{1}
& CV-FCN & {\bf97.26\%} & {\bf96.38\%} & {\bf0.9531} \cr
\hline
\toprule[0.3pt]
\hline
\toprule[0.3pt]
\end{tabular}
\end{table}
\begin{table}[tp]
\centering
\fontsize{5.6}{7}\selectfont
\caption{Overall, Average Accuraties (\%) and Kappa Coefficient of \protect\\the NL$\_$CV-FCN Method and the CV-FCN Method} 
\label{tab:max locations maps} 
\begin{tabular}{c|c|ccc}  
\hline
\toprule[0.3pt]
\hline 
\toprule[0.3pt]
Dataset & Methods & OA & AA & $\kappa$ \\
\hline
\multirow{2}{*}{Flevoland}& NL$\_$CV-FCN & 99.29\% & 96.41\% & 0.9915 \\ \clineB{2-5}{1}
& CV-FCN & {\bf99.72\%} & {\bf98.14\%} & {\bf0.9967} \cr
\hline
\multirow{2}{*}{SanFrancisco}& NL$\_$CV-FCN & 98.97\% & 98.84\% & 0.9857 \\ \clineB{2-5}{1}
& CV-FCN & {\bf99.69\%} & {\bf99.65\%} & {\bf0.9957} \cr
\hline
\multirow{2}{*}{Oberpfaffenhofen}& NL$\_$CV-FCN & 95.80\% & 94.31\% & 0.9273 \\ \clineB{2-5}{1}
& CV-FCN & {\bf97.26\%} & {\bf96.38\%} & {\bf0.9531} \cr
\hline
\toprule[0.3pt]
\hline
\toprule[0.3pt]
\end{tabular}
\end{table}
\subsubsection{Ablation Experiment 1 - Impact of  Fully Skip Connections} 
The fully skip connections is an important part of CV-FCN because it enables the network to enhance more detail. The core idea is to superimpose feature maps of different levels to improve the final classification effect. Here, to evaluate the effectiveness of fully skip connections on classification accuracy, we construct the CV-FCN structure without skip connections. We use NS$\_$CV-FCN to represent this CV-FCN network. Table \ref{tab:skip connection} contains the evaluation indices for classification.

As illustrated in Table \ref{tab:skip connection}, CV-FCN outperforms NS$\_$CV-FCN which reveals that fully skip connections are useful for improving classification accuracy. Compared with NS$\_$CV-FCN, the proposed CV-FCN increases the accuracy by 0.99$\%$ of OA, 3.17$\%$ of AA, and 0.0116 of the Kappa coefficient, respectively, on the Flevoland dataset. Moreover, on the SanFrancisco dataset, CV-FCN is able to achieve the accuracy increments by 1.32$\%$ of OA, 1.53$\%$ of AA, and 0.0183 of Kappa, respectively. In particular, on the Oberpfaffenhofen dataset, CV-FCN increases the accuracy significantly by 3.9$\%$ of OA, 5.87$\%$ of AA, and 0.07 of Kappa, respectively. This superiority can be attributed to the fact that fully skip connections fuse features of different levels to preserve more discriminative information for PolSAR image classification.
\subsubsection{Ablation Experiment 2 - Impact of Max Location Maps}
The most prominent trait in the complex upsampling section is that the max location maps are utilized to perform nonlinear upsampling of the feature maps, which are beneficial for more precise reconstruction output. To examine the effect of max locations maps, we construct a CV-FCN structure wherein complex upsampling layers upsample feature maps without the guidance of max locations maps. We call this network as NL$\_$CV-FCN. The experimental results on all datasets are shown in Table \ref{tab:max locations maps}.

As illustrated in Table \ref{tab:max locations maps}, CV-FCN outperforms NL$\_$CV-FCN on all three datasets. Specifically, CV-FCN is able to achieve the accuracy increments by 0.43$\%$ of OA, 1.73$\%$ of AA, and 0.0052 of Kappa, respectively, on the Flevoland dataset; by 0.72$\%$ of OA, 0.81$\%$ of AA, and 0.001 of Kappa, respectively, on the SanFrancisco dataset; by 1.46$\%$ of OA, 2.07$\%$ of AA, and 0.0258 of Kappa, respectively, on the Oberpfaffenhofen dataset. These results suggest that the max locations maps benefit the classification accuracy since they have the capacity of retrieving more sufficient spatial information.
\begin{figure}[ht]
\centering
\subfloat{\includegraphics[width=7cm,height=5cm]{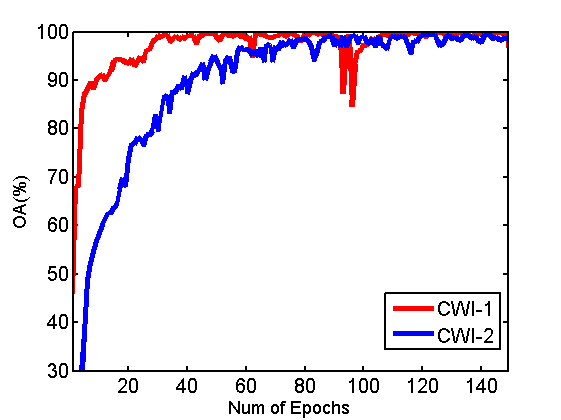}}
\caption{The validation curves of different weight initialization schemes on the Flevoland Benchmark dataset. Red line denotes the proposed complex-valued weight initialization scheme, and blue line denotes the old initialization scheme.}
\label{fig:weight curves}
\end{figure} 
\begin{table*}[tp] 
\centering
\fontsize{5.6}{8}\selectfont 
\caption{Overall, average accuraties (\%) and Kappa coefficient of different complex weight initialization methods on the Flevoland Benchmark PolSAR image.}
\label{tab:Weight initialization} 
\begin{tabular}{c|c|cccccccccccccc}  
\hline
\toprule[0.3pt]
\hline 
\toprule[0.3pt]
\multicolumn{2}{c|}{Epochs}& 10 & 20 & 30 & 40 & 50 & 60 & 70 & 80 & 90 & 100 & 110 & 120 & 130 & 140 \\
\hline
\multirow{3}{*}{CWI-1}& OA & 68.51\% & 84.58\% & 89.05\% & 94.99\% & 96.08\% & 97.56\% & 97.78\% & 98.54\% & 99.03\% & 99.12\% & 98.85\% & {\bf99.25\%} & 98.40\% & 97.60\% \cr
& AA & 28.55\% & 48.55\% & 62.30\% & 75.13\% & 79.90\% & 88.89\% & 89.31\% & 96.34\% & 96.56\% & 96.66\% & 95.82\% & {\bf98.02\%} & 96.30\% & 95.14\% \cr
& $\kappa$ & 0.6163 & 0.8174 & 0.8709 & 0.9410 & 0.9539 & 0.9713 & 0.9738 & 0.9828 & 0.9885 & 0.9897 & 0.9865 & {\bf0.9912} & 0.9812 & 0.9717 \cr
\hline
\multirow{3}{*}{CWI-2}& OA & 91.78\% & 94.89\% & 97.79\% & 99.09\% & 98.93\% & {\bf99.36\%} & 98.07\% & 98.81\% & 98.03\% & 98.77\% & 98.79\% & 97.70\% & 95.17\% & 94.11\% \cr
& AA & 67.31\% & 75.51\% & 90.08\% & 96.59\% & 96.99\% & {\bf98.08\%} & 95.06\% & 96.26\% & 95.64\% & 97.50\% & 97.13\% & 94.11\% & 94.95\% & 73.15\% \cr
& $\kappa$ & 0.9029 & 0.9398 & 0.9740 & 0.9893 & 0.9874 & {\bf0.9925} & 0.9773 & 0.9865 & 0.9768 & 0.9855 & 0.9857 & 0.9729 & 0.9434 & 0.9305 \cr
\hline
\toprule[0.3pt]
\hline
\toprule[0.3pt]
\end{tabular}
\end{table*}
\begin{figure}[H] %
\centering
\subfloat[]{\includegraphics[width=0.45\linewidth,height=3.5cm]{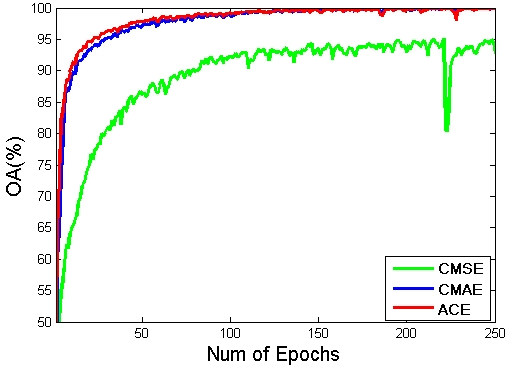}}
\quad
\subfloat[]{\includegraphics[width=0.45\linewidth,height=3.5cm]{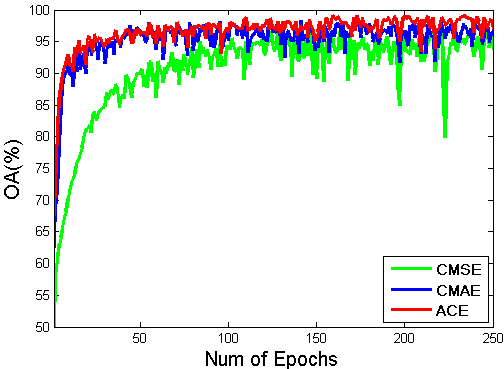}}
\caption{The overall accuracy curves for different loss functions of training and validation on the Oberpfaffenhofen dataset. Red lines denote the proposed loss function, blue lines denote the CMSE loss function, and green lines denote the CMAE loss function. (a) The overall accuracy of training; (b) The overall accuracy of validation.}
\label{fig:curves-loss}
\end{figure} %
\begin{figure*}[t] %
\centering
\subfloat[]{\includegraphics[width=0.28\linewidth,height=4cm]{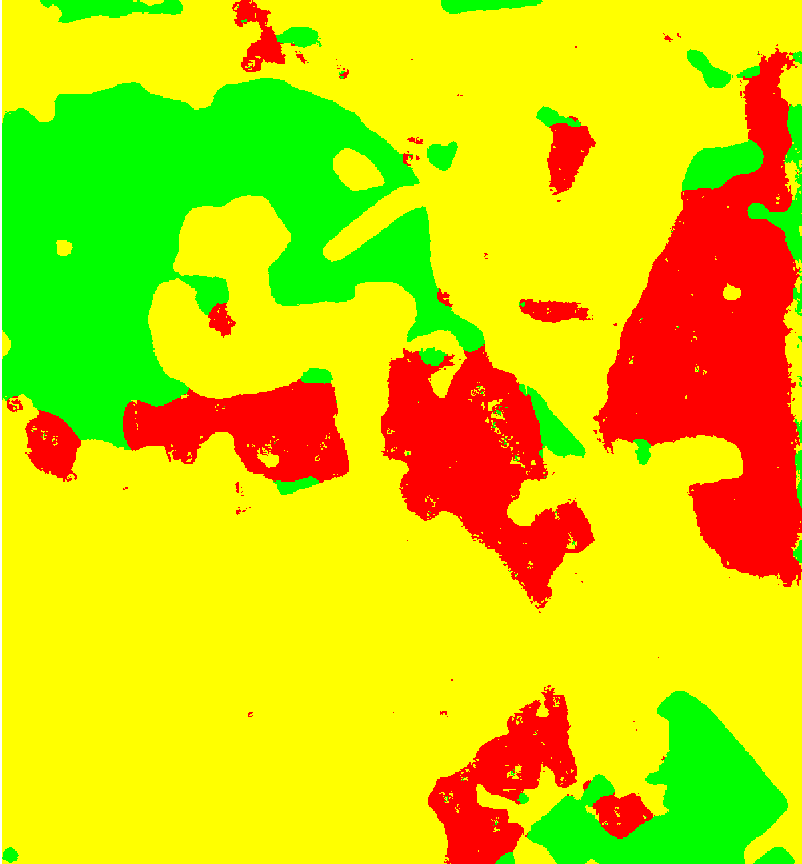}}
\quad
\subfloat[]{\includegraphics[width=0.28\linewidth,height=4cm]{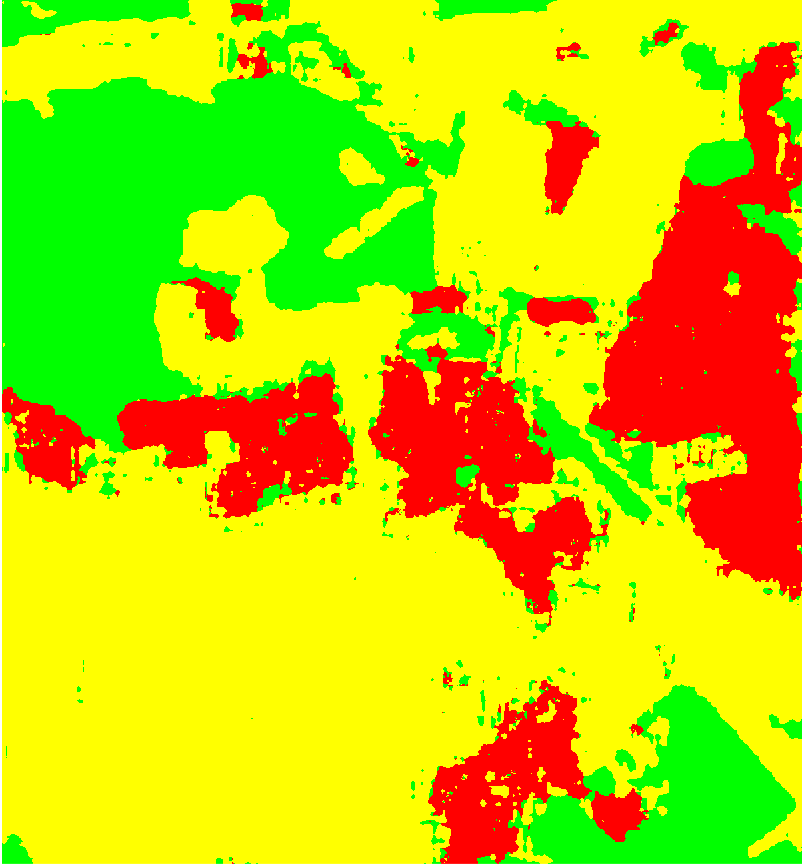}}
\quad
\subfloat[]{\includegraphics[width=0.28\linewidth,height=4cm]{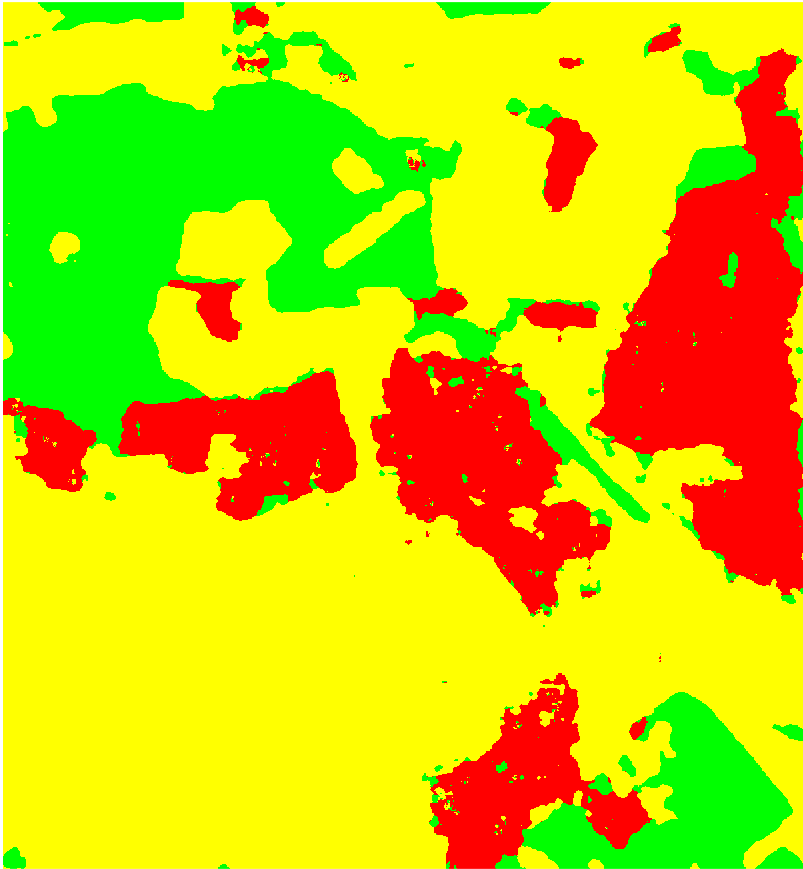}}
\caption{Classification results of Oberpfaffenhofen area with different loss functions.  (a) CMSE loss function; (b) CMAE loss function; (c)  Proposed ACE loss function.}
\label{fig:image-loss}
\end{figure*} %
\subsubsection{Comparison Experiment 1 - Complex Weight Initialization} 
To evaluate the impact of complex-valued weight initialization which is critical for CV-FCN learning, we conduct a comparison experiment on the Flevoland Benchmark dataset. We only utilize the complex weight initialization in \cite{XF-CV-SCNN} as the old CV initialization scheme for comparison. The old CV initialization scheme is to initialize the real and imaginary parts of a CV weight separately with a uniform distribution \cite{XF-CV-SCNN}. Fig. \ref{fig:weight curves} illustrates the difference in the validation curves of one presentative experiment, where the proposed weight initialization and the compared weight initialization are denoted by CWI-1 and CWI-2, respectively. Furthermore, we also report the evaluation indices of two initialization schemes as a function of epoch. Specifically, we first train CV-FCN for 10 epochs and then update classification results every 10 epochs. Table \ref{tab:Weight initialization} contains a comparison of the results. 

As shown in Fig. \ref{fig:weight curves}, both old initialization and proposed initialization lead to convergence, but proposed initialization trains CV-FCN faster and reaches the optimal value earlier. As illustrated in Table \ref{tab:Weight initialization}, proposed initialization achieves the best results when the training epoch is set as around 60, while the old initialization achieves is around 120. These results validate that the proposed initialization not only facilitates faster learning but also improves the classification performance of CV-FCN. This may be attributed to the ability of proposed initialization to reduce risks of vanishing or exploding gradients, which has great significance for deep networks training. Additionally, these phenomenons, in part, illustrate that the proposed initialization scheme is suitable for CV-FCN to achieve given PolSAR image classification.

\subsubsection{Comparison Experiment 2 - Loss Function}
We carry out a comparison experiment on the Oberpfaffenhofen dataset to evaluate the effectiveness of the complex average cross-entropy loss function. The complex-valued mean square error (MSE) in \cite{XF-CV-SCNN} and the complex-valued mean absolute error (MAE) are utilized as compared loss functions, which will be respectively denoted by CMSE and CMAE. The CMSE and the CMAE can be respectively expressed as 
\begin{footnotesize}
\begin{equation}
J^{C M S E}=\frac{1}{N} \cdot \sum_{n=1}^{N}\left[(\mathfrak{R}(R[n])-\mathfrak{R}(O[n]))^{2}+(\mathfrak{I}(R[n])-\mathfrak{I}(O[n]))^{2}\right],
\end{equation}
\end{footnotesize} 
\begin{footnotesize}
\begin{equation}
J^{C M A E}=\frac{1}{N} \cdot \sum_{n=1}^{N}[|\mathfrak{R}(R[n])-\mathfrak{R}(O[n])|+| \mathfrak{I}(R[n])-\mathfrak{I}(O[n])|].
\end{equation}
\end{footnotesize} 

The overall accuracy curves for different loss functions of training and validation illustrated in Fig. \ref{fig:curves-loss}(a) and Fig. \ref{fig:curves-loss}(b), respectively. Moreover, the resulted typical classification maps for different loss functions are shown in Fig. \ref{fig:image-loss}.   
 
As seen from Fig. \ref{fig:curves-loss}, the proposed complex loss function denoted by ACE and the CMAE can converge faster than CMSE. The training and validation accuracies by using the ACE and CMAE remain relatively stable after 120 epochs, while CMSE still does not achieve similar stability until 250 epochs. Additionally, the best accuracy of the proposed loss function is higher than the CMAE when validation. As shown in Fig. \ref{fig:image-loss}, the classification map with CMSE is smoother than CMAE and ACE, this finding is potentially explained that the CMSE loss can mitigate the effects of the speckle noise. However, boundary delineation between different categories are ambiguous due to being too smooth. Although the classification map with CMAE contains clear structural information, it has more misclassification points since affected by the speckle noise. Notably, the proposed loss function can achieve correct boundary localization as well as shows better robustness to the speckle noise. Thus, these above phenomena can partially establish the effectiveness of the proposed loss function employed.
\begin{table*}[tp]   
\centering  
\fontsize{6}{7}\selectfont 
\caption{Individual Class, Overall, Average Accuraties (\%) and Kappa Coefficient of all Competing Methods on \protect\\the Flevoland Benchmark PolSAR Image}  
\label{tab:Flevoland result table}  
\begin{tabular}{c|ccc|cccc|cccc}  
\hline
\toprule[0.3pt]
\hline 
\toprule[0.3pt] 
\multirow{2}{*}{Class}&  
\multicolumn{3}{c|}{Non-deep Methods}&\multicolumn{8}{c}{ DL-based Methods}\cr\cline{2-12}
&SVM  &  Wishart    &  MRF    &  RV-MLP   & RV-SCNN  & RV-DCNN   &  RV-FCN  & CV-MLP  &  CV-SCNN  & CV-DCNN  & CV-FCN \cr
\hline    
Potato &86.29\% & 79.27\% & 99.26\% & 99.01\% & 99.86\% & 99.90\% & 99.89\% & 99.14\% & 99.63\% & 99.69\% & {\bf99.98\%}\cr
Fruit  &89.43\% & 82.24\% & 96.78\% & {\bf99.98\%} & 98.87\% & 98.19\% & 99.05\% & 88.54\% & 97.37\% & 96.53\% & 99.97\%\cr
Oats   &87.23\% & 88.59\% & 96.63\% & 83.14\% & 55.42\% & 99.97\% & 66.94\% & 95.34\% &   {\bf100\%} &   {\bf100\%} & 77.57\%\cr
Beet   &68.49\% & 67.71\% & 82.25\% & 76.91\% & 84.87\% & 83.25\% & 99.28\% & 80.82\% & 85.52\% & 90.42\% & {\bf99.97\%}\cr
Barely &76.37\% & 76.40\% & 93.47\% & 99.91\% &   {\bf100\%} & 99.99\% & 99.67\% & 99.65\% & 99.76\% &   {\bf100\%} &   {\bf100\%}\cr
Onions &32.11\% & 33.66\% & 32.68\% & 56.81\% & 65.36\% & 89.08\% & 69.89\% & 50.80\% & 83.50\% & 89.13\% & {\bf99.24\%}\cr
Wheat  &84.05\% & 86.90\% & 97.27\% & 99.92\% & 99.97\% & 99.99\% & 99.96\% & 99.86\% & 99.99\% & 99.97\% &   {\bf100\%}\cr
Beans  &92.42\% & 88.82\% & {\bf99.54\%} & 81.33\% & 82.57\% & 88.69\% & 90.51\% & 71.16\% & 91.65\% & 97.79\% & 98.79\%\cr
Peas   &93.47\% & 94.81\% & 99.95\% &   {\bf100\%} & 99.96\% &   {\bf100\%} & 99.82\% &   {\bf100\%} &   {\bf100\%} &   {\bf100\%} & 99.90\%\cr
Maize  &62.64\% & 61.09\% & 14.19\% & 98.91\% & 93.23\% & 87.74\% & 91.00\% & {\bf99.53\%} & 90.12\% & 98.15\% & 98.85\%\cr
Flax   &97.19\% & 94.88\% & 99.51\% & 61.78\% & 56.41\% & 69.04\% & 99.79\% & 87.77\% & 64.96\% & 96.65\% & {\bf99.97\%}\cr
	Rapeseed  &87.45\% & 66.63\% & 94.76\% & {\bf100\%} & {\bf100\%}  & {\bf100\%}   & 99.94\% & {\bf100\%}   & 99.98\% & {\bf100\%}   & 99.99\%\cr
Grass  &66.63\% & 65.70\% & 74.22\% & 88.13\% & 98.54\% & 99.12\% & 93.53\% & 83.11\% & 98.35\% & {\bf100\%}   & 99.87\%\cr
Lucerne &79.98\% & 82.72\% & 89.19\% & 98.95\% & 99.93\% & {\bf100\%}  & 99.36\% & 98.17\% & 99.99\% & {\bf100\%}   & 99.97\%\cr\hline
{OA} &81.67\% &76.44\% &92.58\% &95.34\% &96.09\% &97.18\% &98.63\% &96.21\% &97.06\% &98.75\% &{\bf99.72\%}\cr
{AA} &78.84\% &76.39\% &83.55\% &88.91\% &88.21\% &93.92\% &93.47\% &90.35\% &93.63\% &97.74\% &{\bf98.14\%}\cr
{$\kappa$} &0.7880 &0.7287 &0.9131 &0.9450 &0.9538 &0.9668 &0.9837 &0.9554 &0.9654 &0.9853 &{\bf0.9967}\cr
\hline 
\toprule[0.3pt] 
\hline
\toprule[0.3pt]
\end{tabular}  
\end{table*} 
\begin{table}[!t]
\fontsize{6}{7.5}\selectfont 
\centering 
\caption{Detailed Configuration of the RV-DCNN and the CV-DCNN. \protect\\$K$ Denotes the Total Number of Classes. The ReLU Layers \protect\\in RV-DCNN, Complex BN Layers and $\mathbb{C}$ReLU Layers in CV-DCNN are Omitted for Brevity}
\label{table-DCNN}
\begin{tabular}{|c|c|c|c|c|}
\hline
\bf Network & \bf Module \bf type & \bf dimension & \bf stride & \bf pad \\
\hline
\multirow{18}{*}{RV-DCNN} & Convolution & 3$\times$3$\times$9$\times$9 & 1 &1 \\ \clineB{2-5}{1}
& Max-Pooling & 2$\times$2 & 2 & 0  \\ \clineB{2-5}{1}

& Convolution & 3$\times$3$\times$9$\times$18 & 1 &1 \\ \clineB{2-5}{1}
& Max-Pooling & 2$\times$2 & 2  & 0  \\ \clineB{2-5}{1}

& Convolution & 3$\times$3$\times$18$\times$36 & 1 &1 \\ \clineB{2-5}{1}
& Max-Pooling & 2$\times$2 & 2  & 0  \\ \clineB{2-5}{1}

& Convolution & 3$\times$3$\times$36$\times$72 & 1 &1 \\ \clineB{2-5}{1}
& Max-Pooling & 2$\times$2 & 2  & 0  \\ \clineB{2-5}{1}

& Convolution & 3$\times$3$\times$72$\times$144 & 1 &1 \\ \clineB{2-5}{1}
& Max-Pooling & 2$\times$2 & 2  & 0 \\ \clineB{2-5}{1}

& Fully Connection & 144$\times$144 &    &1 \\ \clineB{2-5}{1}
& Fully Connection & 144$\times$$K$   &  &1 \\ \clineB{2-5}{1}
& Softmax &  &  &  \\

\hline

\multirow{18}{*}{CV-DCNN} & Complex Convolution & 3$\times$3$\times$12$\times$12 & 1 &1 \\ \clineB{2-5}{1}
& Complex Max-Pooling & 2$\times$2 & 2  & 0 \\ \clineB{2-5}{1}

& Complex Convolution & 3$\times$3$\times$12$\times$24 & 1 &1 \\ \clineB{2-5}{1}
& Complex Max-Pooling & 2$\times$2 & 2  & 0 \\ \clineB{2-5}{1}

& Complex Convolution & 3$\times$3$\times$24$\times$48 & 1 &1 \\ \clineB{2-5}{1}
& Complex Max-Pooling & 2$\times$2 & 2  & 0 \\ \clineB{2-5}{1}

& Complex Convolution & 3$\times$3$\times$48$\times$96 & 1 &1 \\ \clineB{2-5}{1}
& Complex Max-Pooling & 2$\times$2 & 2  & 0 \\ \clineB{2-5}{1}

& Complex Convolution & 3$\times$3$\times$96$\times$192 & 1 &1 \\ \clineB{2-5}{1}
& Complex Max-Pooling & 2$\times$2 & 2  & 0 \\ \clineB{2-5}{1}

& Complex Fully Connection & 192$\times$192 &  &1  \\ \clineB{2-5}{1}
& Complex Fully Connection & 192$\times$2$K$   &  &1  \\ \clineB{2-5}{1}
& Complex Softmax &  &  & \\ \clineB{2-5}{1}
\hline
\end{tabular}
\end{table} 
\subsection{ Comapring Models}
We demonstrate the effectiveness of the proposed method by comparison with some state-of-the art methods including SVM \cite{SVM09}, Wishart classifier \cite{Wishart04}, Markov random field (MRF) \cite{MRF08}, MLP \cite{RV-MLP(18)}, CVNN \cite{CV-MLP}, CNN \cite{XF-RV-SCNN}, CV-CNN \cite{XF-CV-SCNN}, and FCN \cite{LY(18)-FCN}. In the previous second part, we have already introduced the structure of CV-FCN. The specific settings of comparison methods are briefly described as follows.
\begin{itemize}
\item	{\bf Non-deep methods}: The non-deep methods include SVM \cite{SVM09}, Wishart classifier \cite{Wishart04}, and MRF \cite{MRF08}. They all adopt the input feature vector shown in Equation (\ref{RV-data}). For SVM-based methods, the radial basis function (RBF) kernel is chosen advised by \cite{SVM09}. For MRF, parameters are set according to the original publication \cite{MRF08}. 
\item	{\bf RV-FCN}: To compare with CV-FCN, we operate the RV-FCN \cite{LY(18)-FCN} that is the same as that it does in CV-FCN. Since the dimension of the input patch for RV-FCN is 9 and the dimension for CV-FCN is 6. For a fair comparison, we adjust parameter settings in RV-FCN to have the same degree of freedom (DoF) as CV-FCN. We mainly adjust the number of kernels in every convolutional layer of RV-FCN, which are about 0.8 times that in CV-FCN. 
\item	{\bf RV-MLP/CV-MLP}: Referring to \cite{RV-MLP(18)}, we choose a three-layer CV-MLP network, which consists of 128 neurons in the first complex hidden layer and 256 in the second complex hidden layer. The last layer is a softmax classifier to predict the probability distribution. To have the same DoF, for RV-MLP, there are 96 neurons in the first hidden layer and 180 in the second hidden layer. For MLPs, we choose a 32$\times$32 neighborhood of each pixel as the patch fed into networks to consider more contextual information.
\item {\bf RV-SCNN/CV-SCNN}: We use SV-SCNN and CV-SCNN to represent networks in \cite{XF-RV-SCNN} and \cite{XF-CV-SCNN}, respectively. According to SV-SCNN in \cite{XF-RV-SCNN}, the architecture of CV-SCNN is adjusted, which contains the input layer, two convolution layers interleaved with two pooling layers, two fully connected layers, and the softmax layer. For SCNNs, a 32$\times$32 neighborhood of each pixel is employed as the patch for training.
\item {\bf RV-DCNN/CV-DCNN}: The downsampling section in FCN is transformed from a CNN structure. Therefore, for a fair comparison between FCN and CNN, we construct a new CNN structure represented by DCNN according to CV-FCN structure. Table \ref{table-DCNN} reports the detail configuration of SV-DCNN and CV-DCNN. Compared with CNNs in \cite{XF-CV-SCNN}, DCNNs contain more convolutional layers. For DCNNs, we have the same operation as SCNNs to generate patches for training. 
\end{itemize}
\subsection{ Classification Performance Evaluation}
To evaluate the effectiveness of CV-FCN, comparisons with above models on three PolSAR datasets are presented as follows.
\begin{figure*}[t]
\centering
\subfloat[GroundTruth]{\includegraphics[width=0.24\linewidth,height=3.8cm]{./GroundTruth-Flevoland.png}}
\label{Flevoland-GT}
\subfloat[SVM]{\includegraphics[width=0.24\linewidth,height=3.8cm]{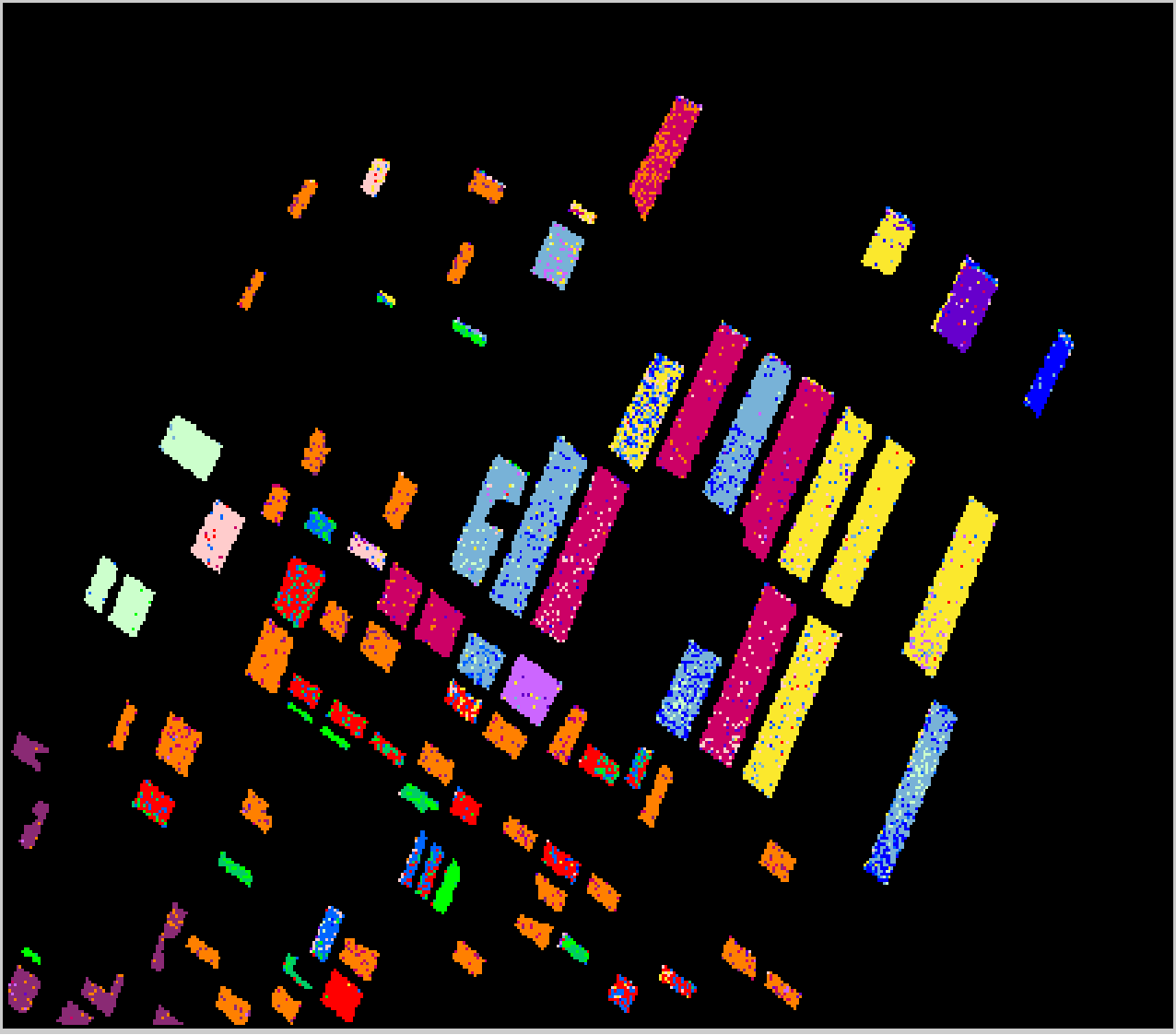}}
\label{SanFrancisco-SVM}
\subfloat[Wishart]{\includegraphics[width=0.24\linewidth,height=3.8cm]{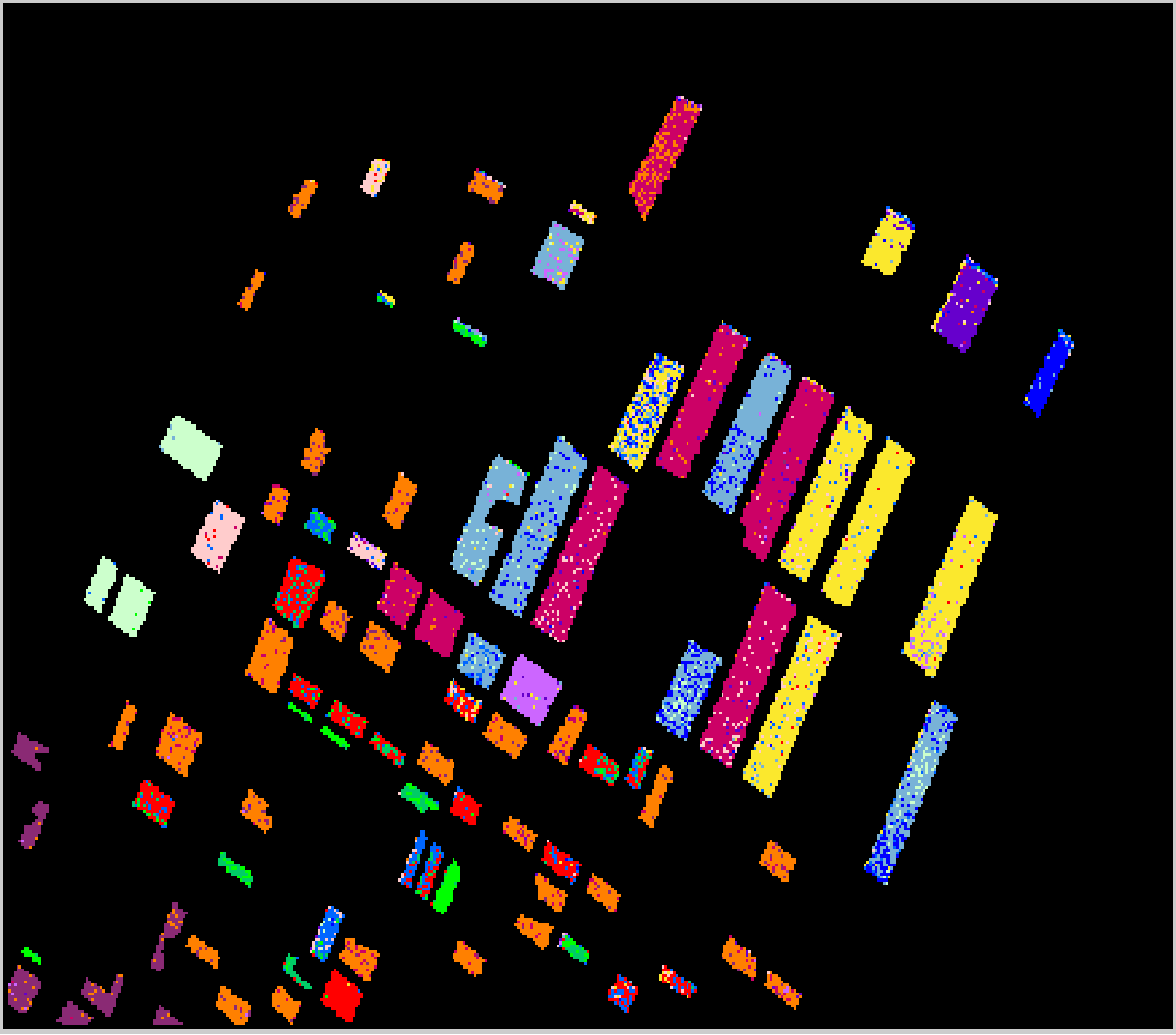}}
\label{SanFrancisco-Wishart}
\label{SanFrancisco-MRF}
\subfloat[MRF]{\includegraphics[width=0.24\linewidth,height=3.8cm]{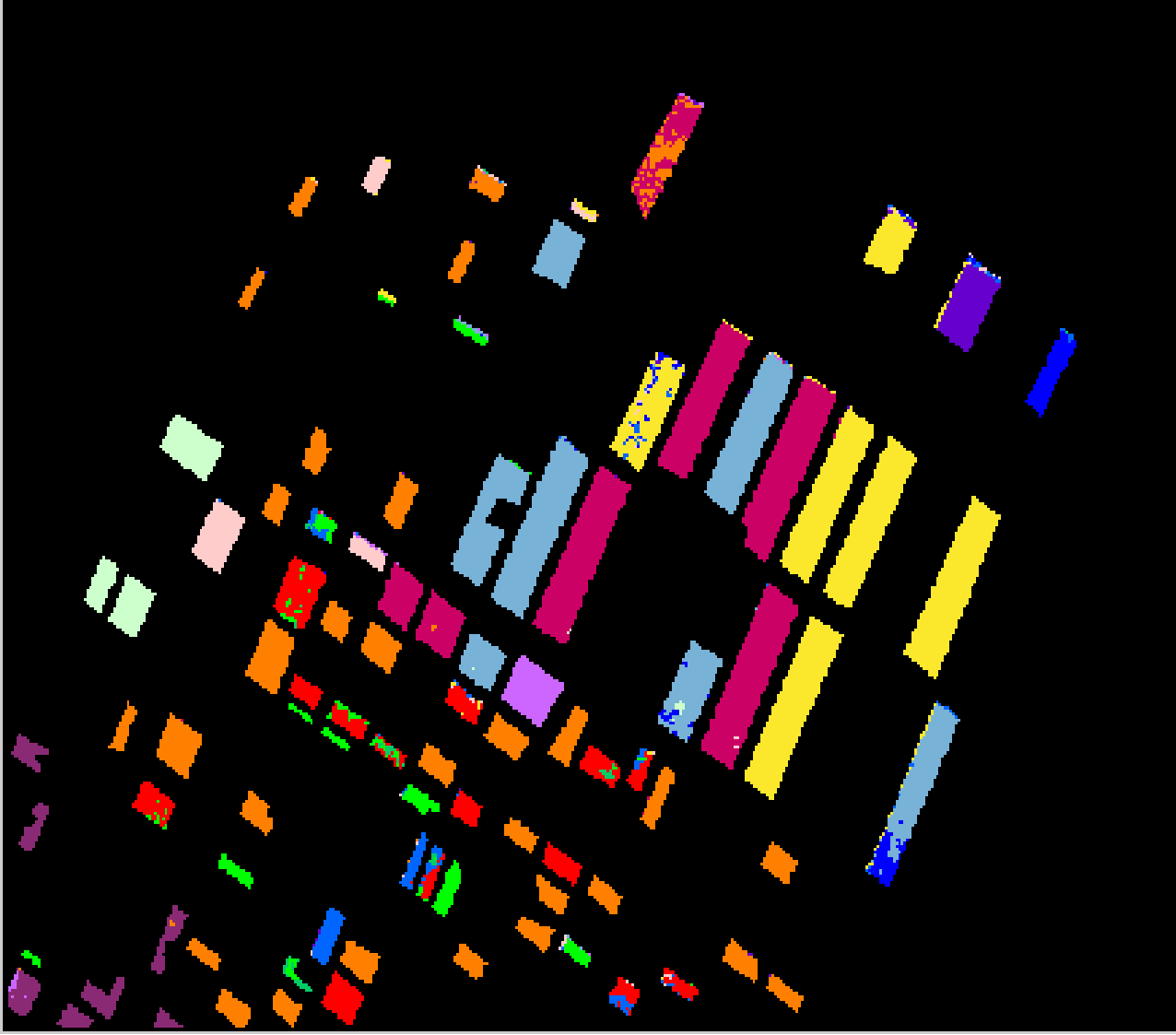}}

\label{SanFrancisco-RV-MLP}
\subfloat[RV-MLP]{\includegraphics[width=0.24\linewidth,height=3.8cm]{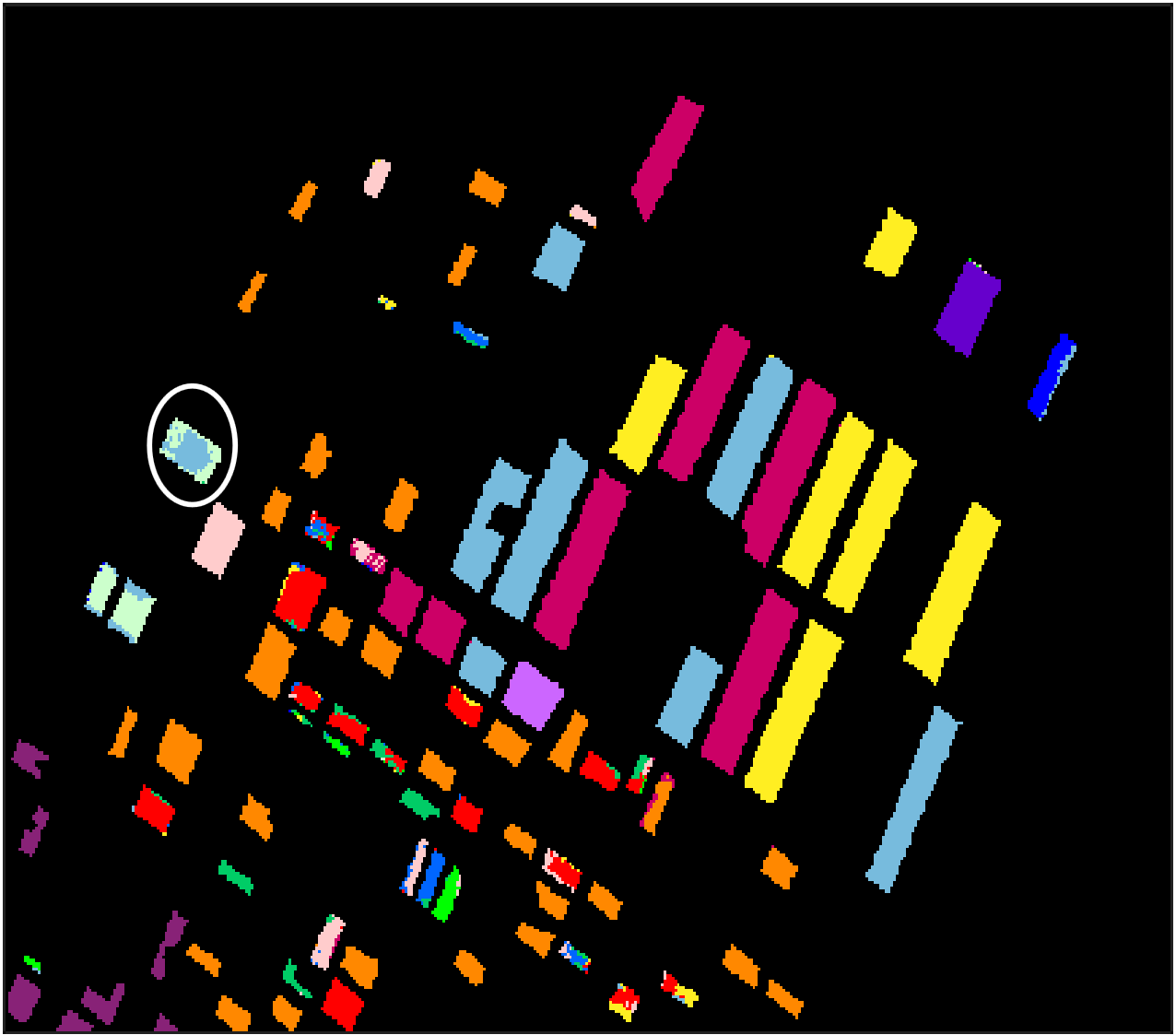}}
\label{SanFrancisco-RV-SCNN}
\subfloat[RV-SCNN]{\includegraphics[width=0.24\linewidth,height=3.8cm]{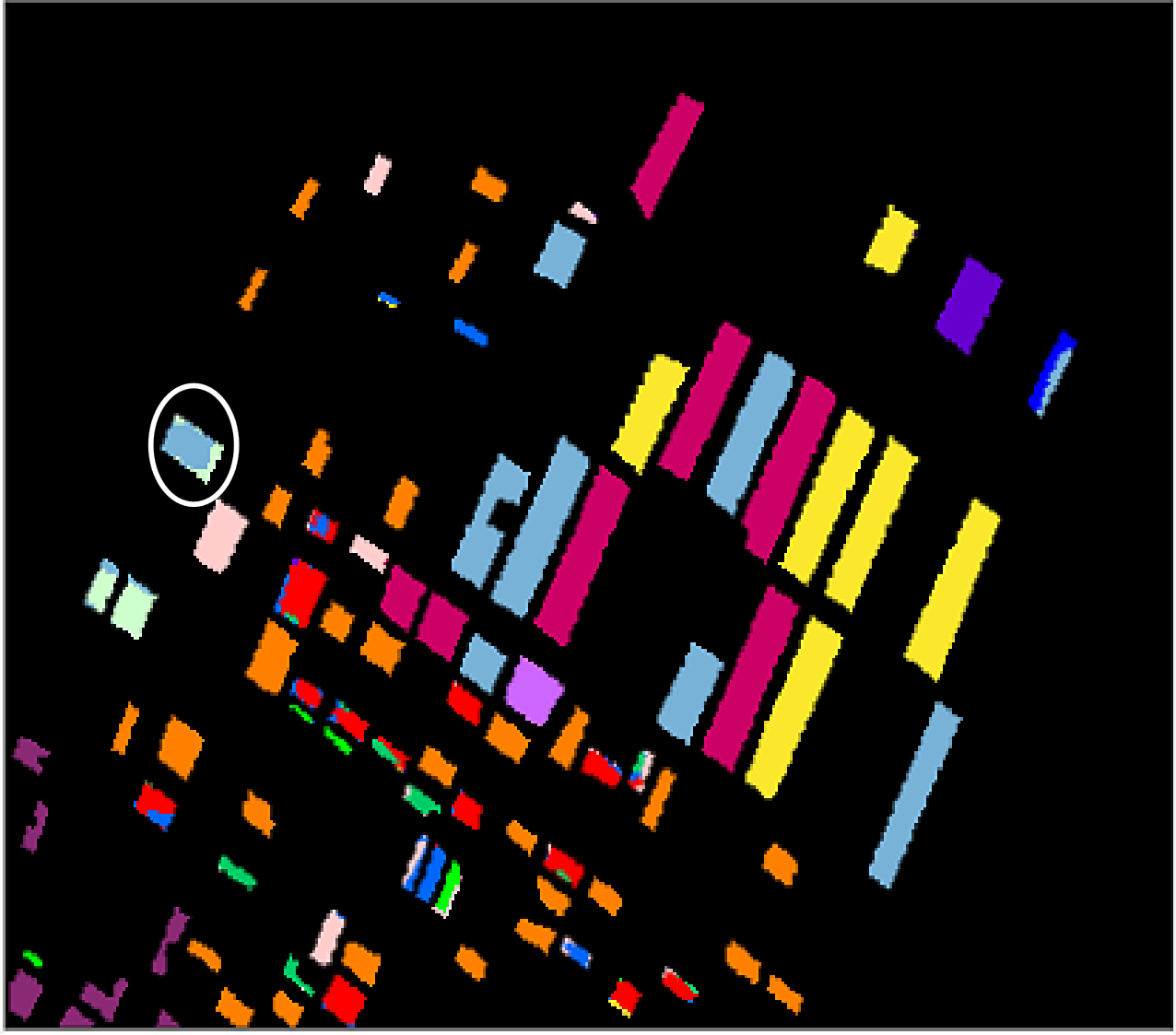}}
\label{SanFrancisco-RV-DCNN}
\subfloat[RV-DCNN]{\includegraphics[width=0.24\linewidth,height=3.8cm]{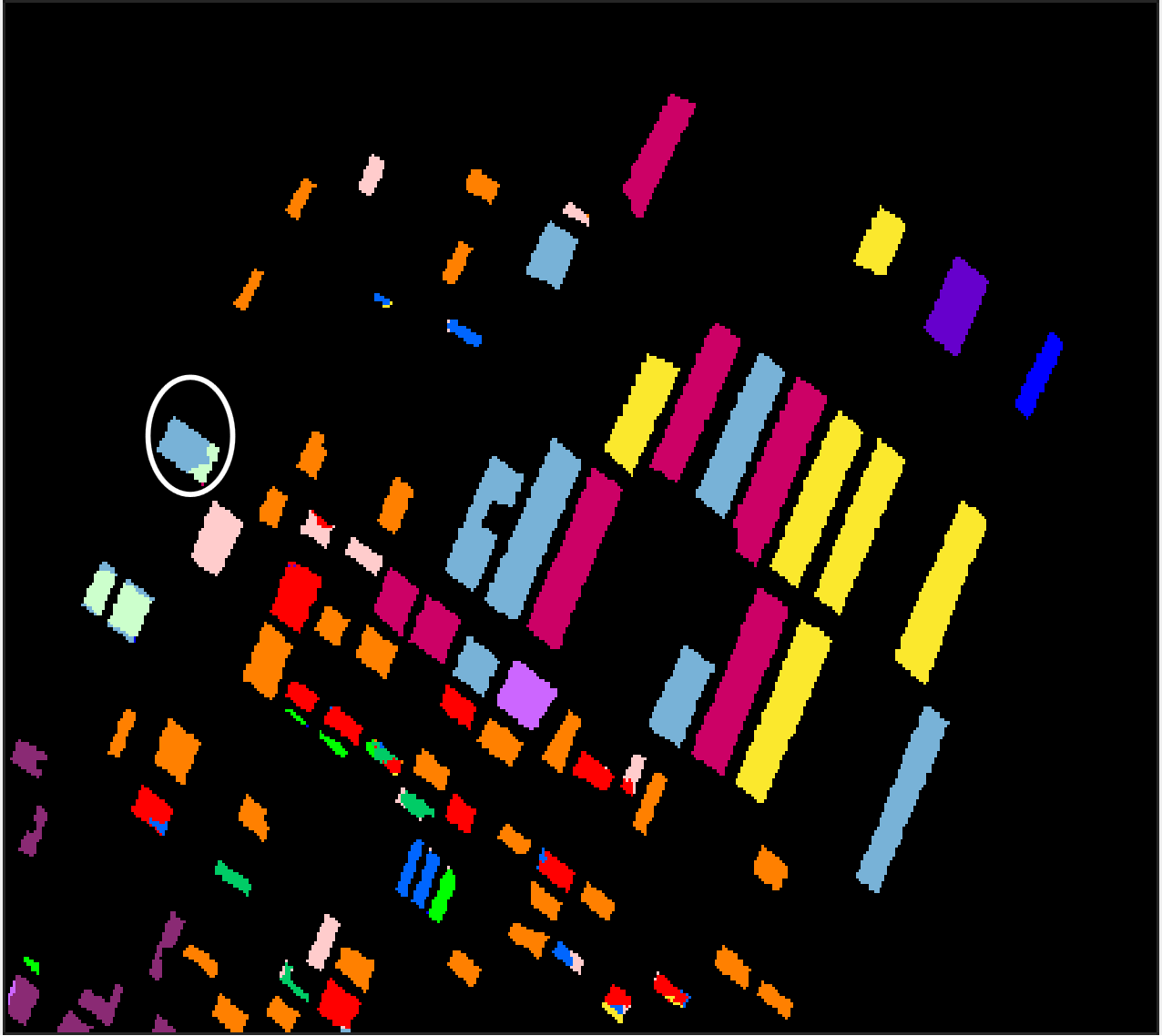}}
\label{SanFrancisco-RV-FCN}
\subfloat[RV-FCN]{\includegraphics[width=0.24\linewidth,height=3.8cm]{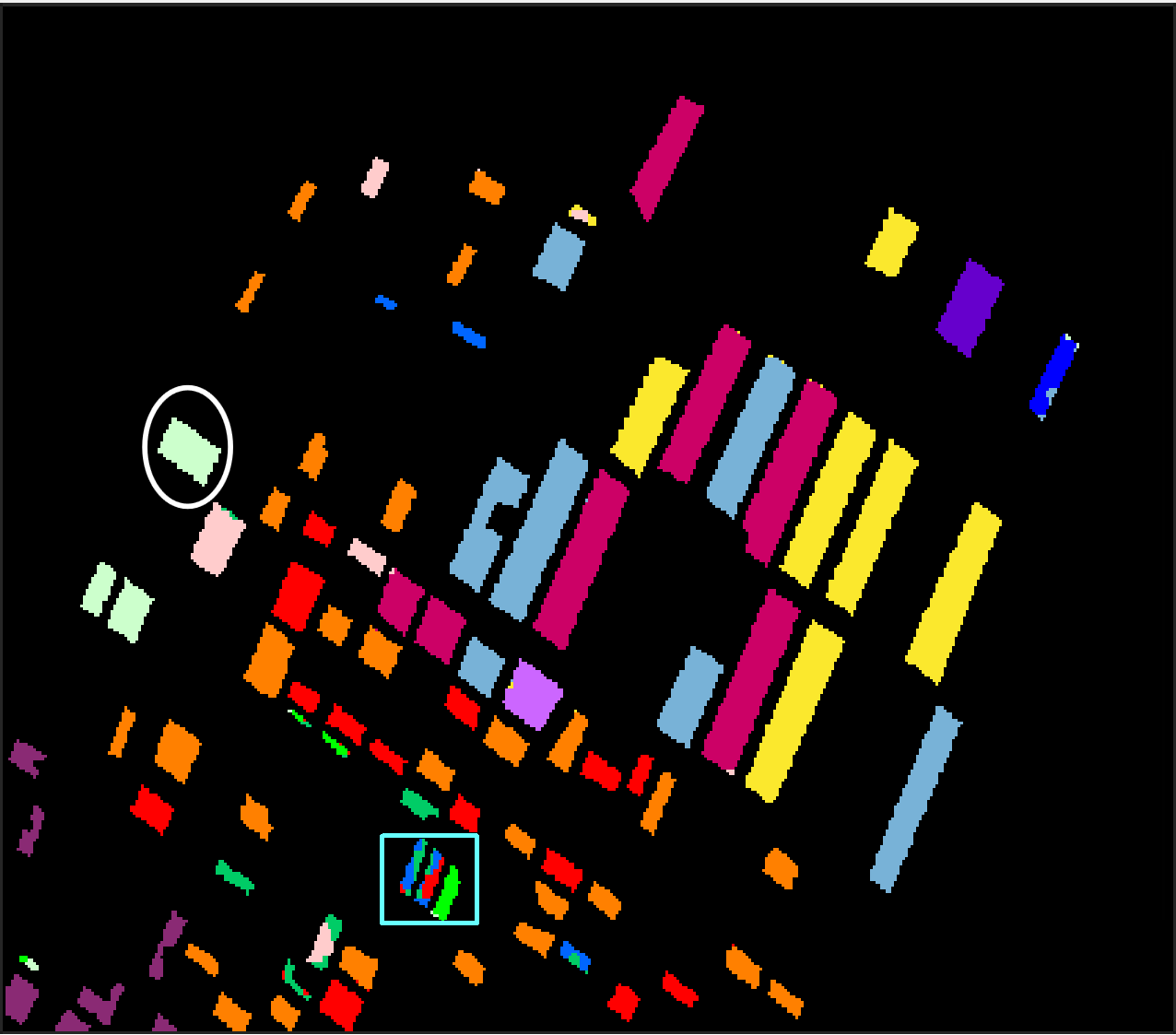}}

\label{SanFrancisco-CV-MLP}
\subfloat[CV-MLP]{\includegraphics[width=0.24\linewidth,height=3.8cm]{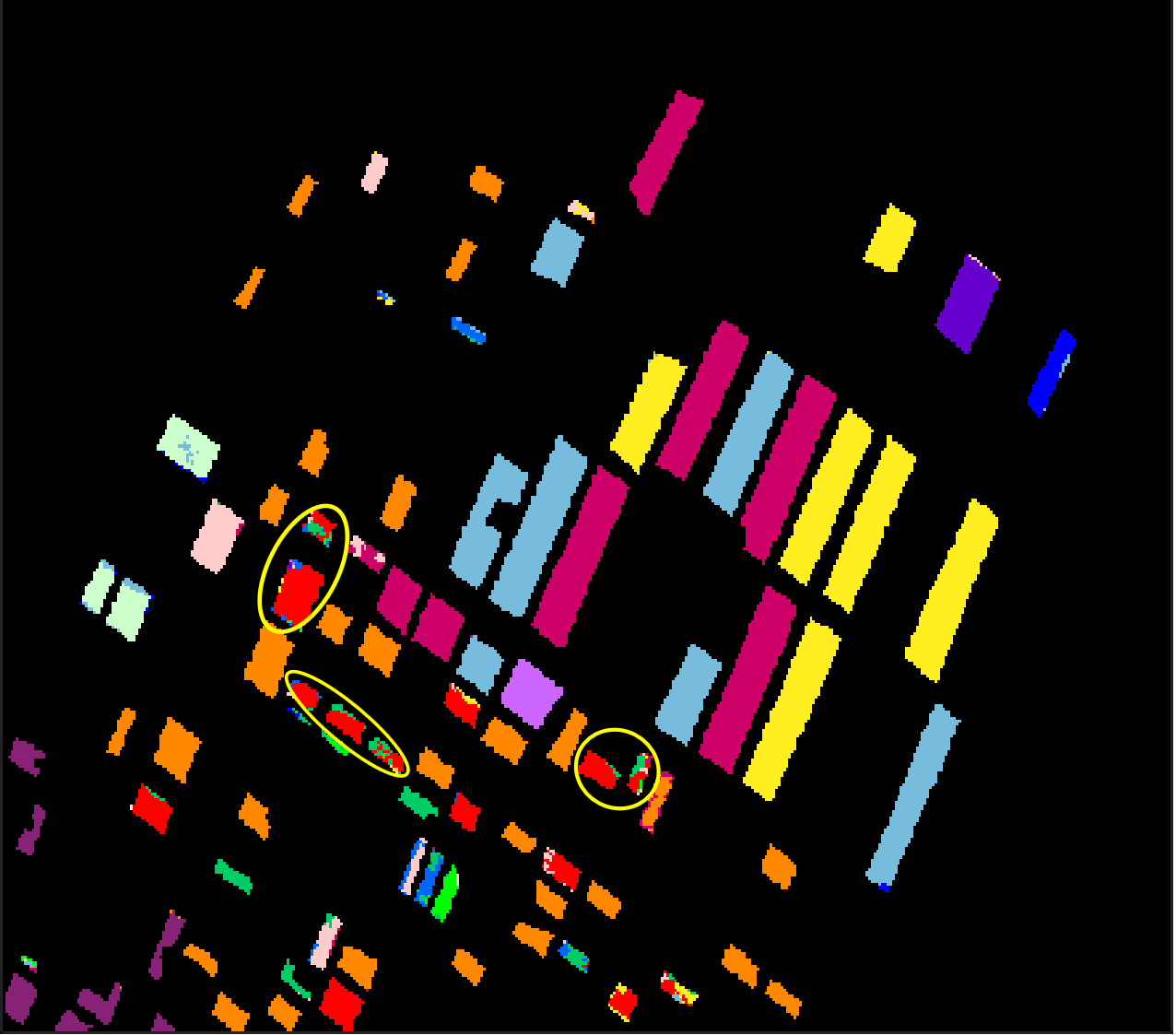}}
\label{SanFrancisco-CV-SCNN}
\subfloat[CV-SCNN]{\includegraphics[width=0.24\linewidth,height=3.8cm]{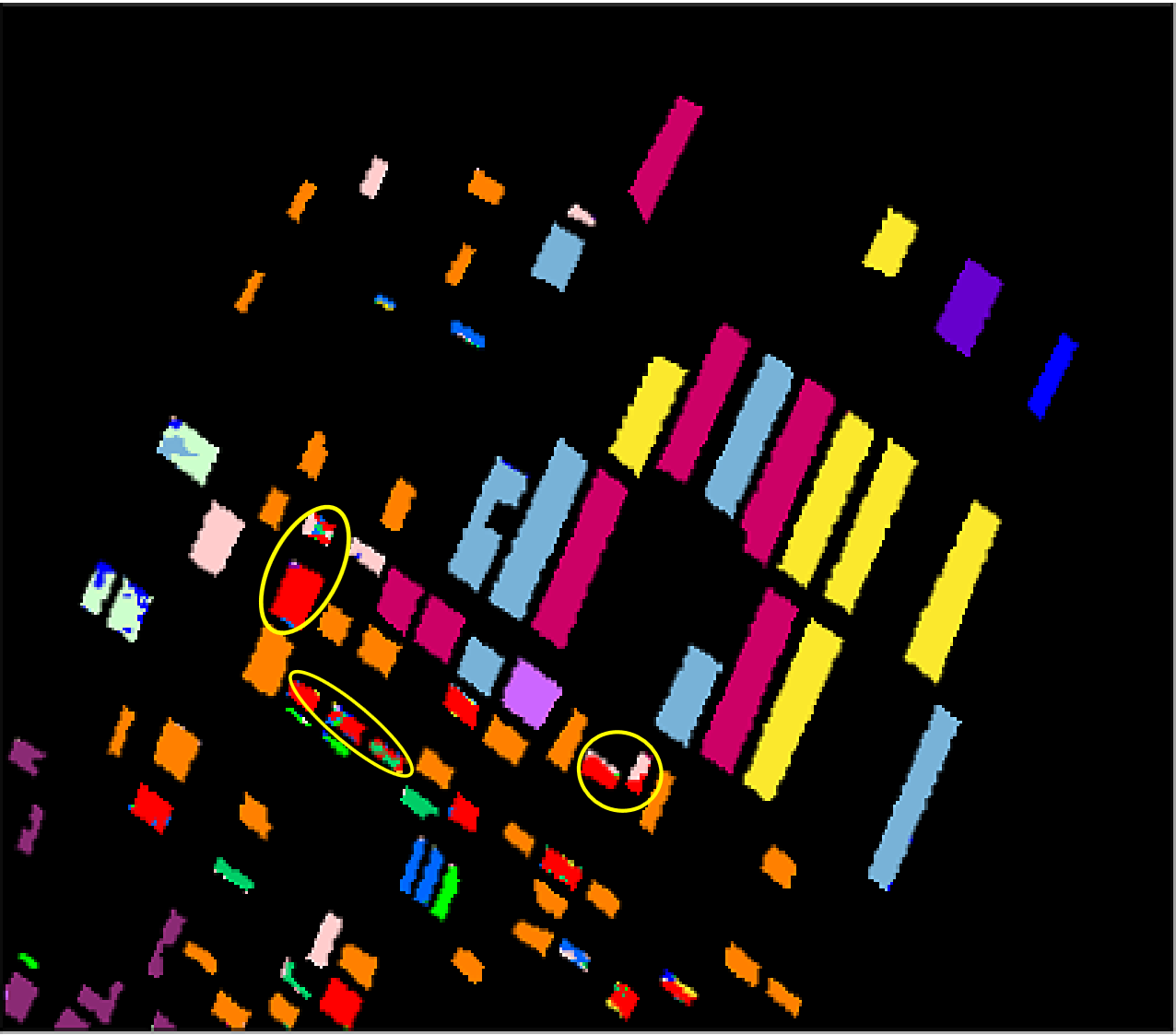}}
\label{SanFrancisco-CV-DCNN}
\subfloat[CV-DCNN]{\includegraphics[width=0.24\linewidth,height=3.8cm]{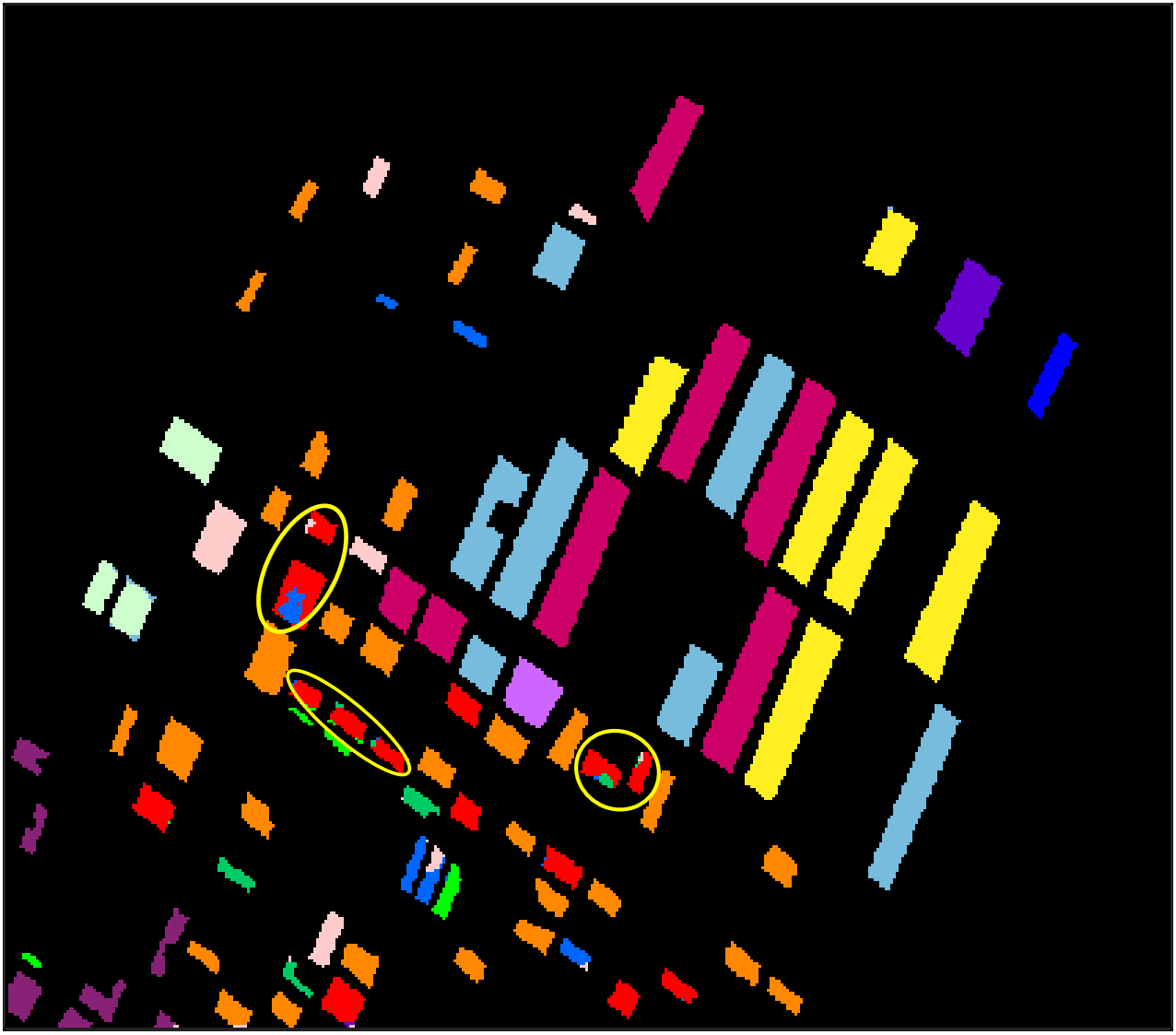}}
\label{SanFrancisco-CV-FCN}
\subfloat[CV-FCN]{\includegraphics[width=0.24\linewidth,height=3.8cm]{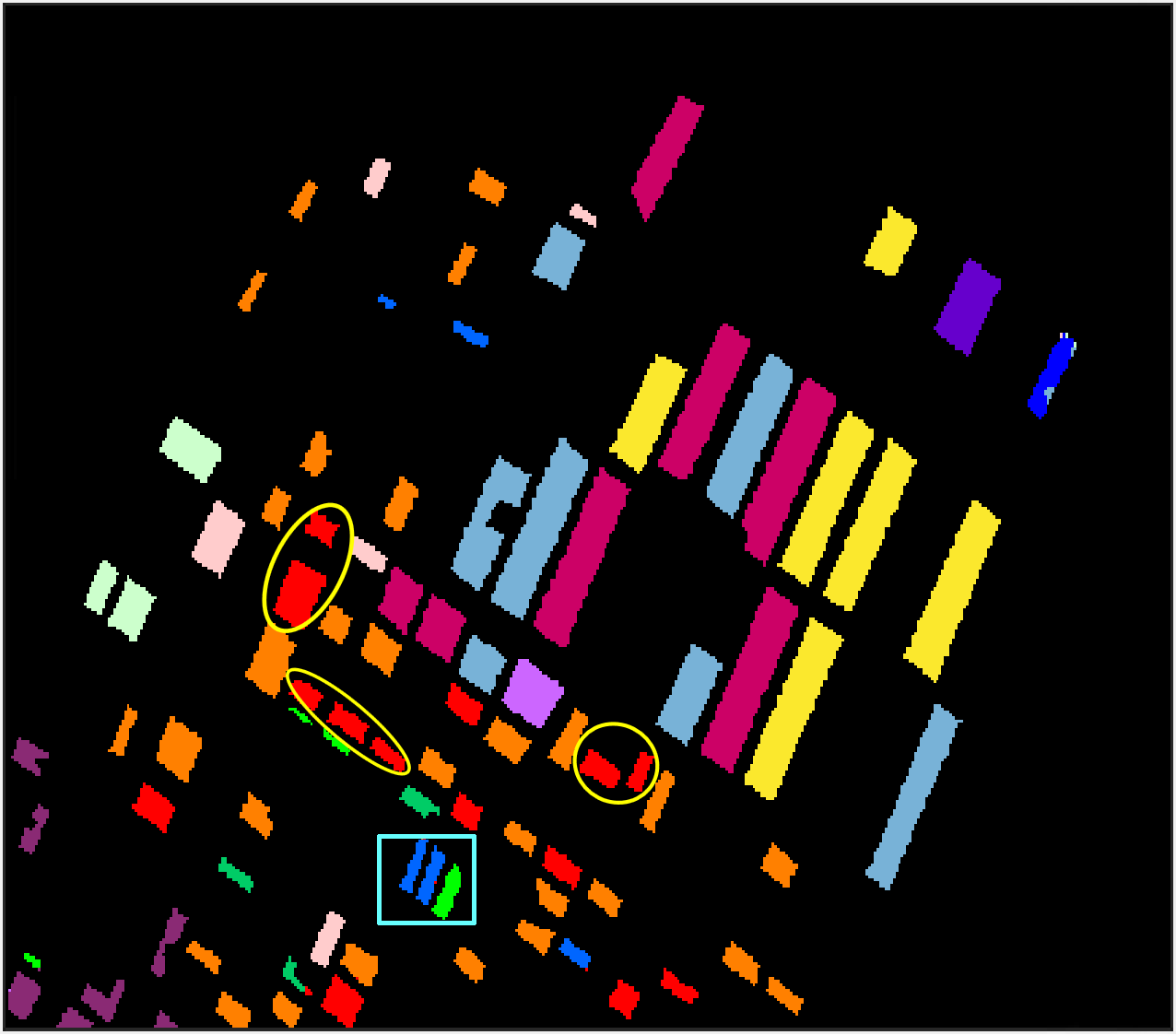}}

\caption{\protect\ Classification results of Flevoland Benchmark area data with different methods. }
\label{figure-reslut-Flevoland}
\end{figure*} 
\begin{figure*}[t] 
\centering
\label{SanFrancisco-GT}
\subfloat[GroundTruth]{\includegraphics[width=0.24\linewidth,height=3.8cm]{./Groundtruth-SanFrancisco.png}}
\label{SanFrancisco-SVM}
\subfloat[SVM]{\includegraphics[width=0.24\linewidth,height=3.8cm]{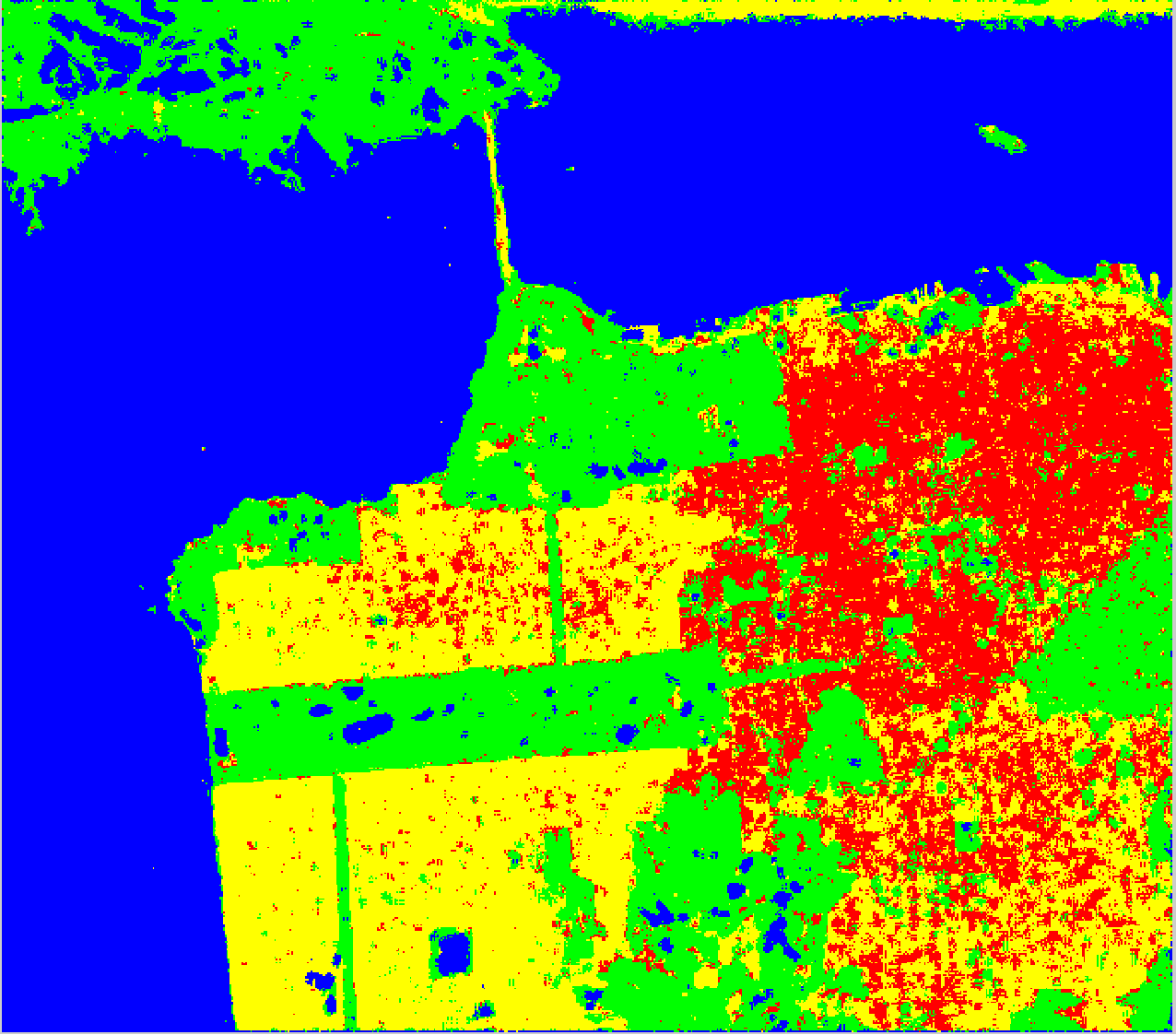}}
\label{SanFrancisco-Wishart}
\subfloat[Wishart]{\includegraphics[width=0.24\linewidth,height=3.8cm]{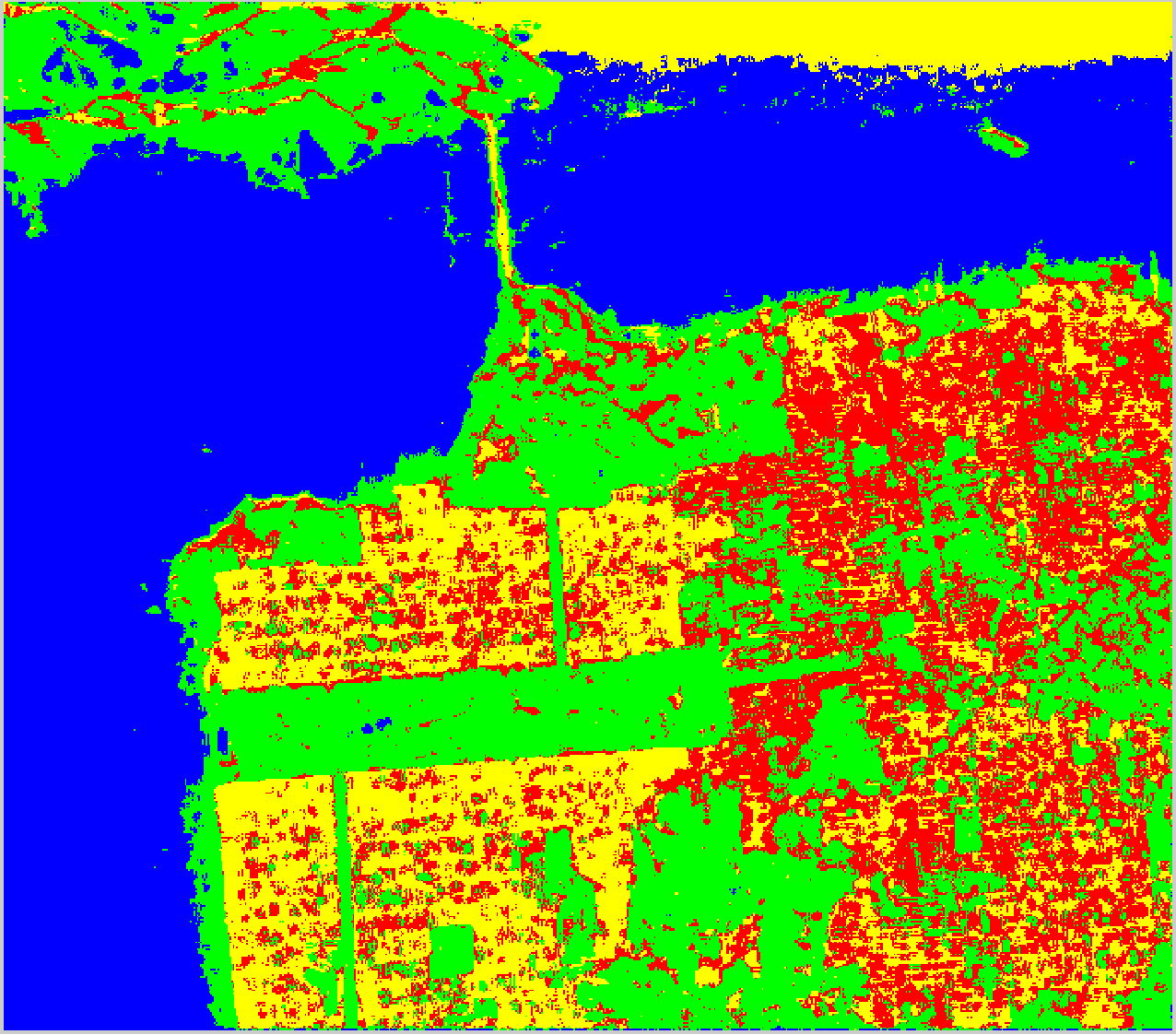}}
\label{SanFrancisco-MRF}
\subfloat[MRF]{\includegraphics[width=0.24\linewidth,height=3.8cm]{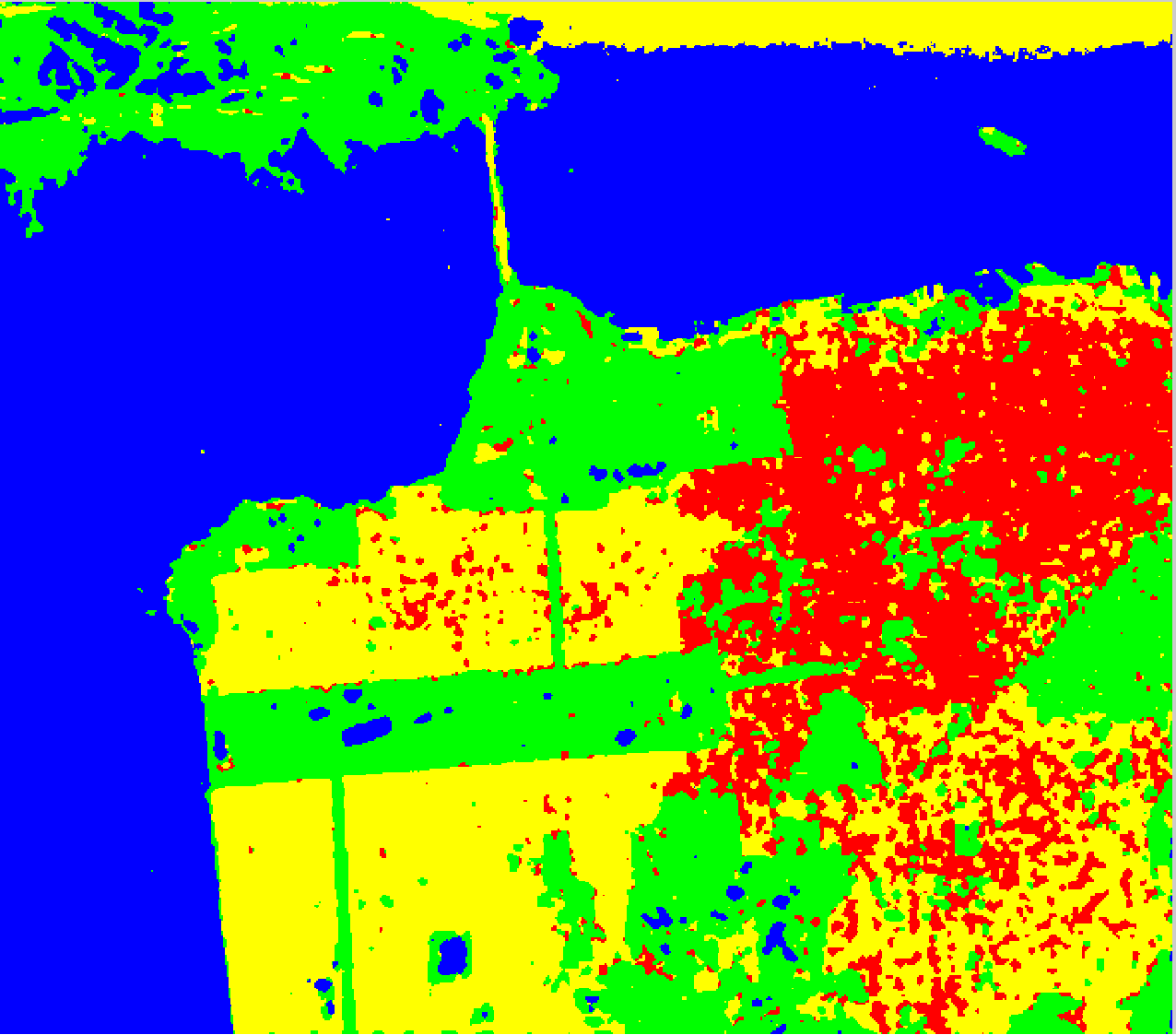}}

\label{SanFrancisco-RV-MLP}
\subfloat[RV-MLP]{\includegraphics[width=0.24\linewidth,height=3.8cm]{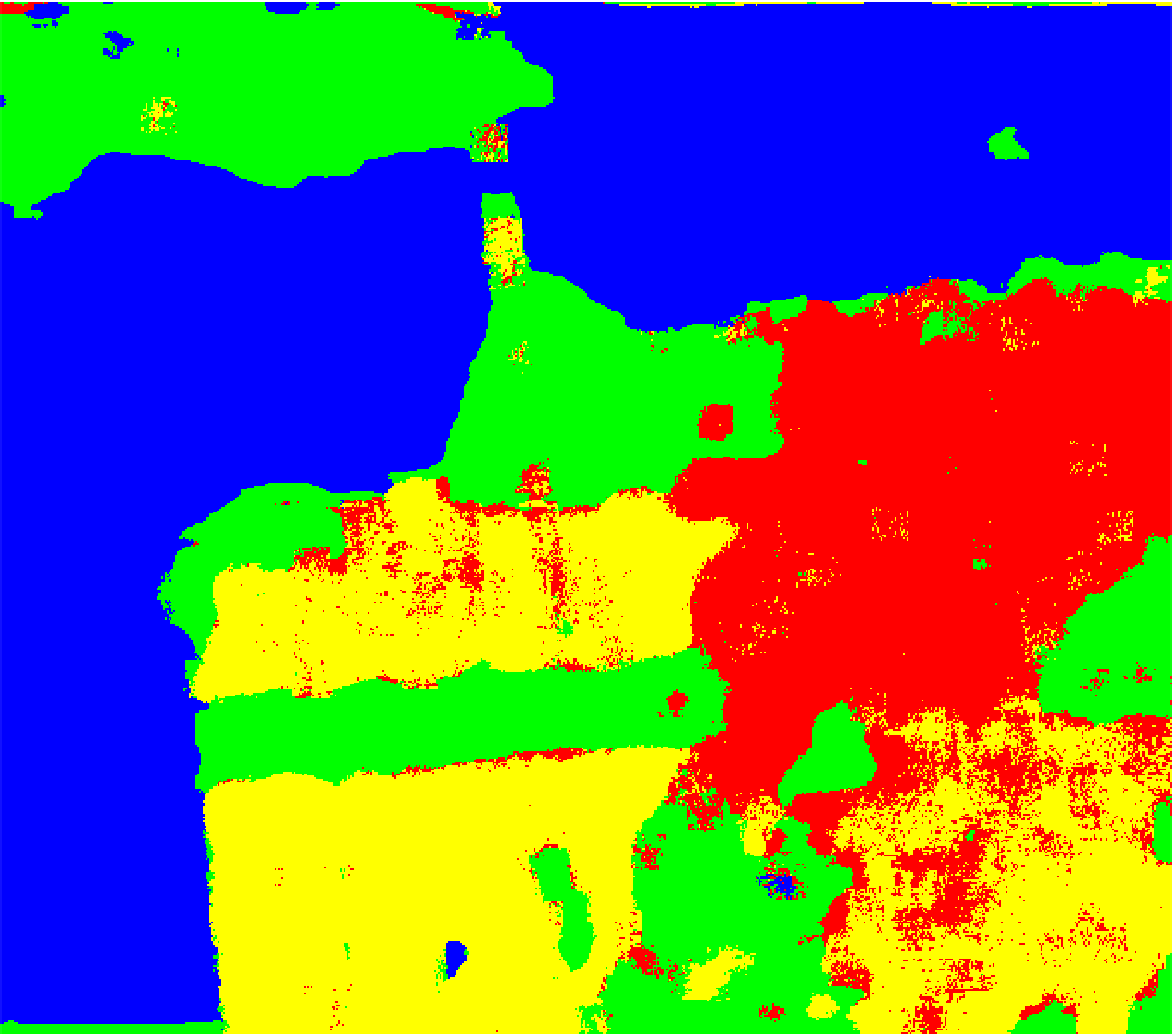}}
\label{SanFrancisco-RV-SCNN}
\subfloat[RV-SCNN]{\includegraphics[width=0.24\linewidth,height=3.8cm]{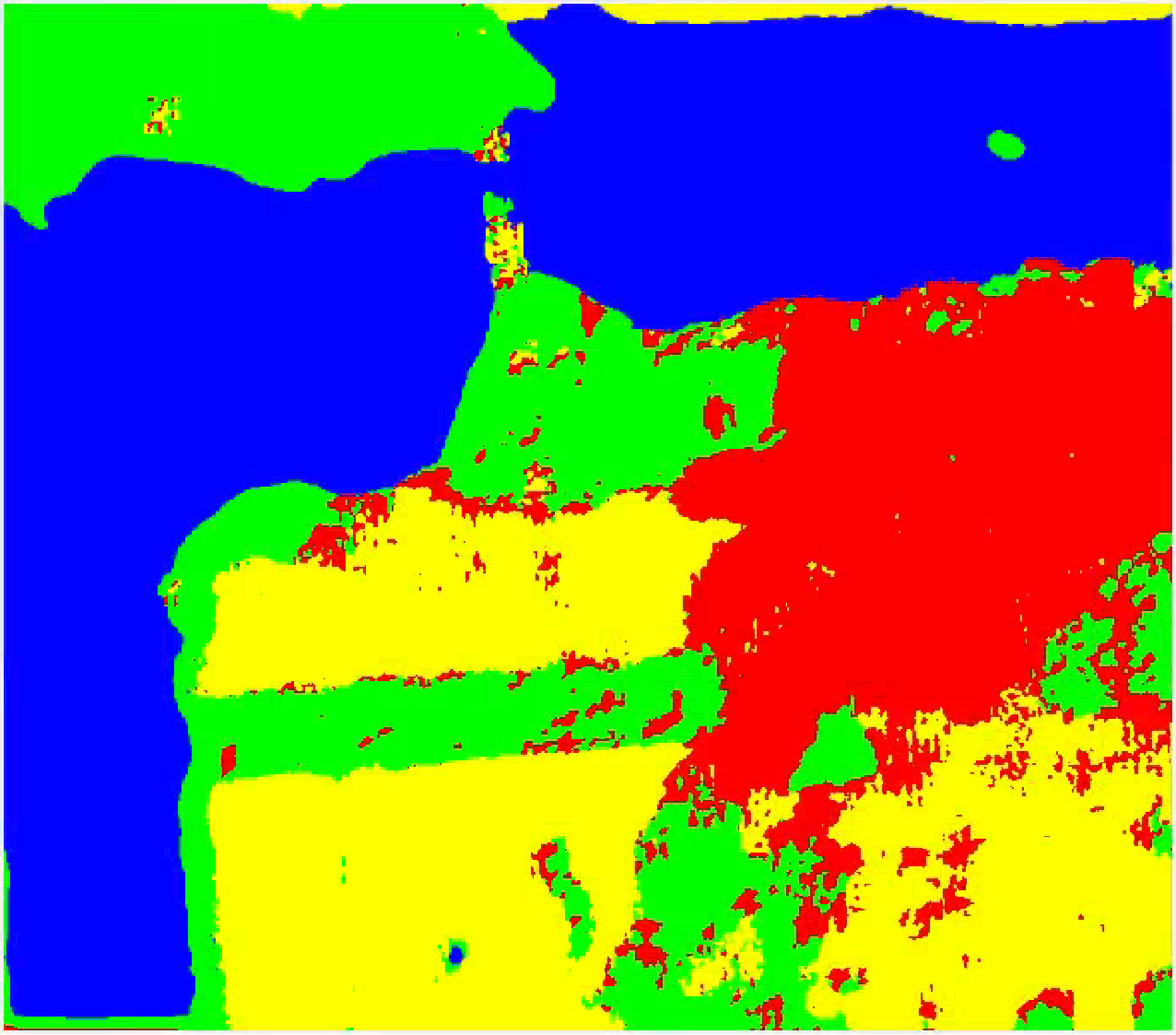}}
\label{SanFrancisco-RV-DCNN}
\subfloat[RV-DCNN]{\includegraphics[width=0.24\linewidth,height=3.8cm]{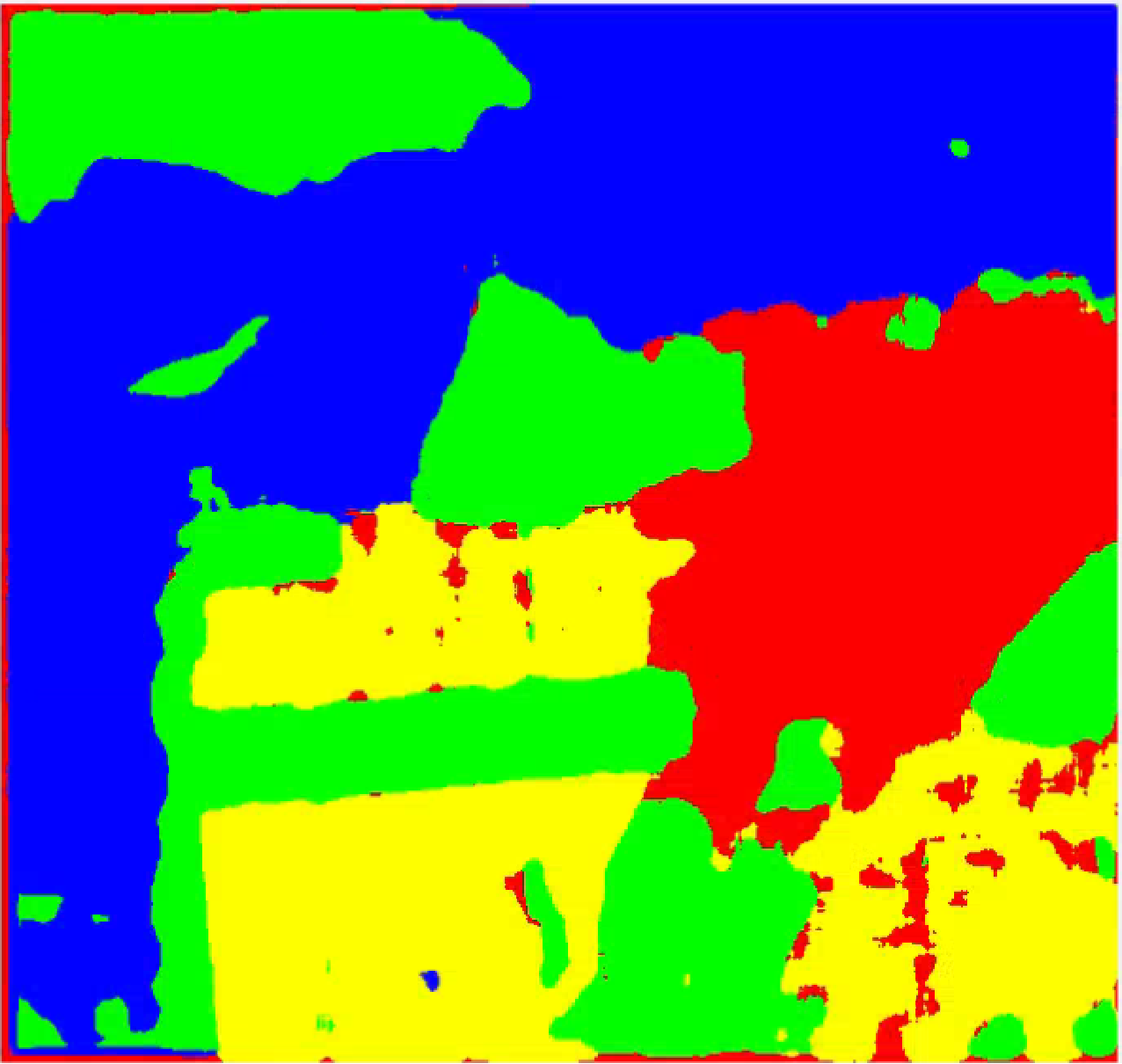}}
\label{SanFrancisco-RV-FCN}
\subfloat[RV-FCN]{\includegraphics[width=0.24\linewidth,height=3.8cm]{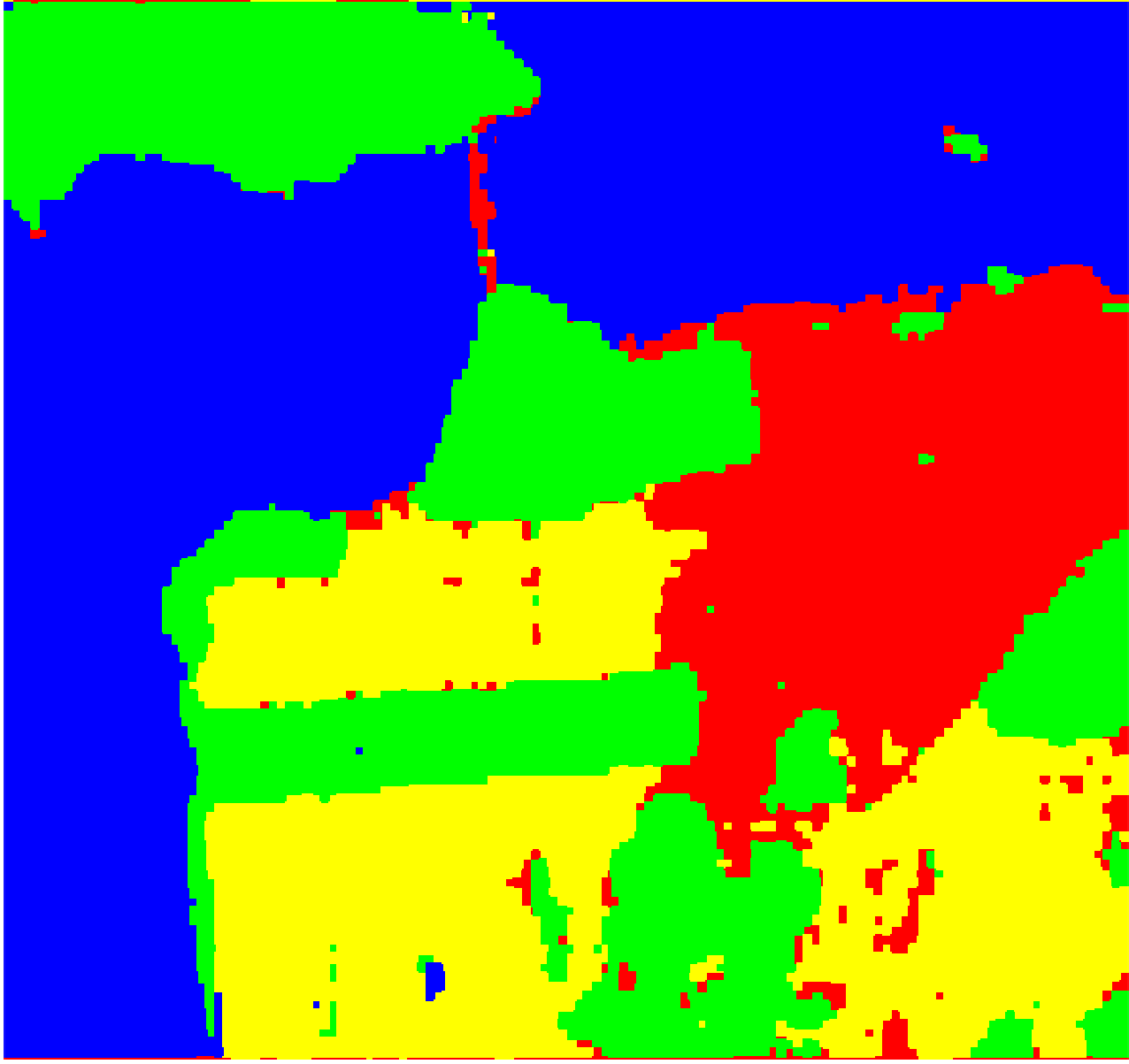}}

\label{SanFrancisco-CV-MLP}
\subfloat[CV-MLP]{\includegraphics[width=0.24\linewidth,height=3.8cm]{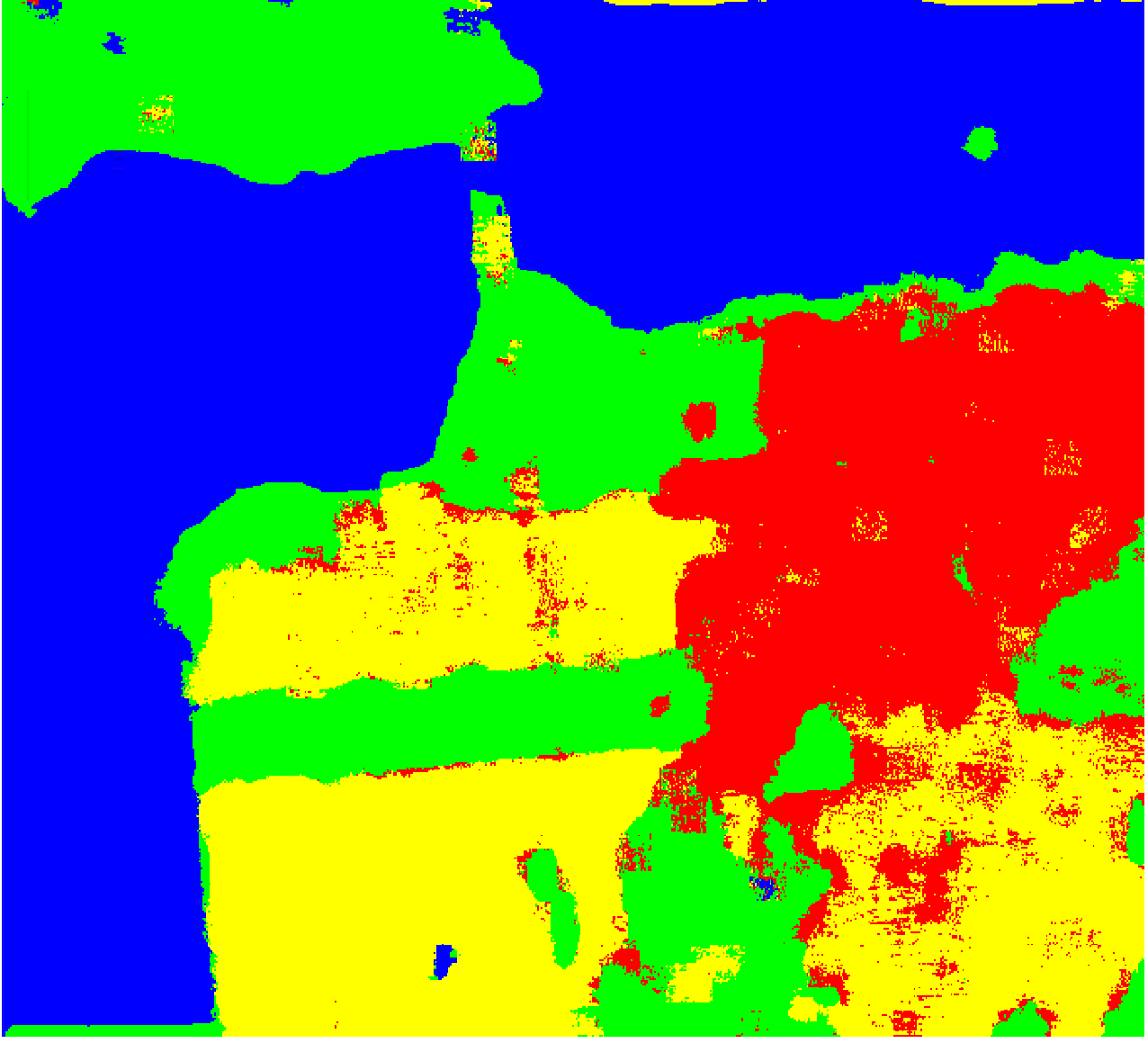}}
\label{SanFrancisco-CV-SCNN}
\subfloat[CV-SCNN]{\includegraphics[width=0.24\linewidth,height=3.8cm]{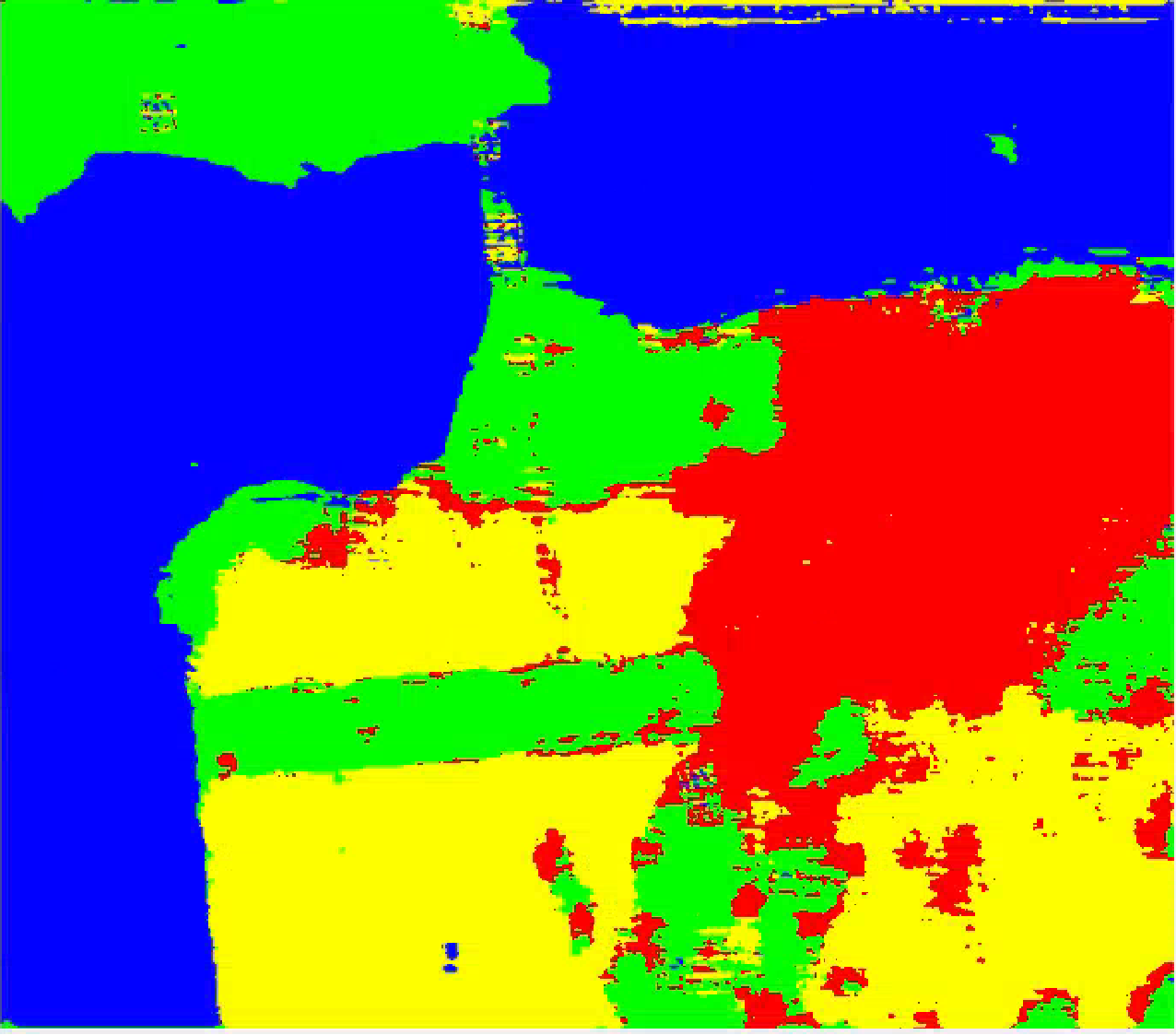}}
\label{SanFrancisco-CV-DCNN}
\subfloat[CV-DCNN]{\includegraphics[width=0.24\linewidth,height=3.8cm]{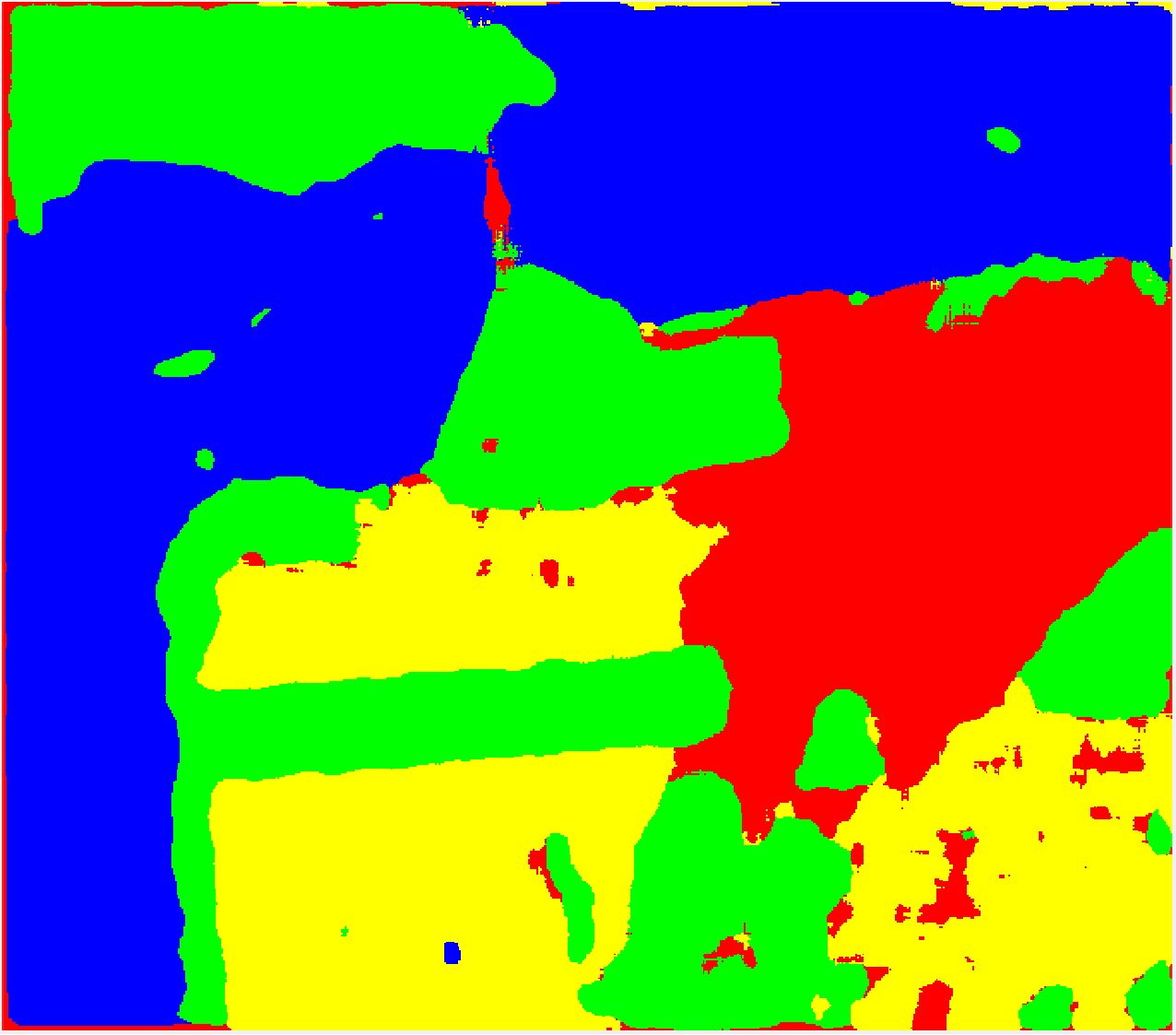}}
\label{SanFrancisco-CV-FCN}
\subfloat[CV-FCN]{\includegraphics[width=0.24\linewidth,height=3.8cm]{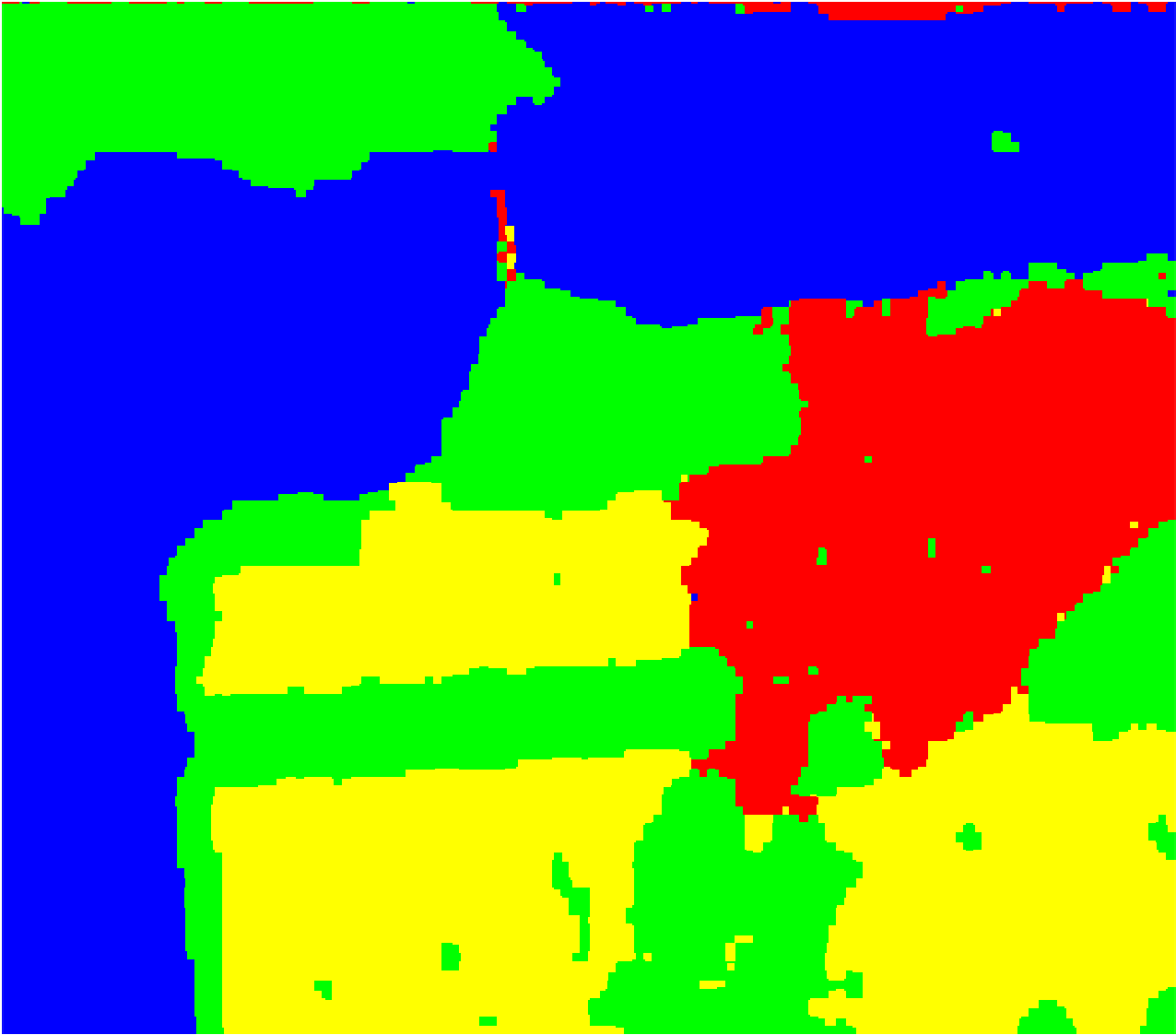}}

\caption{Classification results of San Francisco area data with different methods. }
\label{figure-reslut-SanFrancisco}
\end{figure*} 

\subsubsection{ Flevoland Benchmark Dataset Result}
For this dataset, we randomly choose 5$\%$ of available labeled samples per class for training. The classification maps obtained from all methods are shown in Fig. \ref{figure-reslut-Flevoland}, and the accuracies are reported in Table \ref{tab:Flevoland result table}. 

As shown in Fig. \ref{figure-reslut-Flevoland}(b) and Fig. \ref{figure-reslut-Flevoland}(c), classification maps obtained from SVM and Wishart are seriously affected by speckle noisy points since they only consider polarimetric information. Compared with Fig. \ref{figure-reslut-Flevoland}(b) and Fig. \ref{figure-reslut-Flevoland}(c), the classification map from MRF shown in Figure \ref{figure-reslut-Flevoland}(d) is much clearer in which misclassification pixels are significantly reduced. The reason is that MRF can embed the spatial smoothness information into the classification stage. Fig. \ref{figure-reslut-Flevoland}(e)-(l) demonstrate the classification resluts from all DL-based methods, where Fig. \ref{figure-reslut-Flevoland}(e)-(h) are the results of different RV-NNs and Fig. \ref{figure-reslut-Flevoland}(i)-(l) give the results of different CV-NNs. It can be seen that all DL-based methods outperform non-deep methods, which indicates that learning features have stronger discriminative ability than traditional features.

When comparing RV-NNs, it can be seen that RV-FCN performs best for the classification of flax class [marked by white ovals in Fig. \ref{figure-reslut-Flevoland}(e)-(h)]. In addition, among CV-NNs, CV-FCN has the highest classification accuracy on the beet class [marked by yellow ovals in Fig. \ref{figure-reslut-Flevoland}(i)-(l)] and the whole class label map of CV-FCN is much clearer than others. The above two results indicate that proposed FCN architecture is advantageous for PolSAR classification compared to other network structures, especially CNNs.

Moreover, comparing RV-NNs and CV-NNs directly, we can observe that CVNNs have better performance than their RV counterparts. For example, Fig. \ref{figure-reslut-Flevoland}(h) and Fig. \ref{figure-reslut-Flevoland}(l) are classification results from RVFCN and CV-FCN, respectively. The confusion between oats class and beet class is severe in Fig. \ref{figure-reslut-Flevoland}(h), but does not appear in Fig. \ref{figure-reslut-Flevoland}(l) [marked by sky-blue rectangles]. This confirms the effectiveness of complex-valued features with phase information for the classification of PolSAR imagery. From the overall effects depicted in Fig. \ref{figure-reslut-Flevoland}, the classification map of CV-FCN is noticeably closer to the ground truth map.
\begin{table*}[tp] 
\centering
\fontsize{6}{7.5}\selectfont 
\caption{Individual Class, Overall, Average Accuraties (\%) and Kappa Coefficient of all Competing Methods on \protect\\the San Francisco PolSAR Image}   
\label{tab:SanFran result table}  
\begin{tabular}{c|ccc|cccc|cccc}  
\hline
\toprule[0.3pt]
\hline 
\toprule[0.3pt] 
\multirow{2}{*}{Class}&  
\multicolumn{3}{c|}{Non-deep Methods}&\multicolumn{8}{c}{ DL-based Methods}\cr\cline{2-12}

&SVM  &  Wishart    &  MRF    &  RV-MLP   & RV-SCNN  & RV-DCNN   &  RV-FCN  & CV-MLP  &  CV-SCNN  & CV-DCNN  & CV-FCN \cr
\hline    
Class1 &{\bf99.99\%} & 96.47\% & 99.37\% & 99.05\% & 97.78\% & 93.35\% & 99.62\% & 99.12\% & 99.68\% & 97.59\% & 99.98\%\cr
Class2 &87.49\% & 83.23\% & 89.09\% & 91.65\% & 84.77\% & 97.10\% & 98.91\% & 93.28\% & 89.68\% & 98.69\% & {\bf99.38\%}\cr
Class3 &77.51\% & 56.45\% & 82.85\% & 90.51\% & 96.88\% & 95.16\% & 96.68\% & 92.34\% & 96.50\% & 95.75\% & {\bf99.54\%}\cr
Class4 &77.66\% & 64.54\% & 82.16\% & 96.91\% & 98.09\% & {\bf99.81\%} & 98.79\% & 96.65\% & 97.62\% & 99.45\% & 99.72\%\cr\hline 
{ OA} &88.46\% &79.13\% &90.53\% &94.96\% &94.23\% &95.74\% &98.61\% &95.77\% &96.26\% &97.74\% &{\bf99.69\%}\cr
{ AA} &85.66\% &75.17\% &88.37\% &94.53\% &93.96\% &96.36\% &98.49\% &95.35\% &95.87\% &97.87\% &{\bf99.65\%}\cr
{ $\kappa$} &0.8394 &0.7128 &0.8683 &0.9303 &0.9203 &0.9416 &0.9807 &0.9415 &0.9483 &0.9687 &{\bf0.9957}\cr
\hline 
\toprule[0.3pt]
\hline
\toprule[0.3pt] 
\end{tabular}  
\end{table*}  

The evaluation indices of all methods are listed in Table \ref{tab:Flevoland result table}. As shown in Table \ref{tab:Flevoland result table}, MRF and all DL-based methods achieve OA exceeding 90$\%$. All CV-NNs methods achieve better performance than their RV counterparts in terms of all evaluation metrics. In particular, the largest part of changes in metrics is AA values. Furthermore, CV-FCN outperforms other compared methods in terms of three quantitative criteria. Although compared with CV-DCNN, CV-FCN attains only 0.97$\%$ improvement in terms of OA, all classes besides oats show comparable or higher accuracy which is consistent with the results shown in Fig. \ref{figure-reslut-Flevoland}. In summary, from Fig. \ref{figure-reslut-Flevoland} and Table \ref{tab:Flevoland result table}, for Flevoland Benchmark dataset, CV-FCN achieves the best performance compared with other methods and has powerful ability to distinguish different terrain categories.
\begin{figure*}[t] 
\centering
\label{Germany-GT}
\subfloat[GroundTruth]{\includegraphics[width=0.24\linewidth,height=3.8cm]{./GroundTruth-Germany.png}}
\label{Germany-SVM}
\subfloat[SVM]{\includegraphics[width=0.24\linewidth,height=3.8cm]{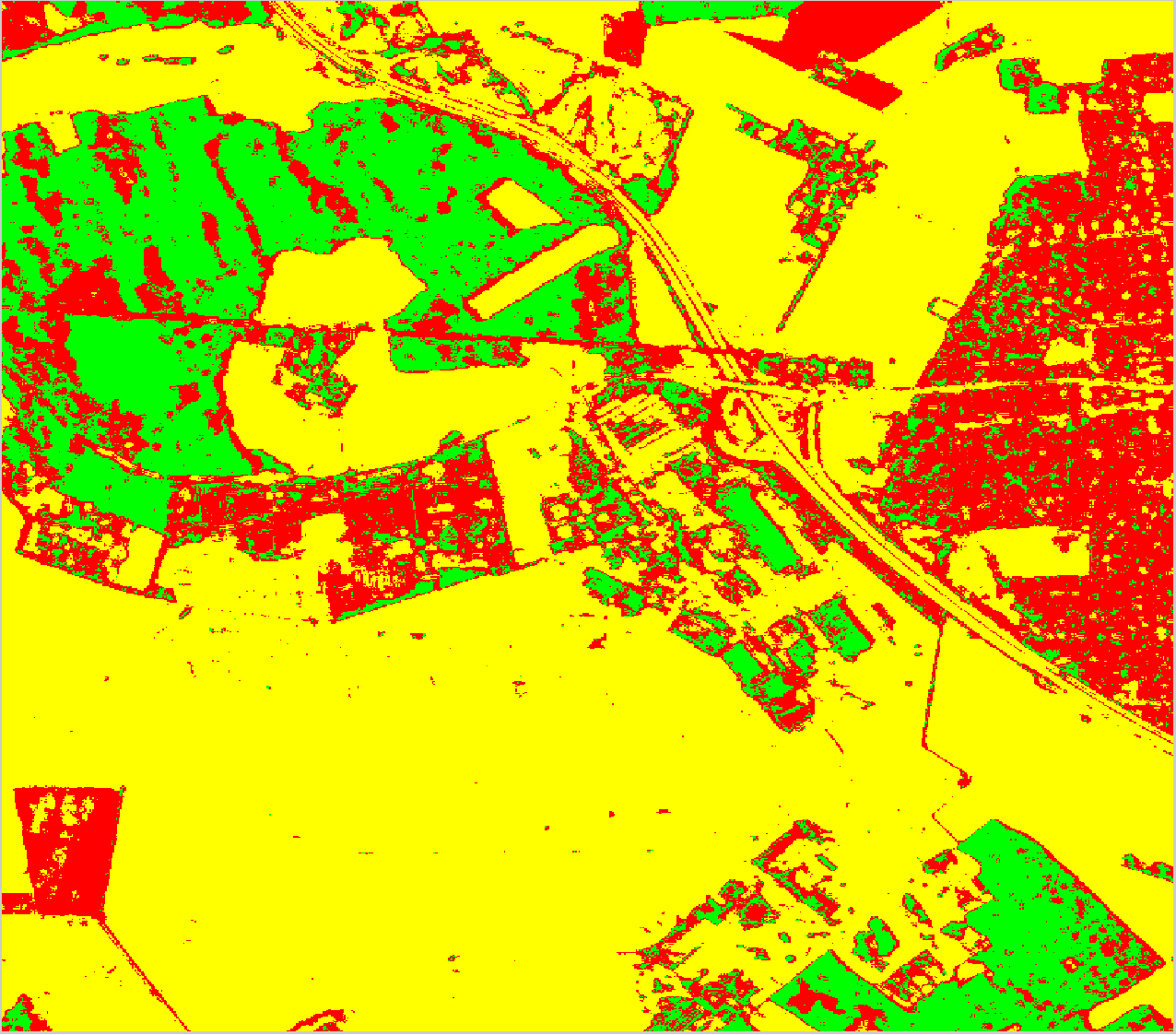}}
\label{Germany-Wishart}
\subfloat[Wishart]{\includegraphics[width=0.24\linewidth,height=3.8cm]{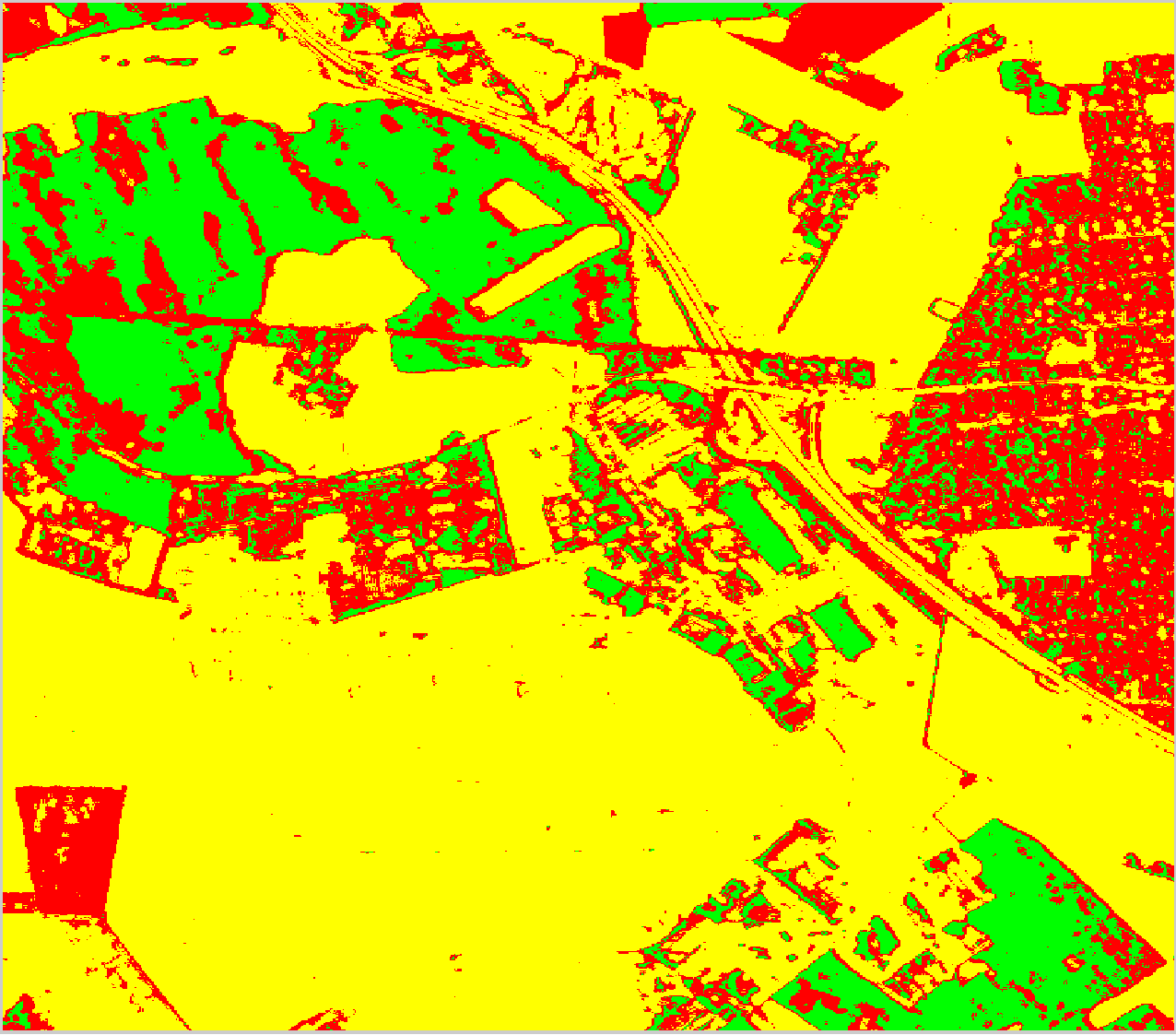}}
\label{Germany-MRF}
\subfloat[MRF]{\includegraphics[width=0.24\linewidth,height=3.8cm]{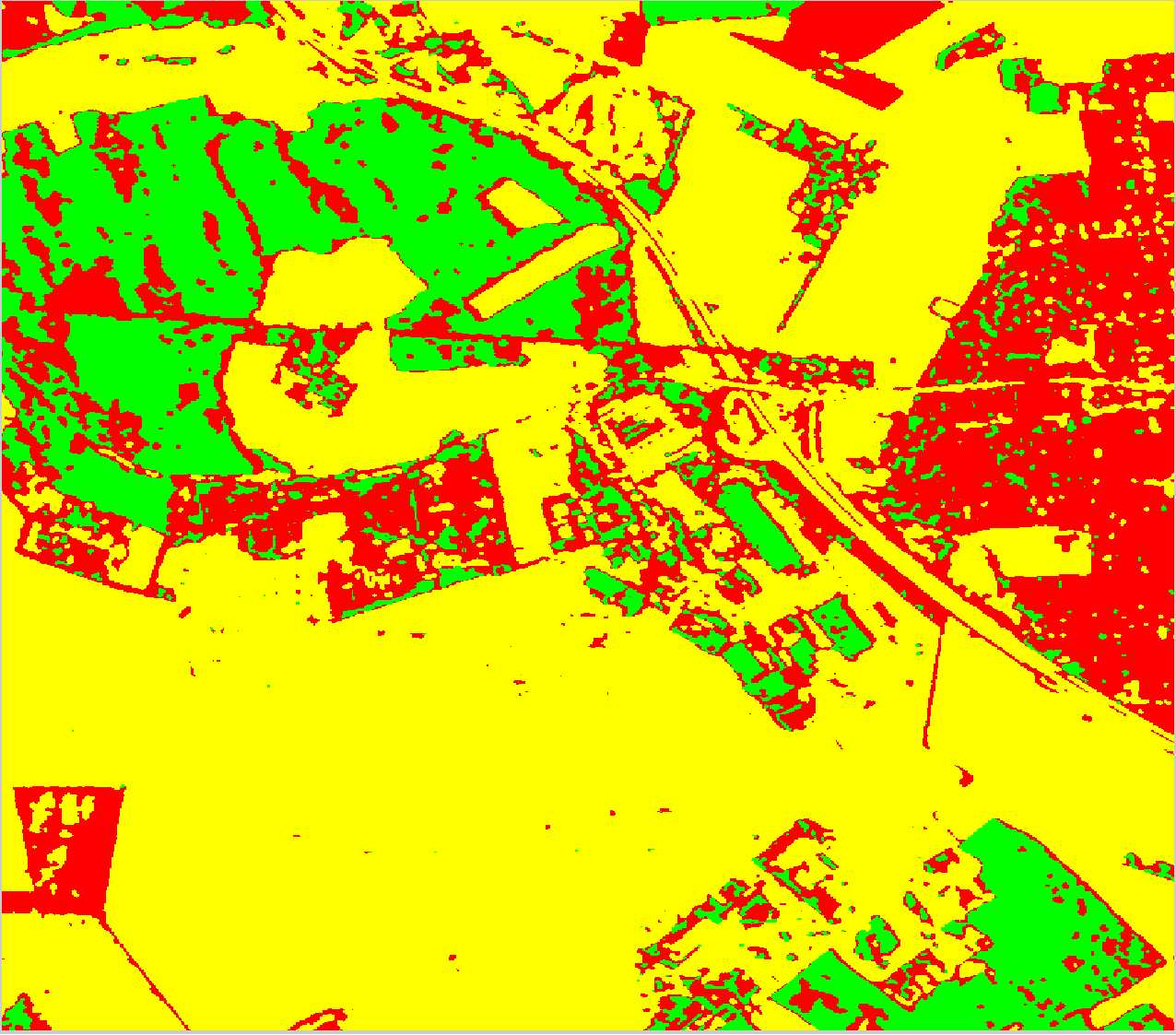}}

\label{Germany-RV-MLP}
\subfloat[RV-MLP]{\includegraphics[width=0.24\linewidth,height=3.8cm]{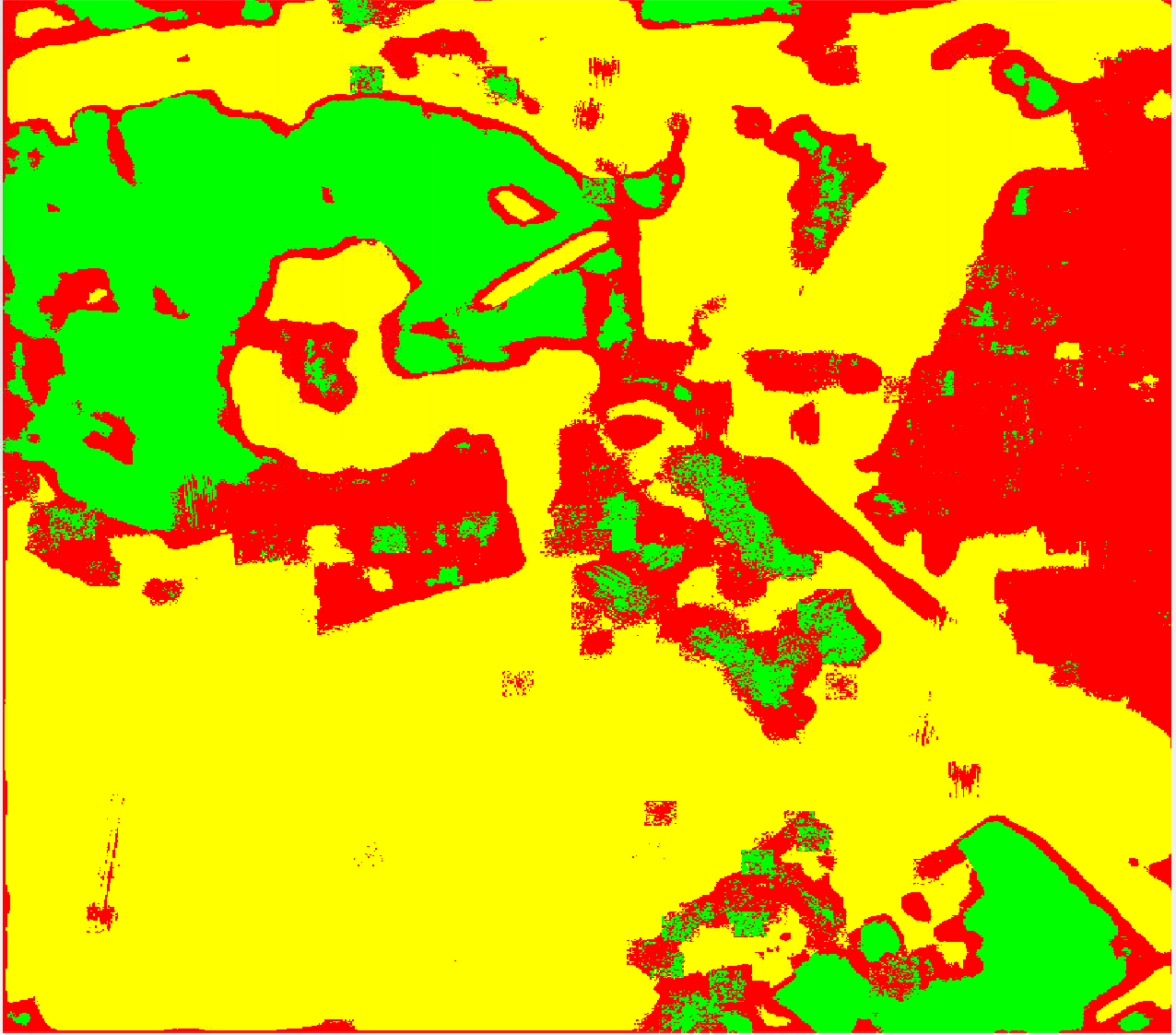}}
\label{Germany-RV-SCNN}
\subfloat[RV-SCNN]{\includegraphics[width=0.24\linewidth,height=3.8cm]{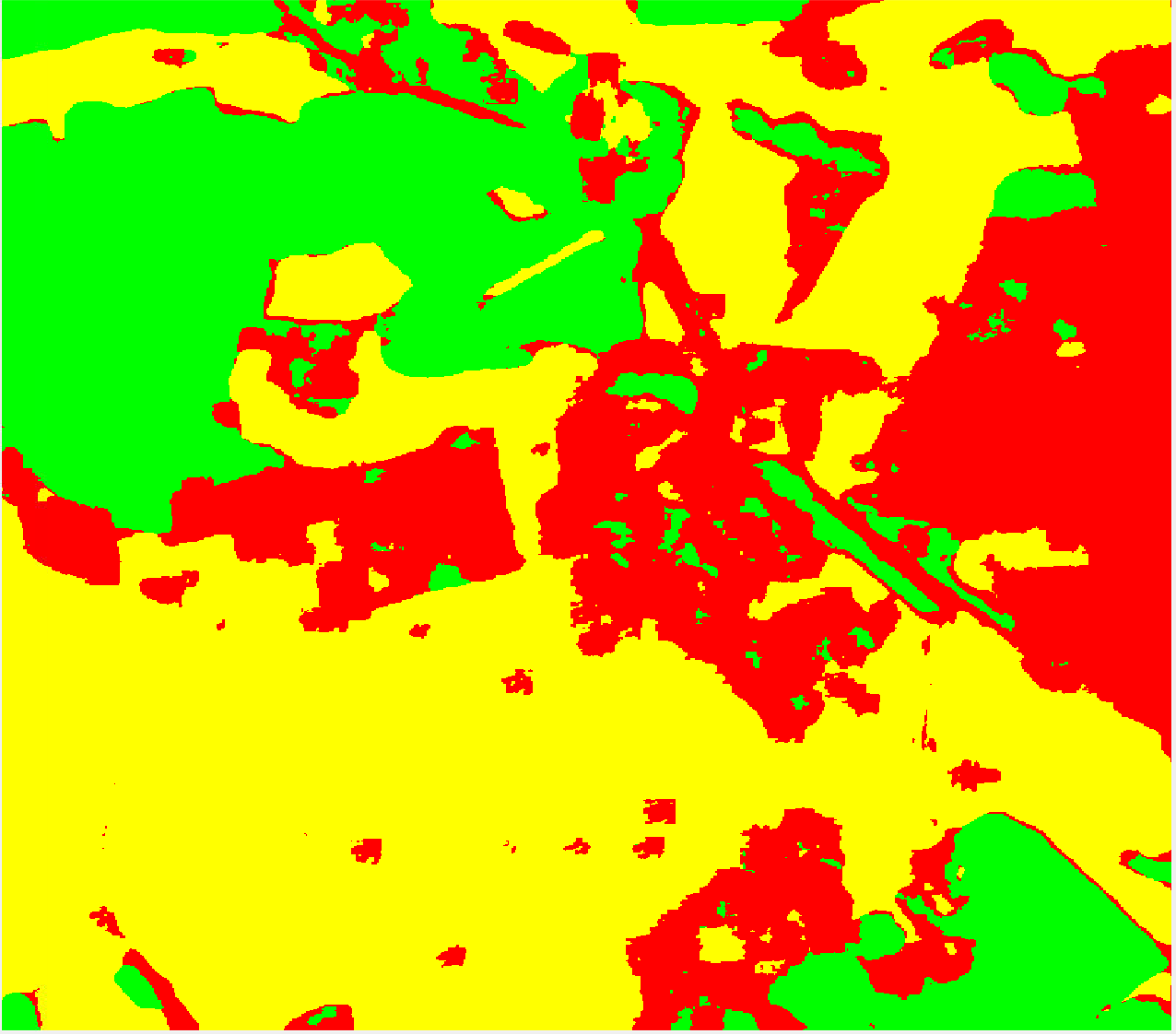}}
\label{Germany-RV-DCNN}
\subfloat[RV-DCNN]{\includegraphics[width=0.24\linewidth,height=3.8cm]{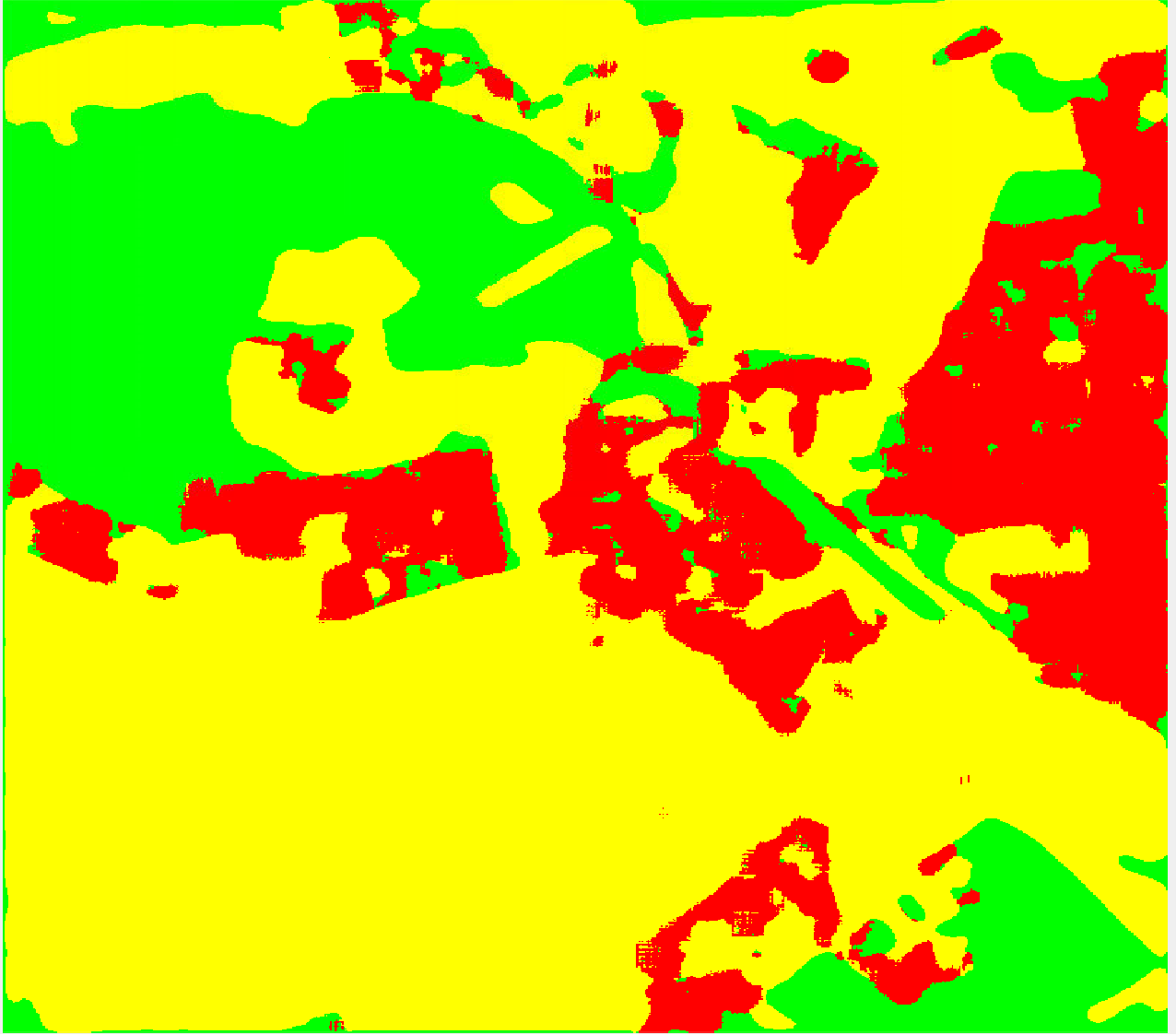}}
\label{Germany-RV-FCN}
\subfloat[RV-FCN]{\includegraphics[width=0.24\linewidth,height=3.8cm]{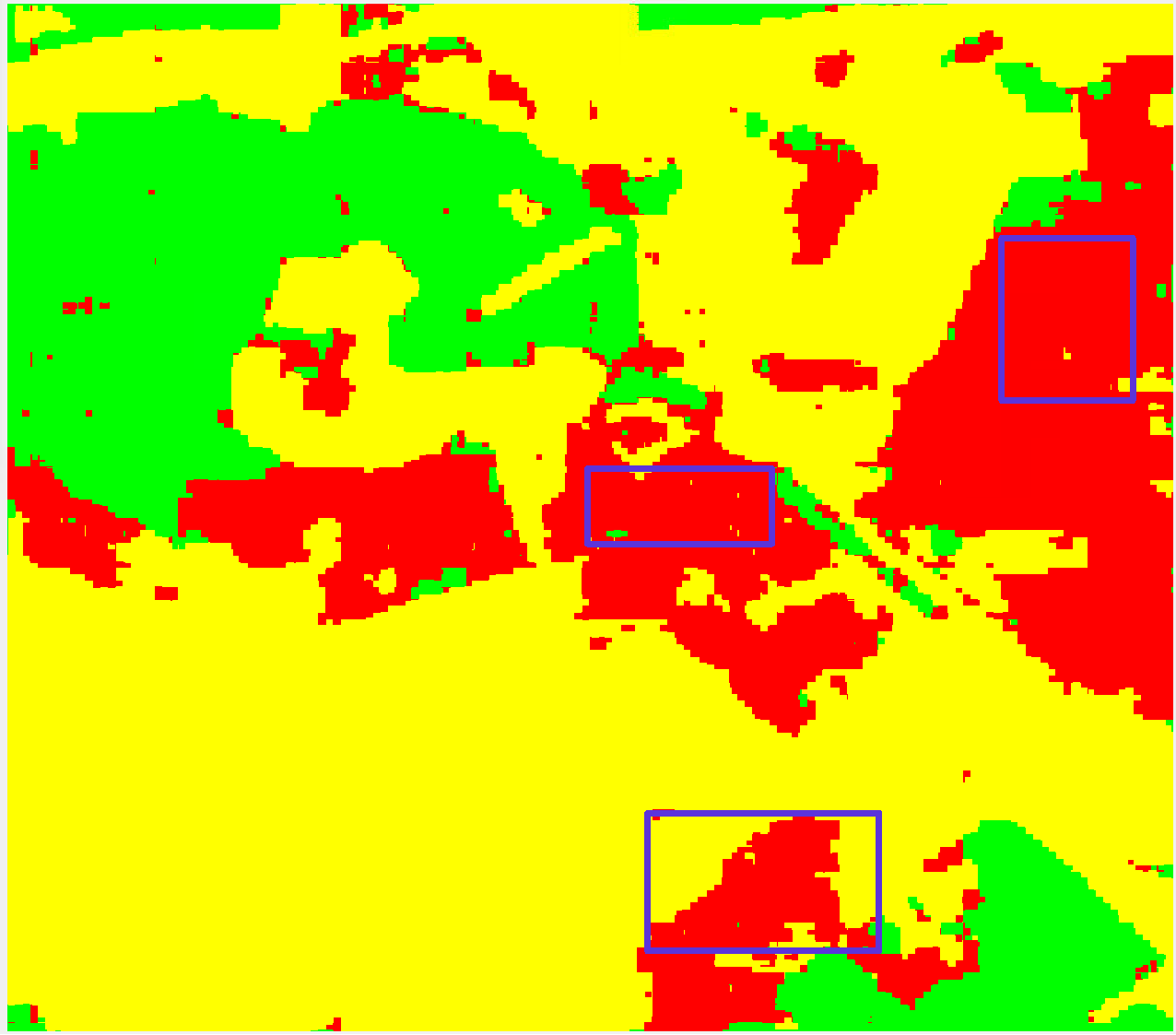}}

\label{Germany-CV-MLP}
\subfloat[CV-MLP]{\includegraphics[width=0.24\linewidth,height=3.8cm]{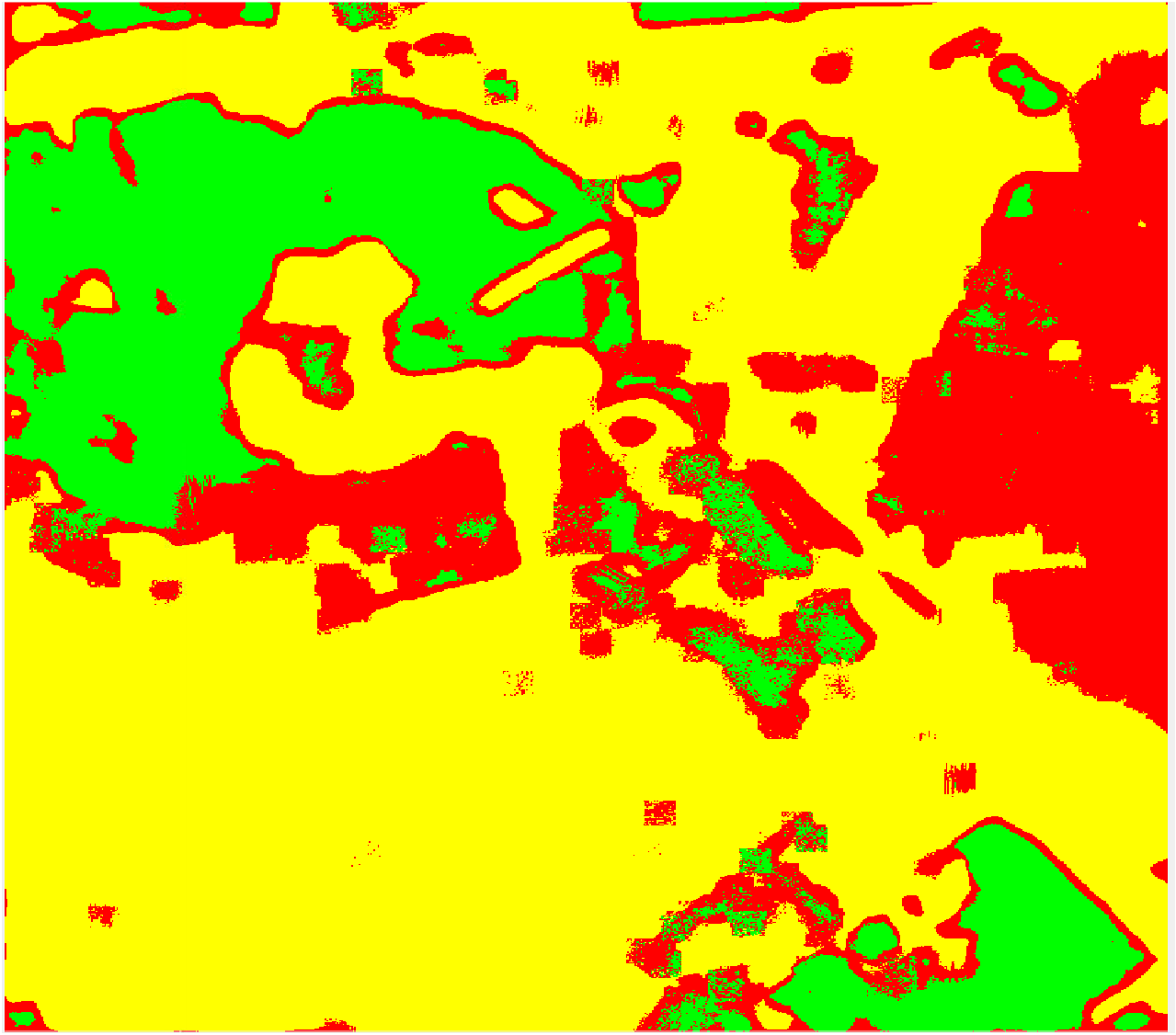}}
\label{Germany-CV-SCNN}
\subfloat[CV-SCNN]{\includegraphics[width=0.24\linewidth,height=3.8cm]{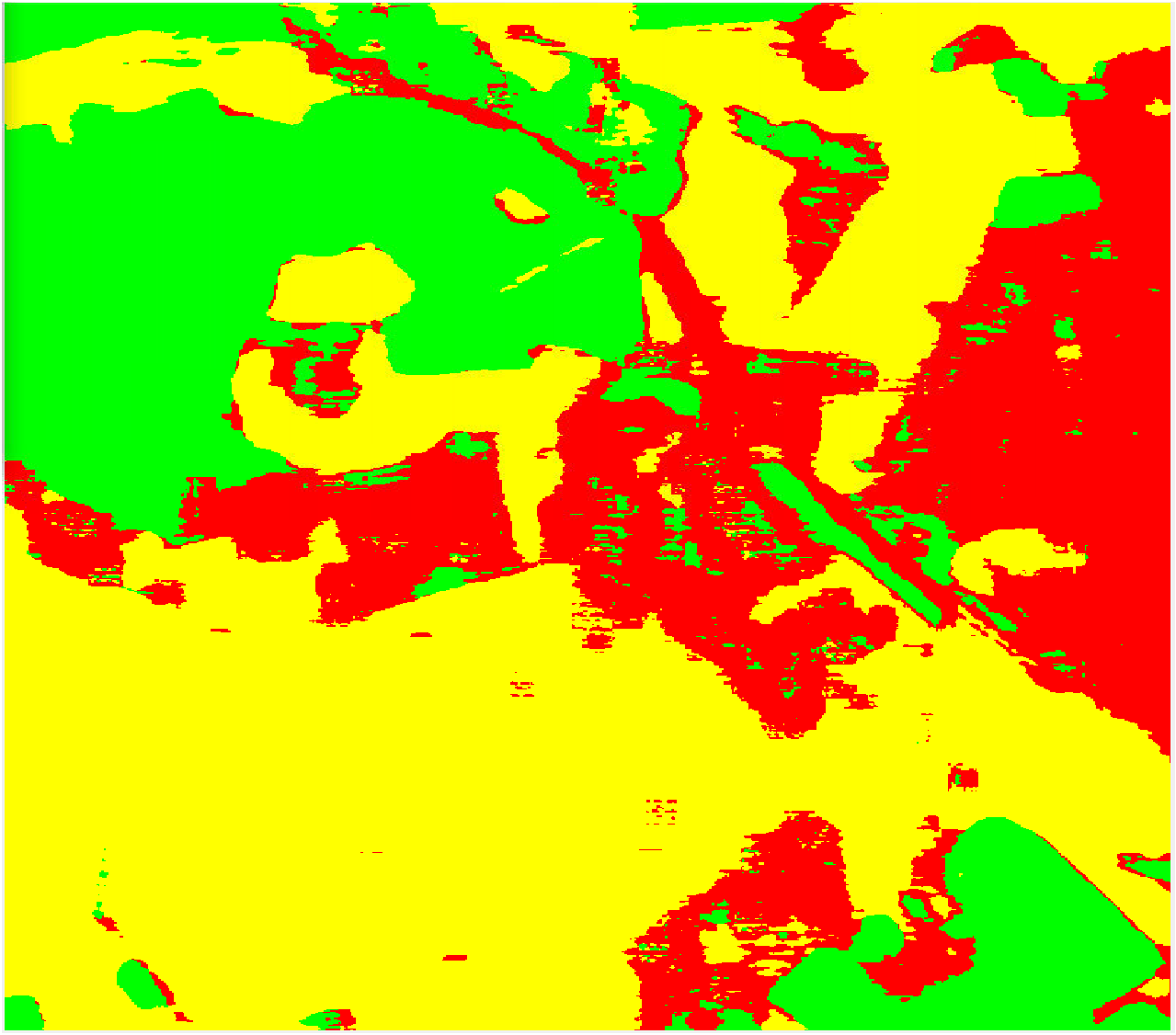}}
\label{Germany-CV-DCNN}
\subfloat[CV-DCNN]{\includegraphics[width=0.24\linewidth,height=3.8cm]{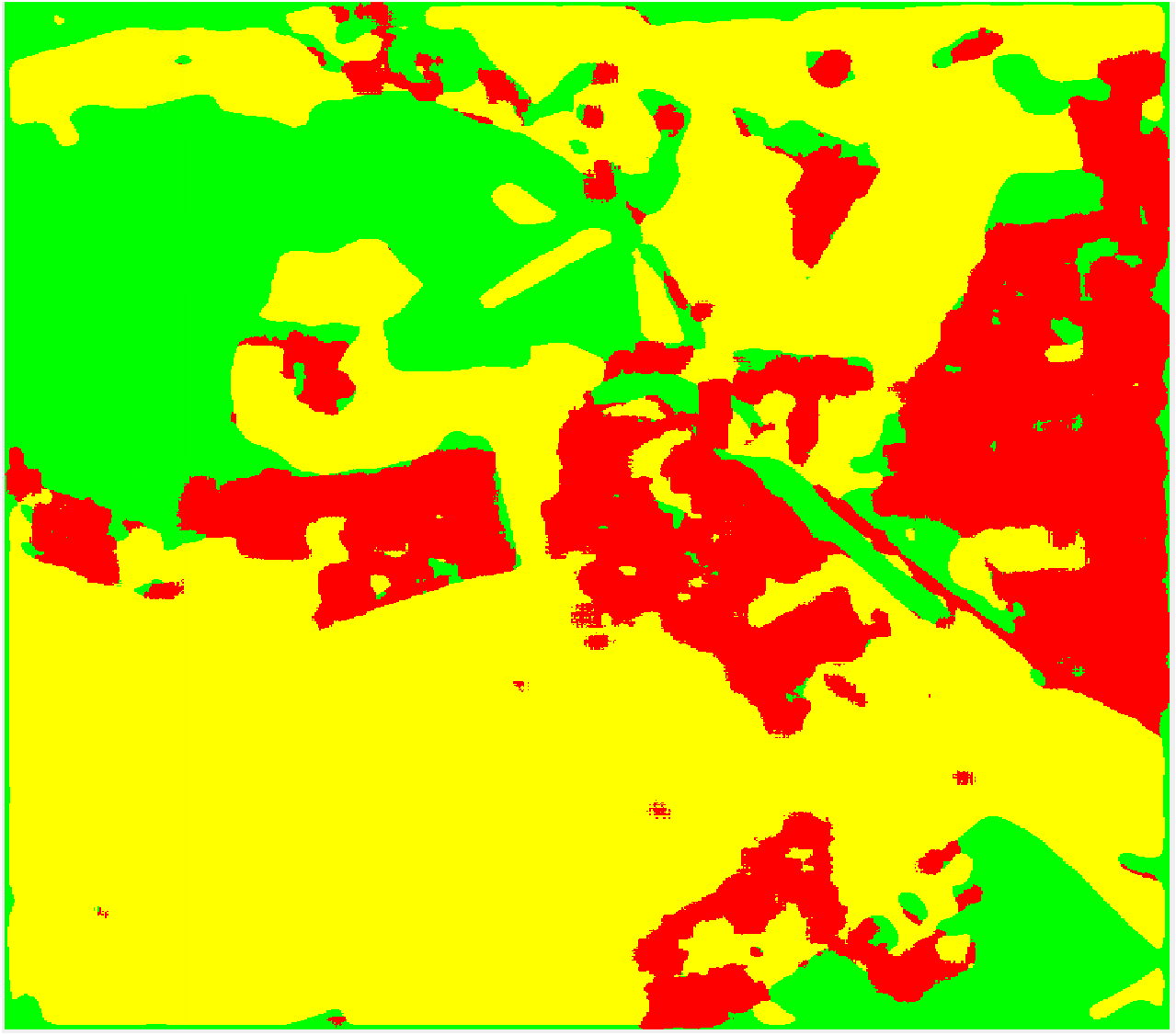}}
\label{Germany-CV-FCN}
\subfloat[CV-FCN]{\includegraphics[width=0.24\linewidth,height=3.8cm]{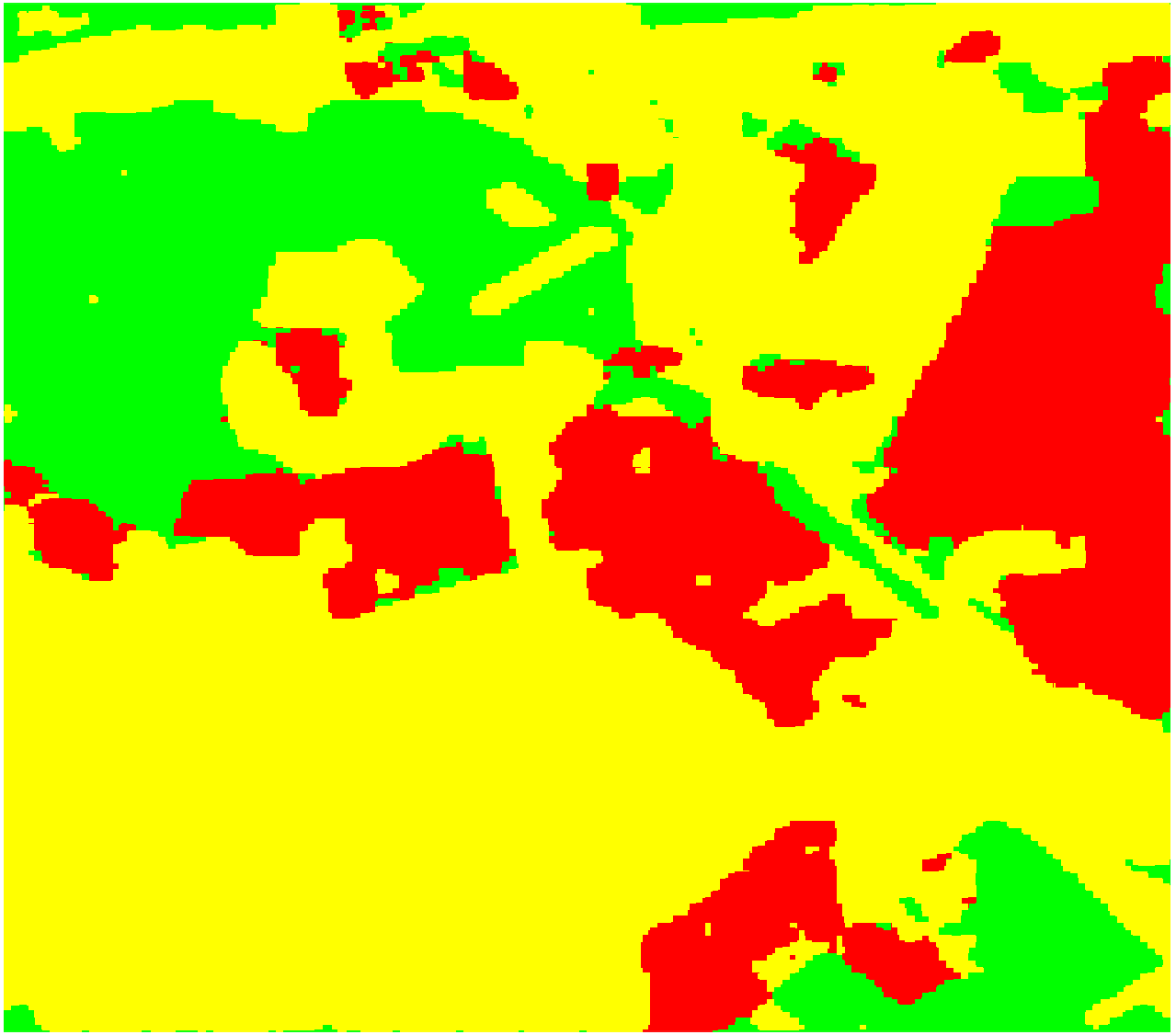}}

\caption{Classification results of Oberpfaffenhofen area data with different methods.}
\label{figure-reslut-Germany}
\end{figure*} 
\begin{table}[tp] 
\centering
\fontsize{7}{8.5}\selectfont 
\caption{Individual Class, Overall, Average Accuraties (\%) and \protect\\Kappa Coefficient of all Competing Methods on \protect\\the Oberfaffenhofen PolSAR Image}  
\label{tab:Germany result table}
\begin{tabular}{c|c|ccc}
\hline
\toprule[0.3pt]
\hline
\toprule[0.3pt]
\multicolumn{2}{c|}{ Methods} &  OA &  AA & \bf $\kappa$ \\
\hline
\multirow{3}{*}{Non-deep} & SVM  & 82.36\% & 76.10\% & 0.6927 \\
& Wishart  & 80.90\% & 74.11\% & 0.6671 \\
& MRF  & 83.70\% & 77.83\% & 0.7156 \\
\hline
\multirow{8}{*}{DL-based} & RV-MLP  & 89.36\% & 86.27\% & 0.8186 \\
& RV-SCNN  & 93.35\% & 94.16\% & 0.8889 \\
& RV-DCNN & 94.75\% & 93.42\% & 0.9097 \\
& RV-FCN  & 95.51\% & 94.38\% & 0.9227 \\ \clineB{2-5}{1}
& CV-MLP  & 90.02\% & 86.59\% & 0.8279 \\
& CV-SCNN  & 93.52\% & 93.46\% & 0.8909 \\
& CV-DCNN & 95.76\% & 94.87\% & 0.9274 \\
& CV-FCN  & {\bf97.26\%} & {\bf96.38\%} & {\bf0.9531} \\
\hline
\toprule[0.3pt]
\hline
\toprule[0.3pt]
\end{tabular}
\end{table} 
\begin{figure*}[ht]
\centering
\subfloat{\includegraphics[width=16.5cm,height=5.3cm]{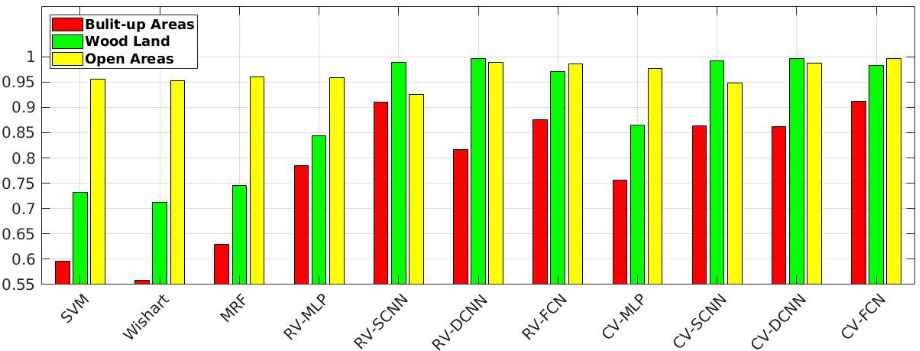}}
\caption{Classification accuracies of Oberpfaffenhofen area classes with different methods.}
\label{fig:Germany class}
\end{figure*} 
\subsubsection{ San Francisco Dataset Result} 
For the San Francisco dataset, we randomly choose 1$\%$ labeled pixels per class for training, and the remaining for testing. The classification results obtained from all methods are shown in Fig. \ref{figure-reslut-SanFrancisco}, and Table \ref{tab:SanFran result table} reports the evaluation metrics of them.

Fig. \ref{figure-reslut-SanFrancisco}(b) and Fig. \ref{figure-reslut-SanFrancisco}(c) give classification results using SVM and Wishart classifier. It can be viewed that vegetation, low-density urban and high-density urban are severely mixed and there are many isolated pixels in images. Fig. \ref{figure-reslut-SanFrancisco}(d) shows the classification result obtained from MRF, where the confusion between low-density urban and high-density urban is not severe. In addition, misclassification occurs much slightly than the previous two methods, as MRF can consider the spatial information to obtain a smoother classification map. Nevertheless, due to limited discriminative features, it is difficult for non-deep methods to distinguish complex backscatters, especially for vegetation and urban areas. 

Fig. \ref{figure-reslut-SanFrancisco}(e)-(l) show the classification results of DL-based methods. From Figure \ref{figure-reslut-SanFrancisco}(e)-(h), it is worth noting that RV-FCN outperforms than the other three methods, where boundaries between categories are much clearer. In addition, from Fig. \ref{figure-reslut-SanFrancisco}(i)-(l), CV-FCN yields the optimal visual effect compared with other CV-NNs. All of the above analysis about NNs demonstrate the effectiveness of proposed FCN structure, which can capture more discriminative features and effectively incorporate more spatial information. Furthermore, from Fig. \ref{figure-reslut-SanFrancisco}, CV-FCN achieves the best performance, which illustrates that both the FCN structure and the phase information have contributions to improve classification accuracies.

As Table \ref{tab:SanFran result table} shows, CV-FCN achieves the highest classification accuracy. The OA value of CV-FCN is about 3$\%$, 4$\%$ higher than CV-DCNN and CV-SCNN, respectively, which indicates that proposed FCN structure is suitable for PolSAR data. In addition, the results of CV-FCN are slightly better than RV-FCN. This confirms that the phase information plays an important role in the improvement of classification accuracy. Furthermore, CV-FCN yields the highest accuracies in all evaluation metrics, which is coincident with results in Fig. \ref{figure-reslut-SanFrancisco}.
\subsubsection{ Oberpfaffenhofen Dataset Result}
For the Oberpfaffenhofen dataset, we also choose 1$\%$ of pixels with ground-truth class labels for training. Fig. \ref{figure-reslut-Germany} shows the visual classification results. The overall evaluation indices are given in Table \ref{tab:Germany result table} and Fig. \ref{fig:Germany class} demonstrates the classification accuracies of every class obtained from different methods.

From Table \ref{tab:Germany result table}, CV-FCN achieves the best performance in terms of all metrics. The accuracies of non-deep methods are poor which are all below 85$\%$ in terms of OA. This might be a result of limited labeled pixels as prior information and little discriminative features. It can be also seen that CV-NNs have better performance than their RV counterparts. However, this superiority is not prominent. In terms of OA, CV-MLP, CV-SCNN, CV-DCNN, and CV-FCN are only 0.66$\%$, 0.17$\%$, 1.01$\%$, and 1.75$\%$ higher than RV-MLP, RV-SCNN, RV-DCNN, and RV-FCN, respectively.

Fig. \ref{figure-reslut-Germany}(e)-(h) and Fig. \ref{figure-reslut-Germany}(i)-(l) show classification results of RV-NNs and CV-NNs, respectively. As shown in Fig. \ref{figure-reslut-Germany}(e)-(h), the classification result using RV-FCN is much clear than the other three, especially in the purple boxes which are noticeably closer to the ground truth map. This situation also occurs in the comparison among CV-NNs. Comparing all results shown in Fig. \ref{figure-reslut-Germany}, for this dataset, the classification map of CV-FCN is the best close to the ground truth map.

As shown in Fig. \ref{fig:Germany class}, non-deep methods have poor abilities to distinguish built-up areas and wood land. That can also be observed in Fig. \ref{figure-reslut-Germany}(b)-(d) where the misclassification in whole images is severe and all classification maps have many isolated pixels. However, the accuracies of wood land and open areas using DCNNs and FCNs are all over 95$\%$, which illustrates the discriminative feature learning ability of deep networks. In addition, CV-FCN is advantageous in terms of accuracies for all categories relative to other methods, which demonstrates its effectiveness in extracting more discriminative features. Overall, the above analyses exactly illustrate that CV-FCN can exhibit better contextual consistency and extracts more discriminative features for PolSAR image classification.

As the above comparisons demonstrate, the classification performance of CV-FCN exceeds other methods on all PolSAR datasets. On the one hand, CV-FCN improves the classification accuracy effectively compared to its RV counterpart (i.e., RV-FCN). Meanwhile, this conclusion is also established in other network structures, which confirms the validity of complex-valued features containing the phase information. On the other hand, compared with CNN structures, CV-FCN can perform more coherent labeling and show better robustness to the speckle noise, while resulting in smooth classification with precise location. This demonstrates the effectiveness of CV-FCN architecture in considering more spatial information and extracting more discriminative features for PolSAR image classification. 

\section{Conclusion}
\label{sec:conclusions}
In this paper, a novel complex-valued (CV) pixel-level model called CV-FCN has been proposed for PolSAR image classification, which obtains better performance compared with non-deep methods and other DL-based methods. This model integrates the feature extraction module and the classification module in a unified framework. For learning meaningful features faster, a new complex-valued weight initialization scheme is proposed to initialize CV-FCN. It greatly facilitates faster learning for this network and is beneficial to improve CV-FCN performance. Then, different-level and robust CV features that retain more discriminative information are extracted via CV-FCN. Particularly, a new complex upsampling scheme in CV-FCN is proposed to output CV predicted labeling. It also recovers rich spatial information with max locations maps to alleviate the problem of speckle noise. Furthermore, a novel average cross-entropy loss function is presented for more precise CV-FCN optimization. The proposed CV-FCN model can enable pixel-to-pixel classification results directly using the PolSAR CV data without any data projection. Moreover, it automatically learns a higher-level feature representation and fuses multi-level features for accurate categories identification. Experimental results on real benchmark PolSAR images show that CV-FCN achieves comparable or better results than the comparing models.  

In the future, this work may be continued with the following ideas: 1) Some experiments demonstrated the effectiveness of the new complex-valued weight initialization scheme to initialize CV-FCN for PolSAR image classification. However, it still needs strong cues to prove the superiority and some visualization to observe the difference; 2) With the limitation of available PolSAR datasets and high-quality training datasets, training rather deep complex-valued networks devoted to PolSAR classification is very challenging, often yielding the risk of overfitting and model collapse. Moreover, data augmentation strategies for natural images are generally not suited for PolSAR images to enlarge training datasets because of the difference of imaging mechanisms. So it appears that an available data augmentation strategy is urgently necessary to tackle the above issues.

\raggedend 
\end{document}